\documentclass[preprint,11pt,nofootinbib]{revtex4-1}

\usepackage{graphicx}
\usepackage{dcolumn} 
\usepackage{bm}
\usepackage{bbm}
\usepackage{amssymb}
\usepackage{amsmath}
\usepackage{epsfig}    
\usepackage{color}
\usepackage{slashed}
\usepackage{hyperref}
\usepackage{multirow}
\newcommand{\Tr}{\text{Tr}}

\usepackage{ulem}

\allowdisplaybreaks

\begin{document}

%=================================================================================================
\title{CP Violation in a Model with Higgs Triplets}
\preprint{OU-HET-1168}

\author{Ting-Kuo Chen}
\affiliation{Department of Physics and Center for Theoretical Physics, National Taiwan University, Taipei, Taiwan 10617, ROC}
\affiliation{Department of Physics, University of Wisconsin-Madison, Madison, WI 53706, USA}

\author{Cheng-Wei Chiang}
\affiliation{Department of Physics and Center for Theoretical Physics, National Taiwan University, Taipei, Taiwan 10617, ROC}
\affiliation{Physics Division, National Center for Theoretical Sciences, Taipei, Taiwan 10617, ROC}

\author{Kei Yagyu}
\affiliation{Department of Physics, Osaka University, Toyonaka, Osaka 560-0043, Japan}

\begin{abstract}

We discuss CP-violation in a model with a real and a complex isospin triplet Higgs fields without introducing any symmetries except for the electroweak gauge symmetry. 
This corresponds to the minimal extension of the Higgs sector with the following properties: (i) providing new source of CP violation, 
(ii) absence of quark flavor changing neutral currents at tree level, and (iii) enabling the electroweak rho parameter to be unity at tree level in the scenario without imposing any new symmetries. 
Our model can be regarded as the generalized version of the Georgi-Machacek model, in which the global $SU(2)_L\times SU(2)_R$ symmetry is explicitly broken due to CP-violating terms in the potential. 
We present analytic formulas for theoretical constraints from perturbative unitarity and vacuum stability as well as 
contributions to the electron electric dipole moment (EDM) and the neutron EDM from all the Barr-Zee type diagrams.  
We then examine the parameter space allowed by the constraints mentioned above and also those from the uniqueness of the vacuum, measurements at Tevatron and LHC by using 
\texttt{HEPfit} to perform a global parameter fit. 
We find that the decays of the two lightest extra neutral (singly-charged) scalars, $H_1$ and $H_2$ ($H_1^\pm$), into $hZ$ ($WZ$) can be significant at the same time under the constraints, which 
can serve as direct evidence of CP violation in our model, but not from models with multi-doublet extensions.  

\end{abstract}

\maketitle

\newpage

%=================================================================================================
\section{Introduction}\label{sec:1}

The properties of the 125-GeV Higgs boson measured at LHC are, so far, consistent with those of the Higgs boson predicted in the Standard Model (SM) within the experimental error.  
Although this makes the SM a more reliable theory, various observations suggest the incompleteness of the SM.  
One of the most serious problems in the SM has been known that it cannot explain the origin of matter-antimatter asymmetry of the Universe, i.e., 
the lack of CP-violating (CPV) sources beyond the Kobayashi-Maskawa phase and the absence of departure from thermal equilibrium in the early Universe, e.g., 
a strong first-order electroweak phase transition~\cite{Kajantie:1996mn,Gurtler:1997hr,Csikor:1998eu}.
Therefore, new physics beyond the SM is strongly expected to exist in order to solve such a problem.

It has been known that extensions of the Higgs sector can realize sufficiently strong first-order electroweak phase transitions, 
because of additional bosonic degrees of freedom~\cite{Dolan:1973qd}. 
Furthermore, new CPV phases generally can appear in Yukawa interactions and the Higgs potential. 
These two ingredients lead to a successful scenario of the electroweak baryogenesis~\cite{Kuzmin:1985mm}. 
The two-Higgs doublet model (2HDM) is one of such extended Higgs models which have been most intensively disucssed~\cite{Turok:1990in,Cline:1995dg,Fromme:2006cm,Cline:2011mm,Shu:2013uua,Fuyuto:2017ewj,Modak:2018csw,Enomoto:2021dkl,Enomoto:2022rrl}. 
However, the 2HDM generally induces flavor-changing neutral currents (FCNCs) via Higgs boson exchanges at tree level, which are strictly constrained from 
various flavor experiments such as $B$ factories~\cite{HFLAV:2022pwe}.

In this paper, we discuss a model with a real and a complex isospin triplet Higgs fields as
the minimal extension of the Higgs sector such that it contains new sources of CPV and keeps the electroweak rho parameter to unity at tree level, but not introducing quark FCNCs via tree-level Higgs mediations. 
The field content is actually the same as that of the Georgi-Machacek (GM) model~\cite{Georgi:1985nv,Chanowitz:1985ug} which can realize the strong first-order electroweak phase transition~\cite{Chiang:2014hia,Zhou:2018zli,Chen:2022zsh} under the theoretical and current experimental constraints.  
The original GM model is, however, forbidden to have new CPV sources 
due to the global $SU(2)_L\times SU(2)_R$ symmetry in the potential, imposed to preserve the custodial $SU(2)_V$ symmetry after the spontaneous symmetry breaking.  
Thus, the model discussed in this paper corresponds to the electroweak gauge-invariant extension of the GM model denoting ``the extended GM model'', by which 
physical CPV phases appear in the scalar potential, while the rho parameter can be kept to unity by choosing the vacuum expectation values (VEVs) of triplet fields to be appropriately aligned.
It should also be mentioned that the explicit breaking of the custodial symmetry is eventually required to make the model consistent at the quantum level because the 
custodial symmetry is broken by the hypercharge and Yukawa interactions~\cite{Gunion:1990dt,Blasi:2017xmc,Chiang:2017vvo,Chiang:2018xpl,Keeshan:2018ypw}.

In this work, we aim to explore the CPV phenomenology in the extended GM model.  
For this purpose, we perform a complete analysis of the theoretical constraints on the model, which include the uniqueness of the vacuum, the vacuum stability, and the perturbative unitarity conditions.  
In view of the new CPV source, we take into account the electron electric dipole moment (eEDM) and neutron EDM (nEDM) measurements. 
Combining them with the Higgs measurements and search limits of additional Higgs bosons from the Tevatron and the LHC, we perform a global fit on model parameters, 
and find that the simultaneous decays of the two lightest neutral scalars, $H_1$ and $H_2$, to the $hZ$ mode can serve as clear and direct evidence of CPV in this model.  
The $gg\to H_{1,2}\to hZ\to bbZ$ processes are shown to have a great potential to be explored at the LHC in the near future.  Furthermore, the decay of the lightest charged scalar, $H_1^\pm$, to the $WZ$ channel~\cite{Grifols:1980uq,Kanemura:2011kc,Bandyopadhyay:2014vma,Bandyopadhyay:2015ifm,Zaro:2002500,Chiang:2015rva} can distinguish this model from a CPV 2HDM that also potentially affords the $hZ$ decay signature.  
On the other hand, the $h\gamma\gamma$ coupling is also very sensitive to the new physics contributions from the new charged scalars as well as the triplet-gauge couplings.  These points are explicitly illustrated using two benchmarks presented in the study.

The structure of this paper is as follows.  In Sec.~\ref{sec:2}, we first define the most general scalar sector under the electroweak symmetry, in which 
the original GM model can be reproduced by taking limits on the parameters. 
We then propose a minimal extension of the GM model to allow CPV.  
In Sec.~\ref{sec:3}, we discuss theoretical constraints on our model, including the uniqueness and stability of the electroweak vacuum, as well as the perturbative unitarity conditions.  In Sec.~\ref{sec:4}, we consider experimental constraints, such as the eEDM, nEDM, and the Tevatron and LHC direct search constraints.  In Sec.~\ref{sec:5}, we present the global fit result and discuss the implications on the eEDM, $H_{1,2}\to hZ$ decays, and the $gg\to H_{1,2}\to hZ\to bbZ$ processes, following which we further present two selected benchmarks that have great potential to be probed at the LHC.   Finally, we conclude the study in Sec.~\ref{sec:6}.

%=================================================================================================
\section{The Extended Georgi-Machacek Model}\label{sec:2}

The Higgs sector of the GM model is composed of an isospin doublet $\phi$ with hypercharge $Y=1/2$, a complex triplet $\chi$ with $Y=1$, and a real triplet $\xi$ with $Y=0$.
It is imposed with a global $SU(2)_L\times SU(2)_R$ symmetry in the Higgs potential such that 
the custodial $SU(2)_V$ symmetry remains after the electroweak symmetry breaking, as a result of which the electroweak rho parameter $\rho$ is unity at tree level. 
Such a custodial symmetric model can be regarded as a special case of our general, extended GM model. 
In the following, we first construct and discuss the general GM model without the custodial symmetry, followed by a comparison to the special case with the custodial symmetry.

\subsection{General Case}\label{sec:2-1}

We introduce the $SU(2)_L$ fundamental (adjoint) representation for $\phi$ ($\chi$ and $\xi$) as : 
\begin{equation}
	\phi=\begin{pmatrix}
		\phi^+ \\
		\phi^0
	\end{pmatrix},~
	\chi = \begin{pmatrix}
		\frac{\chi^+}{\sqrt{2}} & -\chi^{++} \\
		\chi^0 & -\frac{\chi^+}{\sqrt{2}}
	\end{pmatrix} , ~\xi = \begin{pmatrix}
		\frac{\xi^0}{\sqrt{2}} & -\xi^+ \\
		-\xi^- & -\frac{\xi^0}{\sqrt{2}}
	\end{pmatrix}, 
\end{equation}
where the neutral components are parameterized as 
\begin{align}
\phi^0 =\frac{1}{\sqrt{2}}(\phi_r + v_\phi + i\phi_i), \quad
\chi^0 =\frac{1}{\sqrt{2}}(\chi_r + i\chi_i) + v_\chi, \quad
\xi^0  = \xi_r + v_\xi,  
\end{align}
with $v_\phi$, $v_\chi$ and $v_\xi$ denoting the VEVs of the corresponding fields.  Without loss of generality, we can take these VEVs to be real and positive by rephasing the scalar fields.  The Fermi constant $G_F$ and the electroweak rho parameter $\rho$ at tree level can be expressed by these VEVs as: 
\begin{align}
v^2\equiv v_\phi^2+4v_\chi^2+4v_\xi^2 = \frac{1}{\sqrt{2}G_F}, \quad \rho = \frac{v^2}{v^2+4(v_\chi^2-v_\xi^2)} . \label{eq:vev}
\end{align}
The current global fit on $\rho$ given by the Particle Data Group~\cite{ParticleDataGroup:2022pth} is\footnote{If we consider the latest $W$-mass anomaly reported by the CDF-II Collaboration~\cite{CDF:2022hxs}, then $\rho$ could stray a lot from the global fit value depending on the exact model considered, as have been pointed out in Ref.~\cite{Chen:2022ocr}. In this work, we choose to stick to the global fit value.}
\begin{equation}
	\rho=1.00038\pm0.00020 .
\end{equation}
At tree level, this imposes a constraint on the difference between the squared triplet VEVs:\footnote{The $\rho$ parameter receives radiative corrections, particularly from the custodial symmetry breaking sectors such as the hypercharge and Yukawa interactions.  In models with $\rho \neq 1$ at tree level as in the extended GM model, however, a counterterm $\delta \rho$ appears due to the fact that the electroweak parameters cannot be described merely by three inputs as in the SM (e.g., $\alpha_{\rm em}$, $m_W^{}$ and $m_Z^{}$), but should be described by four parameters.  This additional counterterm can be determined by imposing another renormalization condition such that the loop correction to the $\rho$ parameter vanishes~\cite{Chiang:2017vvo,Chiang:2018xpl}.  Although the effect of $\delta \rho$ can appear in various observables such as Higgs boson couplings at loop levels, the concrete analysis requires the renormalization of the Higgs sector in the extended GM model, and is beyond the scope of the present paper.}
\begin{equation}
v_\chi^2-v_\xi^2 = -5.7571\pm3.0289~{\rm GeV}^2.
\end{equation}

The most general Higgs potential that is consistent with the electroweak symmetry is given by~\cite{Blasi:2017xmc,Keeshan:2018ypw} 
\begin{align}
V(\phi,\chi,\xi)
=&\,
m_\phi^2 (\phi^\dagger \phi)
+ m_\chi^2\text{tr}(\chi^\dagger\chi)+m_\xi^2\text{tr}(\xi^2) + 
\left( \mu_{\phi\chi}\phi^\dagger \chi \tilde{\phi} + \text{H.c.} \right) +\mu_{\phi\xi}\phi^\dagger \xi\phi + \mu_{\chi\xi}\text{tr}(\chi^\dagger \chi \xi)
\notag\\
&
+\lambda (\phi^\dagger \phi)^2 
+\rho_1[\text{tr}(\chi^\dagger\chi)]^2+\rho_2\text{tr}(\chi^\dagger \chi\chi^\dagger \chi)
+\rho_3[\text{tr}(\xi^2)]^2
+\rho_4 \text{tr}(\chi^\dagger\chi)\text{tr}(\xi^2)
+\rho_5\text{tr}(\chi^\dagger \xi)\text{tr}(\xi \chi)
\notag\\
&
+\sigma_1\text{tr}(\chi^\dagger \chi)\phi^\dagger \phi+\sigma_2 \phi^\dagger \chi\chi^\dagger \phi
+\sigma_3\text{tr}(\xi^2)\phi^\dagger \phi 
+ \left( \sigma_4 \phi^\dagger \chi\xi \tilde{\phi} + \text{H.c.} \right)
, \label{pot_gen}
\end{align}
where $\mu_{\phi\chi}$ and $\sigma_4$ are generally complex parameters, and $\tilde{\phi} = i\tau^2\phi^*$ is the charge conjugation of $\phi$ with $\tau^a$ ($a=1,2,3$) being the Pauli matrices.

The tadpole conditions, $\partial V/\partial X|_0 = 0$ for $X = \phi_r,~\chi_r,~\xi_r,~\phi_i$, respectively, give the following equations: 
\begin{align}
m_\phi^2 &= -v_\phi^2\lambda - v_\chi[2\Re\mu_{\phi \chi} + v_\chi(\sigma_1 + \sigma_2)] + \frac{v_\xi}{\sqrt{2}}(\mu_{\phi\xi} - \sqrt{2} v_\xi\sigma_3) - \sqrt{2} v_\chi v_\xi\Re\sigma_4, \label{eq:min:general:a}\\
m_\chi^2 &= -\frac{v_\phi^2}{4v_\chi}\Re(2\mu_{\phi\chi}  + \sqrt{2}v_\xi\sigma_4) -\frac{v_\xi}{\sqrt{2}}\mu_{\chi\xi} - \frac{v_\phi^2}{2}(\sigma_1 + \sigma_2) - 2v_\chi^2(\rho_1 + \rho_2) -v_\xi^2\rho_4, 
\label{eq:min:general:b}\\
m_\xi^2 &= \frac{1}{4\sqrt{2}v_\xi}(v_\phi^2\mu_{\phi\xi} -2v_\chi^2\mu_{\chi\xi} - 2v_\phi^2v_\chi\Re\sigma_4) -\frac{v_\phi^2}{2}\sigma_3 - v_\chi^2 \rho_4 - 2v_\xi^2 \rho_3, 
\label{eq:min:general:c}\\
\Im\mu_{\phi\chi} & = -\frac{v_\xi}{\sqrt{2}}\Im \sigma_4, 
\label{eq:min:general:d}
\end{align}
where the condition for $\chi_i$ is equivalent to that for $\phi_i$.  
From the last condition, the two complex phases are reduced to one, so that the extended GM model contains a single independent CP phase, which can be chosen to be arg$(\sigma_4)$.

It is now clear that the above-defined extended GM model is the minimal realization of having a physical CPV phase in the Higgs potential without introducing multiple Higgs doublets or new fermions while keeping $\rho \simeq 1$ with $v_\chi \simeq v_\xi$ at tree level.  
We note that models extended with $SU(2)_L$ scalar singlets can only have nonzero phases in the potential, but such a phase is not related to the CPV because the singlet fields cannot couple to SM fermions.  Nevertheless, the CP properties of the singlets can become definite if they couple to other new fermions~\cite{Chen:2022vac}.

The mass eigenstates for the scalar bosons are defined as follows: 
\begin{align}
\begin{split}
&
\chi^{\pm\pm}
=	H^{\pm\pm}, \quad
{\bm H}_{\rm weak}^+
= O_{G^\pm} \tilde{\bm H}_{\rm mass}^+
= O_{G^\pm} U_{H^\pm} {\bm H}_{\rm mass}^+ ,\\
&
{\bm H}_{\rm weak}^0
= O_{G^0} \tilde{\bm H}_{\rm mass}^0
= O_{G^0}O_{H^0}{\bm H}_{\rm mass}^0 , 
\end{split}
\label{eq:masseigen1}
\end{align}
where 
\begin{align}
\begin{split}
&
{\bm H}_{\rm weak}^\pm 
= 
\begin{pmatrix}
\phi^\pm \\ \chi^\pm \\ \xi^\pm
\end{pmatrix}
,\quad 
\tilde{\bm H}_{\rm mass}^\pm 
= 
\begin{pmatrix}
\tilde{H}_1^\pm \\ \tilde{H}_2^\pm \\ G^\pm
\end{pmatrix}
,  \quad
{\bm H}_{\rm mass}^\pm 
= 
\begin{pmatrix}
H_1^\pm \\ H_2^\pm \\ G^\pm
\end{pmatrix}
,
\\
&
{\bm H}_{\rm weak}^0 
= 
\begin{pmatrix}
\phi_r \\ \chi_r \\ \xi_r \\ \phi_i \\ \chi_i
\end{pmatrix}
,~
\tilde{\bm H}_{\rm mass}^0 
= 
\begin{pmatrix}
\tilde{H}_0 \\ \tilde{H}_1 \\ \tilde{H}_2 \\ \tilde{H}_3 \\ G^0
\end{pmatrix}
,~
{\bm H}_{\rm mass}^0 
= 
\begin{pmatrix}
H_0 \\ H_1 \\ H_2 \\ H_3 \\ G^0
\end{pmatrix}
, 
\end{split} \label{eq:masseigen}
\end{align}
with 
\begin{align}
{\bm H}_{\rm weak}^-=({\bm H}_{\rm weak}^+)^*,~
\tilde{{\bm H}}_{\rm mass}^-=(\tilde{{\bm H}}_{\rm mass}^+)^*,~
{\bm H}_{\rm mass}^-=({\bm H}_{\rm mass}^+)^*.
\end{align}
In Eq.~(\ref{eq:masseigen}), 
$G^\pm$ $(G^0)$ denotes the NGBs to be absorbed into the longitudinal components of $W^\pm$ $(Z)$, and 
$H^{\pm\pm}$, $H^{\pm}_i$ $(i=1,2)$ and $H_j$ $(j=0,\dots,3)$ are the physical doubly-charged, singly-charged and neutral Higgs bosons, respectively, among which 
we identify $H_0\,( \equiv h)$ as the 125-GeV Higgs boson discovered at the LHC.  
The matrices $O_{G^\pm}$, $O_{G^0}$ and $O_{H^0}$ ($U_{H^\pm }$) are orthogonal (unitary) matrices, with the former two separating the NGB modes from the physical Higgs bosons and given simply in terms of the Higgs VEVs as 
\begin{align}
O_{G^\pm} & = \left(
\begin{array}{ccc}
  -\frac{2\sqrt{v_\xi^2+v_\chi^2}}{v} & 0 &  \frac{v_\phi}{v} \\
  \frac{v_\phi v_\chi}{v\sqrt{v_\xi^2+v_\chi^2}} & \frac{v_\xi}{\sqrt{v_\xi^2+v_\chi^2}} &  \frac{2 v_\chi}{v} \\
\frac{v_\xi v_\phi}{v\sqrt{v_\xi^2+v_\chi^2}} & -\frac{v_\chi}{\sqrt{v_\xi^2+v_\chi^2}} &  \frac{2 v_\xi}{v}  
\end{array}
\right)
,\quad
O_{G^0}  =  
\begin{pmatrix}
\mathbbm{1}_{3\times 3} & 0 & 0 \\ 
0& -\frac{2\sqrt{2}v_\chi}{\sqrt{v_\phi^2+8v_\chi^2}} & \frac{v_\phi}{\sqrt{v_\phi^2+8v_\chi^2}} &   \\
0& \frac{v_\phi}{\sqrt{v_\phi^2+8v_\chi^2}}           & \frac{2\sqrt{2}v_\chi}{\sqrt{v_\phi^2+8v_\chi^2}} 
\end{pmatrix}.
\end{align}
On the other hand, the matrices $U_{H^\pm}$ and $O_{H^0}$ are not determined purely by the VEVs, but also depend on the mass matrices for the physical states.  In Appendix~\ref{sec:mass}, we give the explicit forms of the mass matrix for the singly-charged states $M_\pm$ in the basis of $(\tilde{H}_1^\pm,\tilde{H}_2^\pm)$ and that for the neutral scalar states $M_0$ in the basis of $(\tilde{H}_0,\tilde{H}_1,\tilde{H}_2,\tilde{H}_3)$. 
Since there are only two physical states for the singly-charged Higgs bosons, $U_{H^\pm}$ can be expressed in terms of a single mixing angle $\theta_\pm$ and a phase $\phi_\pm$ as 
\begin{align}
U_{H^\pm} = \begin{pmatrix}
1 & 0 & 0 \\
0 & \cos\theta_\pm & -\sin\theta_\pm e^{-i \phi_\pm}\\
0 &  \sin\theta_\pm e^{i \phi_\pm} & \cos\theta_\pm
\end{pmatrix}, 
\end{align}
with 
\begin{align}
\tan2\theta_\pm = \frac{2|(M_\pm)_{12}|}{(M_\pm)_{11} - (M_\pm)_{22}},\quad 
\phi_\pm = -\text{arg}[(M_\pm)_{12}] . 
\end{align}
For the neutral sector, $O_{H^0}$ can generally be expressed in terms of six mixing angles, {\it i.e.}, the parameters of the $O(4)$ group to describe the mixing among the remaining four neutral degrees of freedom.  
We note that in the $\Im \sigma_4 \to 0$ limit, the $(\tilde{H}_0,\tilde{H}_1,\tilde{H}_2)$ states and the $\tilde{H}_3$ state do not mix with one another and are CP eigenstates, with the former multiplet being CP-even and the latter singlet being CP-odd.  
This means that the matrix $O_{H^0}$ in this case has a block diagonal form with a $3\times 3$ part and $\mathbbm{1}_{2\times 2}$.  
Another important thing here is that these mixing matrices take a significantly simpler form if we take the custodial symmetry limit, {\it i.e.}, the original GM model, corresponding to the special case with the CP conservation ($\Im \sigma_4 \to 0$) to be discussed in the next subsection. 
In order to simplify the expression, we rewrite elements of the mixing matrices defined in Eq.~(\ref{eq:masseigen}) as follows:
\begin{align}
\begin{split}
&R_{\phi_r i} \equiv (O_{G^0}O_{H^0})_{1,i+1},~
R_{\chi_r i} \equiv (O_{G^0}O_{H^0})_{2,i+1},~
R_{\xi_r i} \equiv (O_{G^0}O_{H^0})_{3,i+1}, 
\\
&R_{\phi_i i} \equiv (O_{G^0}O_{H^0})_{4,i+1},~
R_{\chi_i i} \equiv (O_{G^0}O_{H^0})_{5,i+1}, 
\\
&R_{\phi^\pm j} \equiv (O_{G^\pm} U_{H^\pm})_{1j},~
R_{\chi^\pm i} \equiv (O_{G^\pm} U_{H^\pm})_{2j},~
R_{\xi^\pm j} \equiv (O_{G^\pm} U_{H^\pm})_{3j}, 
\end{split} \label{eq:mixing}
\end{align}
where $i (=0,1,2,3)$ and $j (=1,2)$ label the physical neutral and singly-charged Higgs bosons, respectively, with $H_0 \equiv h$.
%--------------------------------------------------------------------------------------------------------------------------------------------------------------------------

The most general Yukawa interactions can be divided into the following two parts:
\begin{align}
{\cal L}_Y &= {\cal L}_Y^\phi + {\cal L}_Y^\chi, 
\end{align}
where 
\begin{align}
\begin{split}
{\cal L}_Y^\phi &=-y_u\bar{Q}_L\tilde{\phi}u_R - y_d\bar{Q}_L\phi d_R -y_e\bar{L}_L \phi e_R  + \text{H.c.}, 
\\
{\cal L}_Y^\chi & = -y_\nu \overline{L_L^c}(i\tau_2)\chi L_L + \text{H.c.}. 
\end{split}
\end{align}
The Yukawa interactions for $\phi$, ${\cal L}_Y^\phi$, take the same form as that in the SM to provide mass for the charged fermions, while ${\cal L}_Y^\chi$ provides tiny neutrino mass via the type-II seesaw mechanism~\cite{Cheng:1980qt,Schechter:1980gr,Magg:1980ut,Mohapatra:1980yp}.  
Although the form of ${\cal L}_Y^\phi$ is the same as in the SM, the interaction terms between fermions and the Higgs boson are different from the SM ones due to the Higgs field mixing: 
\begin{align}
{\cal L}_Y^\phi 
\supset& \, 
-\sum_{f = u,d,\ell}\sum_{i = 1}^5\bar{f}\frac{M_f}{v_\phi}\left( R_{\phi_r i} -2iI_f\gamma_5 R_{\phi_i i} \right) f({\bm H}_{\rm mass}^0)_i  
\notag\\
&-\frac{\sqrt{2}}{v_\phi}\left[\bar{u}(V_{ud} M_d P_R - M_u V_{ud}P_L) d 
+ \bar{\nu}_\ell (M_\ell P_R) \ell \right] \sum_{j = 1}^3R_{\phi^\pm j} ({\bm H}_{\rm mass}^+)_j + \text{H.c.} , \label{eq:yuk}
\end{align}
where 
$M_f (\equiv y_f v_\phi/\sqrt{2})$ are the diagonalized mass matrices for the charged fermion $f$, $I_f$ is the third component of the isospin, $i.e.$, $I_{u}\,(I_{d,e}) = 1/2~(-1/2)$, 
and $V_{ud}$ is the Cabibbo-Kobayashi-Maskawa matrix. In Eq.~(\ref{eq:yuk}), the flavor indices are not explicitly shown. 
We note that the CP mixing is introduced via the mixing matrix $O_{H^0}$, with which each of the neutral Higgs bosons can generally have both the scalar-type interaction $\bar{f}f$ and the pseudoscalar-type interaction $\bar{f}i\gamma_5f$.  As alluded to before, in the CP-conserving limit $(\Im \sigma_4 \to 0)$, the $(h,H_1,H_2)$ and $H_3$ bosons only couple respectively to the scalar- and pseudoscalar-type interactions.  
We also note that the mixing matrices do not depend on the flavor structure and the type of fermion, {\it i.e.}, up-type quarks, down-type quarks and charged leptons.  
In this sense, the structure of the Yukawa interactions is similar to that of the type-I 2HDM.

%=================================================================================================
\subsection{Custodial Symmetric Case}\label{sec:original}

The scalar sector of the GM model can be expressed in terms of a bi-doublet $\Phi$ and a bi-triplet $\Delta$ under a global $SU(2)_L\times SU(2)_R$ symmetry as
\begin{equation}
	\Phi=\begin{pmatrix}
		\phi^{0*} & \phi^+ \\
		-\phi^- & \phi^0
	\end{pmatrix} , ~
	\Delta=\begin{pmatrix}
		\chi^{0*} & \xi^+ & \chi^{++} \\
		-\chi^- & \xi^0 & \chi^+ \\
		\chi^{--} & -\xi^- & \chi^0
	\end{pmatrix}. 
\end{equation}
Using $\Phi$ and $\Delta$, it is straightforward to write down the Higgs potential manifestly invariant under the $SU(2)_L\times SU(2)_R$ symmetry as 
\begin{equation}\label{Eq:VGM}
\begin{aligned}
	V(\Phi,\Delta) &= \frac{m_1^2}{2}\Tr(\Phi^\dagger\Phi)+\frac{m_2^2}{2}\Tr(\Delta^\dagger\Delta)+\lambda_1\big[\Tr(\Phi^\dagger\Phi)\big]^2+\lambda_2\big[\Tr(\Delta^\dagger\Delta)\big]^2 \\
	&\quad +\lambda_3\Tr\big[(\Delta^\dagger\Delta)^2\big]+\lambda_4\Tr(\Phi^\dagger\Phi)\Tr(\Delta^\dagger\Delta)+\lambda_5\Tr(\Phi^\dagger\frac{\tau^a}{2}\Phi\frac{\tau^b}{2})\Tr(\Delta^\dagger T^a\Delta T^b) \\
	&\quad +\mu_1\Tr \left(\Phi^\dagger\frac{\tau^a}{2}\Phi\frac{\tau^b}{2} \right)(P^\dagger\Delta P)_{ab}+\mu_2\Tr(\Delta^\dagger T^a\Delta T^b)(P^\dagger\Delta P)_{ab} ,
\end{aligned}
\end{equation}
where $T^a$ are the $3\times 3$ matrix representation of the $SU(2)$ generators, and $P$ is the similarity transformation relating the triplet and adjoint representations of the $SU(2)$ generators given by
\begin{equation}
	P=\frac{1}{\sqrt{2}}\begin{pmatrix}
		-1 & i & 0 \\
		0 & 0 & \sqrt{2} \\
		1 & i & 0
	\end{pmatrix} .
\end{equation}
When the triplet VEVs are aligned as 
\begin{equation}
\langle\Delta\rangle = {\rm diag}(v_\chi,v_\chi,v_\chi), 
\end{equation}
the $SU(2)_L\times SU(2)_R$ symmetry is spontaneously broken down to the custodial $SU(2)_V$ symmetry. 
As has been pointed out in Ref.~\cite{Chen:2022ocr}, the condition $\langle\chi^0\rangle=\langle\xi^0\rangle$ 
is necessary for the case with the $SU(2)_L\times SU(2)_R$ symmetry in order to avoid two additional charged NGB modes associated with the spontaneous breakdown of $SU(2)_V \to U(1)_{\Delta}$, with the latter being an overall phase rotation symmetry of the triplet VEV.

The GM model with the custodial symmetry shows certain characteristic features.  For example, the mass eigenstates of the Higgs bosons are classified into an $SU(2)_V$ quintuplet $H_Q = (H_Q^{\pm\pm},H_Q^{\pm},H_Q^{0})^T$, a triplet $H_T = (H_T^{\pm},H_T^0)^T$ and two singlets $h$, $H$, where the states in the same $SU(2)_V$ multiplet are degenerate in mass due to the custodial symmetry.  While this can be explicitly shown from the potential defined in Eq.~(\ref{Eq:VGM}), here we show it by reducing from the general potential in Eq.~(\ref{pot_gen}) with the following identifications: 
\begin{align}
\begin{split}
&m_\phi^2=m_1^2,~ m_\chi^2 = m_2^2,~ m_\xi^2= \frac{m_2^2}{2},~
\mu_{\phi\xi} = -\frac{\mu_1}{\sqrt{2}},~ \mu_{\phi\chi}= \frac{\mu_1}{2},~
\mu_{\chi\xi}=6\sqrt{2}\mu_2,\\
&\lambda=4\lambda_1,~ 
\rho_1=4\lambda_2+6\lambda_3,~ 
\rho_2=-4\lambda_3,~ 
\rho_3=\lambda_2+\lambda_3,~
\rho_4=4\lambda_2,~
\rho_5=4\lambda_3, \\
& \sigma_1 =4\lambda_4-\lambda_5,~ 
 \sigma_2 =2\lambda_5,~ 
 \sigma_3 =2\lambda_4,~
 \sigma_4 =\sqrt{2}\lambda_5. 
\end{split}\label{rel} 
\end{align}
In this limit, the general potential becomes $SU(2)_L\times SU(2)_R$-symmetric, with the eighteen real parameters in the general potential being rewritten in terms of nine real parameters.  Equivalently, we can impose the following nine relations among the parameters in the general potential to restore the $SU(2)_L\times SU(2)_R$ symmetry:
\begin{align}
\begin{split}
&m_\xi^2=\frac{m_\chi^2}{2},\quad \Re\mu_{\phi\chi} = -\frac{\mu_{\phi\xi}}{\sqrt{2}},\quad \Im\mu_{\phi\chi}=0, 
\\
&\rho_3 = \frac{\rho_1}{4}+\frac{\rho_2}{8},\quad \rho_4=\rho_1+\frac{3}{2}\rho_2,\quad\rho_5=-\rho_2,\quad
\sigma_3 = \frac{\sigma_1}{2}+\frac{\sigma_2}{4},\quad 
\Re\sigma_4 = \frac{\sigma_2}{\sqrt{2}},\quad
\Im\sigma_4 = 0 . 
\end{split}\label{custodial}
\end{align}
When we impose the above conditions with $v_\xi = v_\chi$ in the general potential, the tadpole conditions for $\chi_r$ and $\xi_r$ become equivalent and that for $\chi_i$ becomes trivial, as can be seen in Eqs.~(\ref{eq:min:general:b})-(\ref{eq:min:general:d}).
In addition, the mixing matrices defined in Eq.~(\ref{eq:masseigen1}) are simplified to be  
\begin{align}
O_{H^\pm} = \mathbbm{1}_{3\times 3},\quad 
O_{H^0} = 
\begin{pmatrix}
1 & 0 & 0 & 0\\
0 & \sqrt{\frac{2}{3}} & -\sqrt{\frac{1}{3}} & 0\\
0 & \sqrt{\frac{1}{3}} &  \sqrt{\frac{2}{3}} & 0 \\
0 & 0 & 0 & \mathbbm{1}_{2\times 2}
\end{pmatrix}
\begin{pmatrix}
 \cos\alpha & -\sin\alpha & 0 & 0\\
 \sin\alpha &  \cos\alpha & 0 & 0 \\ 
0  &  0 & 1 & 0    \\
0 & 0 & 0 & \mathbbm{1}_{2\times 2}
\end{pmatrix}. 
\end{align}
Thus, it is clear that the mass eigenstates are classified into the $SU(2)_V$ multiplets: 
\begin{align}
&H^{\pm\pm} = H_Q^{\pm\pm} ,\quad 
H_2^{\pm} = H_Q^{\pm} ,\quad 
H_2 = H_Q^0,\quad 
H_1^{\pm} = H_T^{\pm} ,\quad 
H_3 = H_T^0. 
\end{align}
An explicit calculation shows the following mass relations:
\begin{align}
\begin{split}
m_{H_Q}^2 & \equiv m_{H^{\pm\pm}}^2 = (M_\pm)_{22} =  [O_{H^0}^T(M_0)O_{H^0}]_{33}, 
\\  %= -\frac{v_\phi^2 (\mu_1+6\lambda_5 v_\chi)}{4 v_\chi}-12 \mu_2 v_\chi+8\lambda_3 v_\chi^2 \\
m_{H_T}^2 & \equiv (M_{\pm})_{11} = (M_0)_{55}. % = -\frac{v^2 (\mu_1+2 \lambda_5 v_\chi)}{4v_\chi}.
\end{split}
\end{align}

We note in passing that the GM model does not afford any CPV source, a natural result derived from the symmetry structure.  Furthermore, it has been known for a long while that the custodial symmetry would be broken at the loop level due to the hypercharge interaction and/or fermion loops~\cite{Gunion:1990dt,Blasi:2017xmc,Keeshan:2018ypw}.  For a consistent renormalization prescription of the scalar potential, one has to explicitly break the custodial symmetry from the very beginning~\cite{Chiang:2017vvo,Chiang:2018xpl}.

%=================================================================================================
\subsection{Minimal Extension with CPV}\label{sec:minimal}

As seen in the previous subsection, new CPV phases vanish in the custodial symmetric limit.  We here propose a minimal extension of the original GM model discussed in Sec.~\ref{sec:original} that allows the introduction of a new CPV phase, instead of studying the most general case discussed in Sec.~\ref{sec:2-1}. 
The Higgs potential in the minimally extended model is defined as 
\begin{align}
V_{\rm Min} &= V(\Phi,\Delta) + V_{\rm Soft} + (\sigma_4 \phi^\dagger \chi\xi \tilde{\phi} + \text{H.c.}), \label{pot_min}
\end{align}
where the first term is the $SU(2)_L\times SU(2)_R$-invariant potential given in Eq.~(\ref{Eq:VGM}) and the second term explicitly given by
\begin{align}
V_{\rm Soft} = m_\chi^2\text{tr}(\chi^\dagger\chi)+m_\xi^2\text{tr}(\xi^2) + 
(\mu_{\phi\chi}\phi^\dagger \chi \tilde{\phi} + \text{H.c.}) +\mu_{\phi\xi}\phi^\dagger \xi\phi + \mu_{\chi\xi}\text{tr}(\chi^\dagger\chi\xi)
\end{align}
contains all the possible soft-breaking terms for the $SU(2)_L\times SU(2)_R$ symmetry.  The last $\sigma_4$ term is the hard-breaking term of the $SU(2)_L\times SU(2)_R$ symmetry, and is required to keep the non-zero CPV phase after solving the tadpole condition [see Eq.~(\ref{eq:min:general:d})].  As the dimension-2 and -3 terms are essentially equivalent to the most general case defined in Eq.~(\ref{pot_gen}), we can reparameterize the coefficients of these vertices as in Eq.~(\ref{pot_gen}), {\it e.g.}, $(m_2^2 + m_\chi^2)\text{tr}(\chi^\dagger\chi) \to m_\chi^2\text{tr}(\chi^\dagger\chi)$. 
This minimally extended model is obtained by taking the following limits of the most general case: 
\begin{align}
\begin{split}
&\lambda=4\lambda_1,~ 
\rho_1=4\lambda_2+6\lambda_3,~ 
\rho_2=-4\lambda_3,~ 
\rho_3=\lambda_2+\lambda_3,~
\rho_4=4\lambda_2,~
\rho_5=4\lambda_3, \\
& \sigma_1 =4\lambda_4-\lambda_5,~ 
 \sigma_2 =2\lambda_5,~ 
 \sigma_3 =2\lambda_4. 
\end{split}\label{rel} 
\end{align}
The mass formulas for the Higgs bosons can be obtained by substituting the above equations to those given in the general case discussed in Sec.~\ref{sec:2-1}.

In our global fit and benchmark studies, we choose the following as the input parameters:
\begin{equation}
	\{v_\chi,v_\xi,\lambda_2,\lambda_3,\lambda_4,\lambda_5,\Re\sigma_4,\Im\sigma_4,\mu_{\phi\xi},\mu_{\chi\xi},\Re\mu_{\phi \chi}\} ~.
\end{equation}
We note that $v_\phi$ and $\lambda_1$ are taken such that the VEV $v\simeq 246$~GeV and the Higgs boson mass $m_h \simeq 125$~GeV. 
The masses of the additional Higgs bosons are determined by fixing the above Lagrangian parameters, while we define
the hierarchies: $m_{H_1^\pm}\leq m_{H_2^\pm}$ and $m_h \leq m_{H_1} \leq m_{H_2} \leq m_{H_3}$.

%=================================================================================================
\section{Theoretical constraints}\label{sec:3}

The parameters in the Higgs potential can be further constrained by considering the consistency of the model, such as the uniqueness and stability of the vacuum, and the perturbative unitarity.  In the original GM model, the conditions for vacuum stability and perturbative unitarity have been respectively derived in Ref.~\cite{Hartling:2014zca} and Ref.~\cite{Aoki:2007ah}.  In Refs.~\cite{Keeshan:2018ypw,Blasi:2017xmc}, scenarios with custodial symmetry breaking have been discussed, with these theory bounds being taken into account and the running couplings evaluated by solving renormalization group equations.  To our knowledge, the constraints from the vacuum stability and the perturbative unitarity have not been derived in the most general Higgs potential.  We give analytic formulas of these theory constraints for the most general CPV potential, and confirm that these expressions are successfully reduced to those given in the custodial symmetric case derived in the above-mentioned references.

%=================================================================================================
\subsubsection{Unique Vacuum}\label{sec:3:1:1}

In general, it is possible that the desired electroweak vacuum $\vec{v}=(v_\phi,v_\chi,v_\xi)$ satisfying Eq.~(\ref{eq:vev}) may not be the global minimum of the Higgs potential and some other deeper minima exist.  This will result in the instability of the electroweak vacuum and its decay to the true vacuum by tunneling.  To avoid such a meta-stable situation, one should solve for all the possible VEVs that satisfy the tadpole conditions 
and check whether $\vec{v}$ is indeed the global minimum. 
All the possible VEVs can be found by solving two cubic equations of $v_\chi$ simultaneously, which are obtained from Eqs.~(\ref{eq:min:general:b}) and (\ref{eq:min:general:c}) with $v_\phi$ and $v_\xi$ expressed in terms of the other parameters using Eqs.~(\ref{eq:min:general:a}) and (\ref{eq:min:general:d}). 
We then check whether $\vec{v}$ is indeed the global minimum of the scalar potential by comparing it with all the other solutions.

%=================================================================================================
\subsubsection{Vacuum Stability}\label{sec:3:1:2}

The Higgs potential has to be bounded from below in any direction of the field space with large field values. 
Such stability of the potential is ensured by the following conditions: 
\begin{align}
%\begin{split}
&	\lambda > 0,~~\rho_3 > 0,~~\rho_1 + \text{min}(\rho_2/2,\rho_2) > 0, \\
&	4\lambda \rho_3 > \sigma_3^2 ,~~~4\lambda(\rho_1 + \rho_2\zeta) >  \left(\sigma_1 +  \frac{\sigma_2-\vert\sigma_2\vert\sqrt{2\zeta-1}}{2}\right)^2, \\
&	4(\rho_1 + \rho_2\zeta + \rho_3 - \rho_4 -\eta\rho_5)\rho_3 > (2\rho_3 - \rho_4 - \eta\rho_5)^2, \label{eq:vc3}\\
&	G(t,\zeta,\eta) > 0, 
%\end{split}
\end{align}
where 
\begin{align}
G(t,\zeta,\eta) 
\equiv & \,
4\left[ \left(\rho_1 + \rho_2\zeta + \rho_3 - \rho_4 -\eta\rho_5 \right)t^4 - (2\rho_3 - \rho_4 - \eta\rho_5)t^2 + \rho_3 \right]\lambda 
\notag\\
 & -\left[ \left(\sigma_1 + \omega\sigma_2  -\sigma_3 \right)t^2 - 2|\sigma_4|\sqrt{\frac{1-\eta}{2}} t\sqrt{1-t^2} + \sigma_3 \right]^2, 
\end{align}
with the domains $t\in[0,1],~\zeta\in[1/2,1],~\eta\in[0,1]$.  We note that the condition in Eq.~(\ref{eq:vc3}) can be redundant if 
\begin{align}
\rho_1 + \rho_2\zeta + \rho_3 - \rho_4 -\eta\rho_5 > 0~~\text{or}~~ 2\rho_3 - \rho_4 - \eta\rho_5 > 0~\text{or}~ 0\leq \frac{ 2\rho_3 - \rho_4 - \eta\rho_5}{2(\rho_1 + \rho_2\zeta + \rho_3 - \rho_4 -\eta\rho_5) } \leq 1. 
\end{align}
A detailed derivation of the above conditions is given in Appendix~\ref{sec:stability}.

%=================================================================================================
\subsubsection{Perturbative Unitarity}\label{sec:3:1:3}

\begin{table}[htbp]
\centering
\begin{tabular}{>{\centering\arraybackslash}p{1cm}|>{\centering\arraybackslash}p{1cm}||>{\centering\arraybackslash}p{10cm}}
\toprule
$\vert Q\vert$ & $\vert Y\vert$ & Two-Body States \\
\toprule
& 0 & $\phi^+\phi^-$, $\phi^0\phi^{0*}$, $\chi^{++}\chi^{--}$, $\chi^+\chi^-$, $\chi^0\chi^{0*}$, $\xi^+\xi^-$, $\frac{\xi^0\xi^{0}}{\sqrt{2}}$ \\ \cline{2-3}
& $\frac{1}{2}$ & $\phi^{0*}\chi^0$, $\phi^-\chi^+$, $\phi^{0}\xi^0$, $\phi^-\xi^+$ \\ \cline{2-3}
0& 1 & $\frac{\phi^0\phi^0}{\sqrt{2}}$, $\chi^0\xi^0$, $\chi^+\xi^-$ \\ \cline{2-3}
& $\frac{3}{2}$ & $\phi^0\chi^0$ \\ \cline{2-3}
& 2 & $\frac{\chi^0\chi^0}{\sqrt{2}}$ \\
\colrule
 & 0 & $\phi^+\phi^{0*}$, $\chi^{++}\chi^-$, $\chi^+\chi^{0*}$, $\xi^+\xi^0$ \\ \cline{2-3}
& $\frac{1}{2}$ & $\phi^{0*}\chi^+$, $\phi^-\chi^{++}$, $\phi^0\xi^+$, $\phi^+\xi^0$, $\phi^-\chi^0$, $\phi^0\xi^-$ \\ \cline{2-3}
1& 1 & $\phi^+\phi^0$ , $\chi^{++}\xi^-$, $\chi^+\xi^{0}$, $\chi^0\xi^+$, $\chi^{0*}\xi^+$ \\ \cline{2-3}
& $\frac{3}{2}$ & $\phi^0\chi^+$, $\phi^+\chi^0$ \\ \cline{2-3}
& 2 & $\chi^+\chi^0$ \\
\colrule
 & 0 & $\frac{\xi^+\xi^+}{\sqrt{2}}$, $\chi^{++}\chi^{0*}$ \\ \cline{2-3}
& $\frac{1}{2}$ & $\phi^{0*}\chi^{++}$, $\phi^+\xi^+$ \\ \cline{2-3}
2& 1 & $\frac{\phi^+\phi^+}{\sqrt{2}}$, $\chi^{++}\xi^0$, $\chi^+\xi^+$ \\ \cline{2-3}
& $\frac{3}{2}$ & $\phi^+\chi^+$, $\phi^0\chi^{++}$ \\ \cline{2-3}
& 2 & $\chi^{++}\chi^0$, $\frac{\chi^+\chi^+}{\sqrt{2}}$ \\
\colrule
  & 1 & $\chi^{++}\xi^+$ \\ \cline{2-3}
3& $\frac{3}{2}$ & $\phi^+\chi^{++}$ \\ \cline{2-3}
 & 2 & $\chi^{++}\chi^+$ \\
\colrule
4 & 2 & $\frac{\chi^{++}\chi^{++}}{\sqrt{2}}$ \\
\botrule
\end{tabular}
\caption{\label{perturb:basis} The singlet and symmetric two-body final states formed from the doublet and triplet fields, grouped by the total electric charge ($\vert Q\vert$) and the total hypercharge ($\vert Y\vert$). A symmetry factor of $1/\sqrt{2}$ is included for the states involving identical fields.}
\end{table}

We now consider the perturbative unitarity conditions from all the high-energy 2-to-2 elastic bosonic scattering processes.  The longitudinal modes of the weak vector bosons are taken into account as the NGB modes by using the equivalence theorem~\cite{Cornwall:1974km}. 
In Table~\ref{perturb:basis}, we list all the considered 2-to-2 scattering states, classified according to the total electric charge $Q$ and the total hypercharge $Y$.  We note that scatterings between states with different hypercharges (not only the electric charge) do not happen because the hypercharge should be conserved in the high-energy limit. 
We impose the following criteria for each eigenvalue $x_i$ of the $s$-wave amplitude matrix: 
 \begin{equation}
	\vert\Re x_i\vert < 8\pi.
 \end{equation}
We find the nineteen independent eigenvalues as follows:
\begin{align}
x_1&=2 (\rho_1+\rho_2), \\
x_2&= 2\rho_1 - \rho_2, \\
x_3&= 2 \rho_4+\rho_5,\\
x_4&= 2 (\rho_4 + 2 \rho_5), \\
x_5&= \sigma_1+\sigma_2,\\
x_6&= \sigma_1 - \frac{\sigma_2}{2}, \\
x_7^\pm&=\rho_1+4 \rho_3 \pm \sqrt{(\rho_1-4 \rho_3)^2+2 \rho_5^2},\\
x_8^\pm&=\lambda +\rho_1+2 \rho_2 \pm\sqrt{(\lambda - \rho_1 - 2 \rho_2)^2+\sigma_2^2},\\
x_9^\pm&= \frac{\sigma_1}{2} + \sigma_3 \pm\frac{1}{2}\sqrt{(\sigma_1-2 \sigma_3)^2+4 |\sigma_{4}|^2},\\
x_{10}^\pm&= \lambda +\rho_4-\frac{\rho_5}{2} \pm \frac{1}{2}\sqrt{(2 \lambda -2 \rho_4+\rho_5)^2+8 |\sigma_{4}|^2},\\
x_{11}^\pm&=\frac{\sigma_1}{2}+\frac{3}{4} \sigma_2+\sigma_3 \pm\frac{1}{4}\sqrt{(2 \sigma_1+3 \sigma_2-4 \sigma_3)^2+64 |\sigma_{4}|^2},
\end{align}
and $x_{12}^{i}$ $(i=1,2,3)$ being the eigenvalues of the following matrix
\begin{align}
\left(
\begin{array}{ccc}
 20 \rho_3 & 2 \sqrt{3} \sigma_3 & \sqrt{2} (3\rho_4 +\rho_5) \\
 2 \sqrt{3}\sigma_3 & 6 \lambda  & \sqrt{\frac{3}{2}} (2\sigma_1 + \sigma_2 ) \\
 \sqrt{2} (3\rho_4 + \rho_5 ) & \sqrt{\frac{3}{2}} (2\sigma_1 + \sigma_2 ) & 8\rho_1 +6\rho_2 \\
\end{array}
\right). 
\end{align}
We note that the complex parameter $\sigma_4$ appears in the form of its magnitude in the above eigenvalues.  This is because the scattering amplitudes are evaluated in the high-energy limit, where only the quartic couplings in the potential are relevant, and the CPV phase can be removed by rephasing the scalar fields.

%=================================================================================================
\section{Experimental Constraints}\label{sec:4}

We discuss constraints from the eEDM, the nEDM, the Higgs measurements and the additional Higgs searches at the Tevatron and the LHC in this section.

%=================================================================================================
\subsection{eEDM}\label{sec:4:1}

\begin{figure}[t]
\centering
\includegraphics[width=0.35\textwidth]{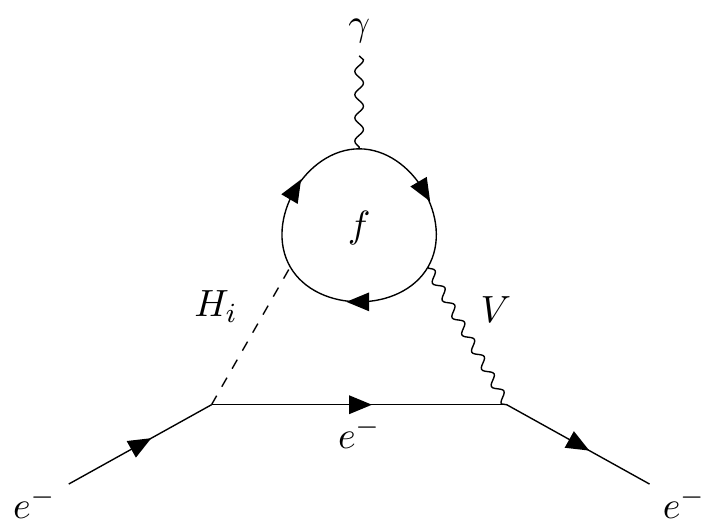}\hspace{1cm}
\includegraphics[width=0.35\textwidth]{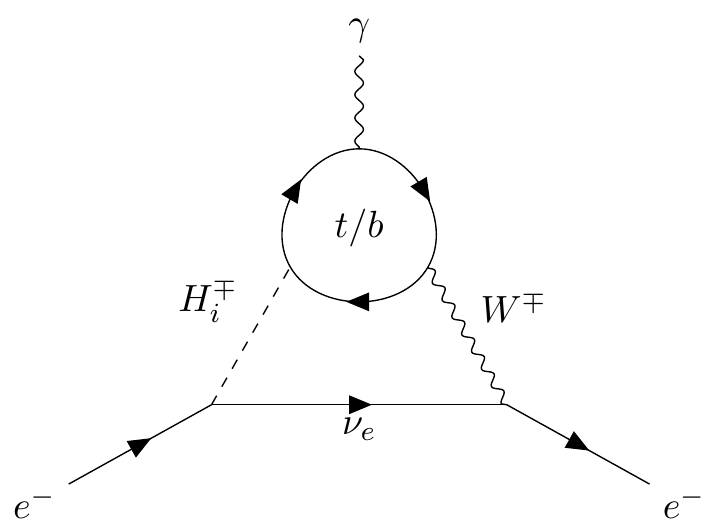}
\\
\vspace{-0.4cm} (a)
\hspace{6.5cm} (b)
\caption{\label{BZ:fermion} Fermion-loop BZ diagrams with $V=\gamma,Z$.}
\end{figure}

\begin{figure}[t]
\centering
\includegraphics[width=0.32\textwidth]{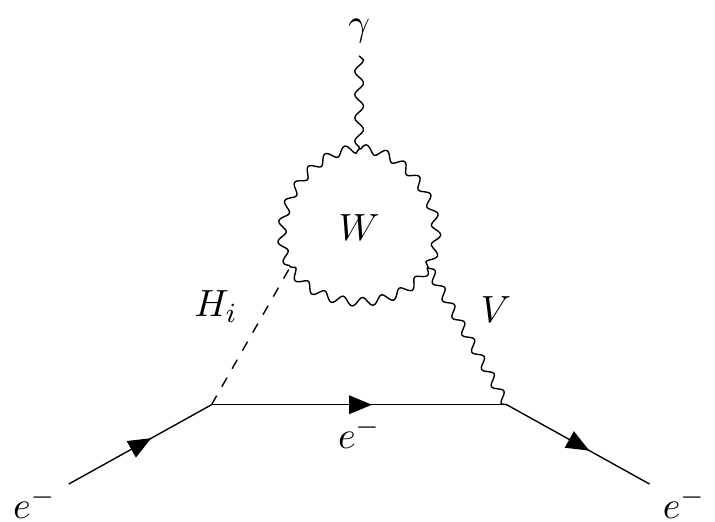}\hspace{0.2cm}
\includegraphics[width=0.32\textwidth]{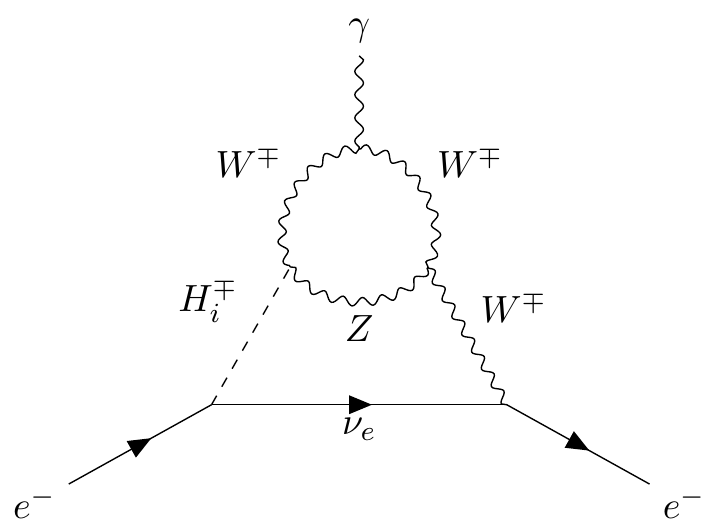}\hspace{0.2cm}
\includegraphics[width=0.32\textwidth]{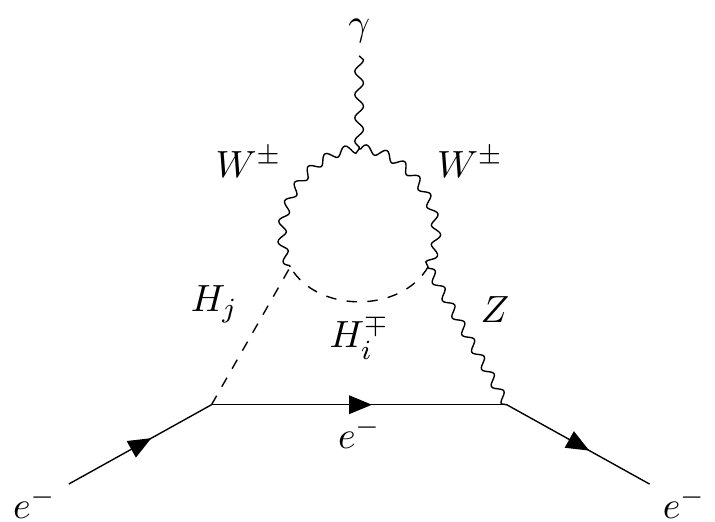}
\\ \vspace{-0.4cm}
(a)
\hspace{5cm} (b)
\hspace{5cm} (c)
\\
\includegraphics[width=0.32\textwidth]{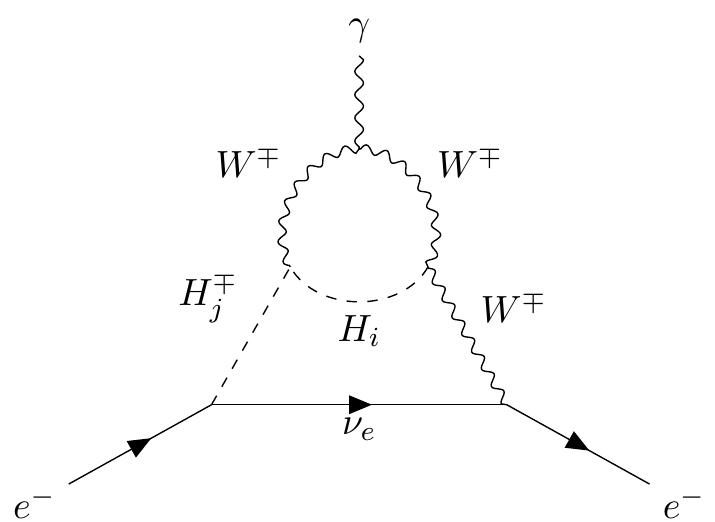}\hspace{0.2cm}
\includegraphics[width=0.32\textwidth]{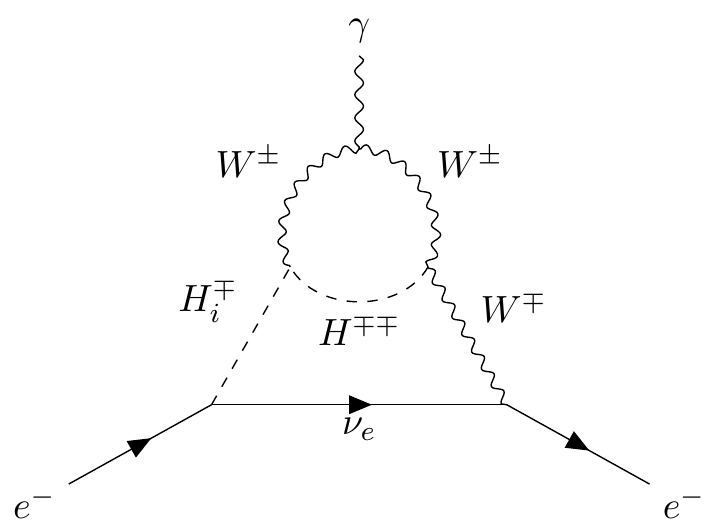}\hspace{0.2cm}
\includegraphics[width=0.32\textwidth]{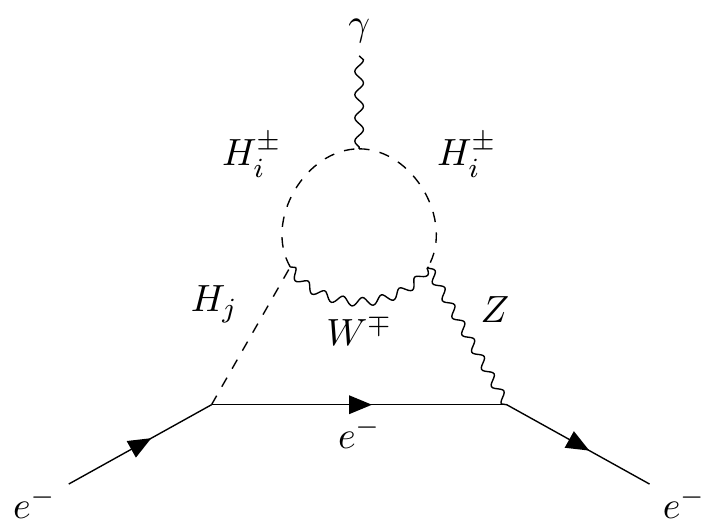}
\\
\vspace{-0.4cm} 
(d)
\hspace{5cm} (e)
\hspace{5cm} (f)
\\
\includegraphics[width=0.32\textwidth]{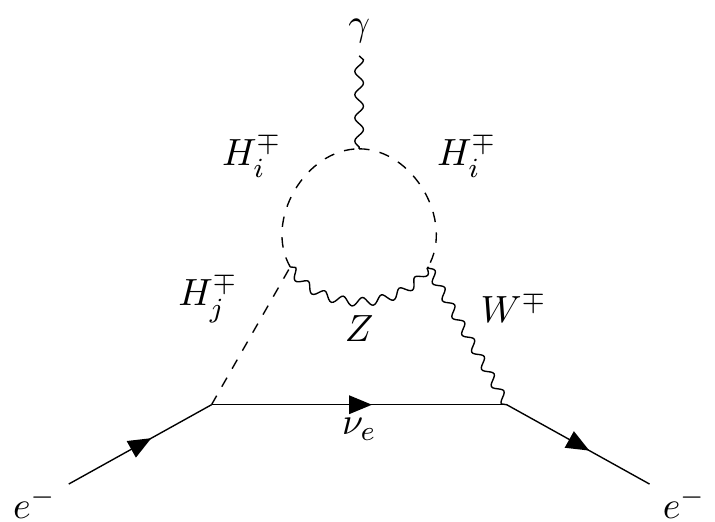}\hspace{0.2cm}
\includegraphics[width=0.32\textwidth]{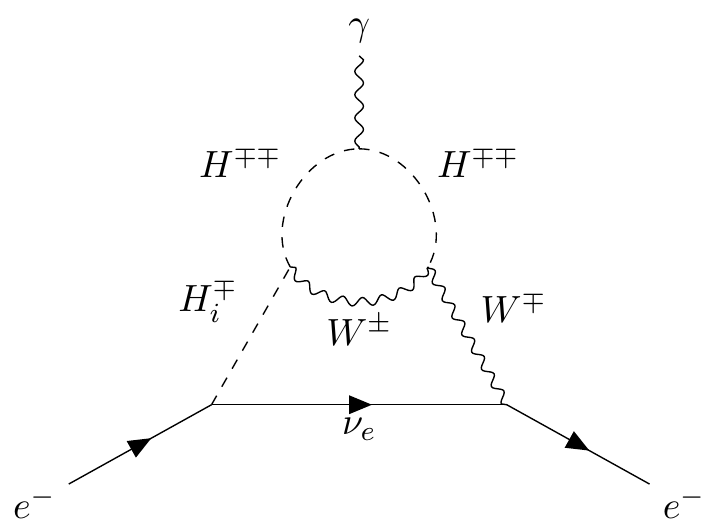}
\\
\vspace{-0.4cm} 
(g)
\hspace{5cm} (h)
\caption{\label{BZ:gauge} Gauge-loop BZ diagrams with $V=\gamma,Z$.}
\end{figure}

\begin{figure}[t]
\centering
\includegraphics[width=0.32\textwidth]{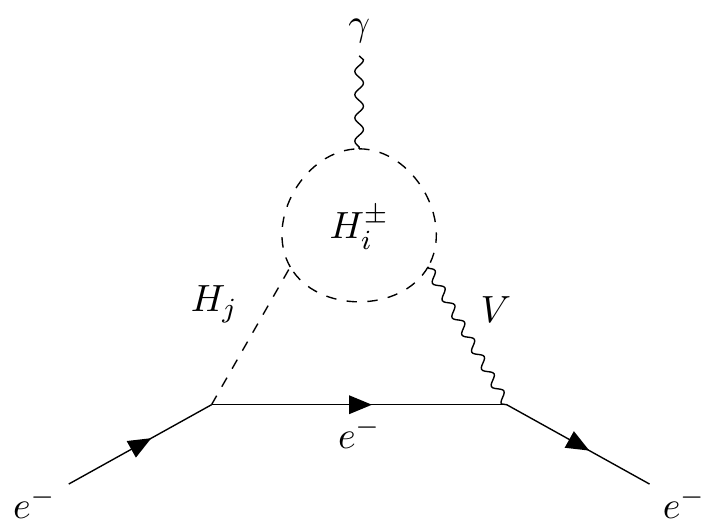}\hspace{0.2cm}
\includegraphics[width=0.32\textwidth]{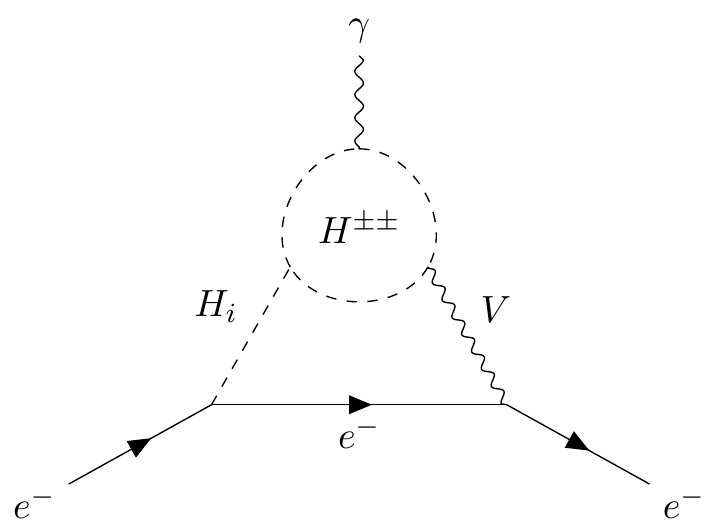}\hspace{0.2cm}
\includegraphics[width=0.32\textwidth]{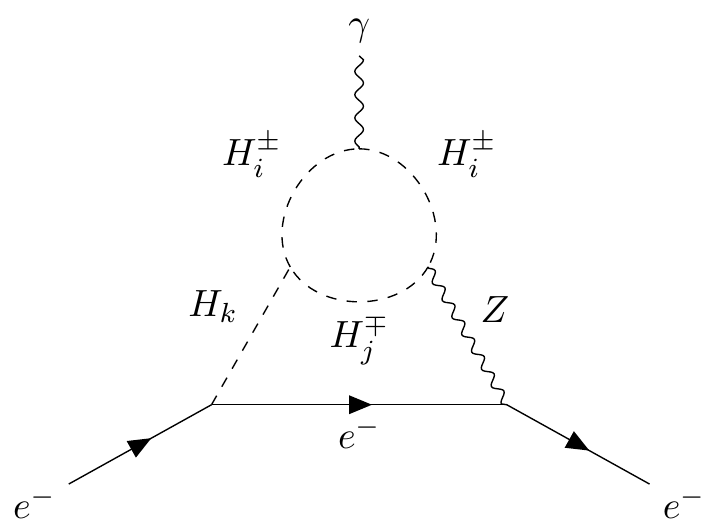}
\\ \vspace{-0.4cm}
(a)
\hspace{5cm} (b)
\hspace{5cm} (c)
\\
\includegraphics[width=0.32\textwidth]{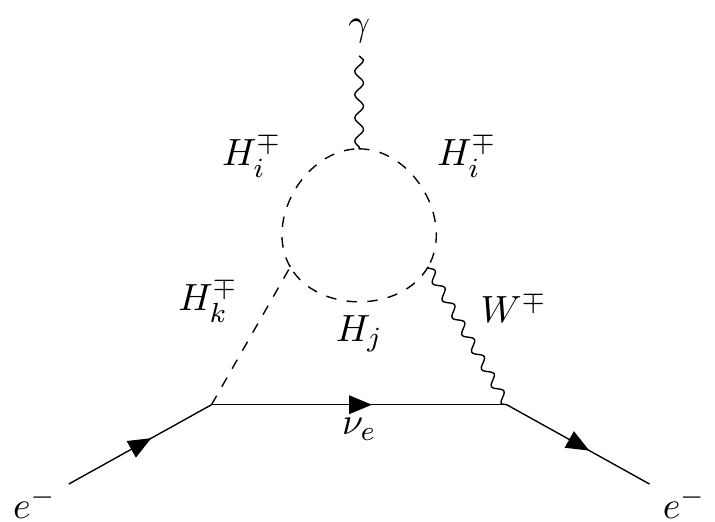}\hspace{0.2cm}
\includegraphics[width=0.32\textwidth]{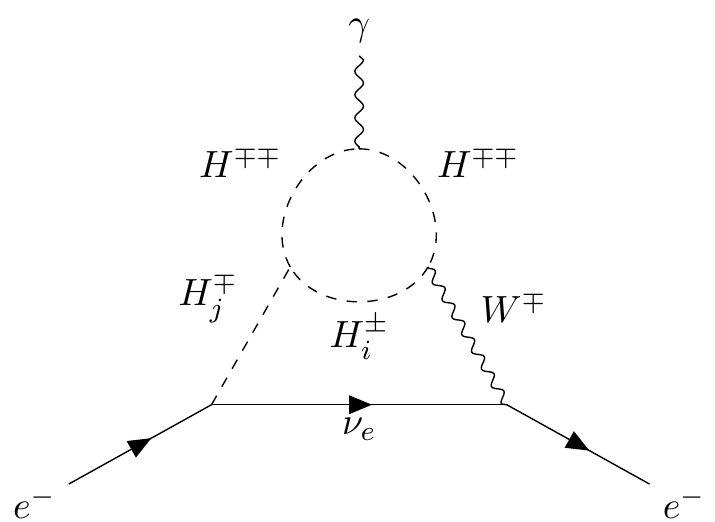}\hspace{0.2cm}
\includegraphics[width=0.32\textwidth]{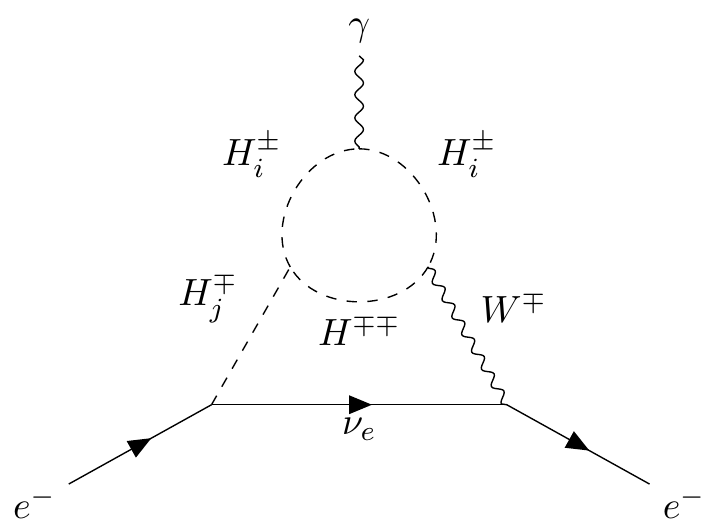}
\\
(d)
\hspace{5cm} (e)
\hspace{5cm} (f)
\caption{\label{BZ:higgs} Scalar-loop BZ diagrams with $V=\gamma,Z$.}
\end{figure}

We define the effective Lagrangian for the EDM operator for a fermion $f$ as 
\begin{align}
{\cal L}_{\rm EDM} = -\frac{d_f}{2} \bar{f}\sigma_{\mu\nu}(i\gamma_5)f F^{\mu\nu}, 
\end{align}
where $F^{\mu\nu}$ is the electromagnetic field strength tensor and $\sigma_{\mu\nu} \equiv \frac{i}{2}[\gamma_\mu,\gamma_\nu]$.

The most stringent bound on the eEDM is reported in Ref.~\cite{Roussy:2022cmp} as
\begin{align}
\vert d_e\vert <4.1\times10^{-30}~e\, \text{cm} \label{eq:edm}
\end{align}
at 90\% confidence level (CL).  As typical new physics models, one-loop contributions to the eEDM in our model are significantly suppressed by the square of the small electron Yukawa coupling with respect to the contributions from two-loop Barr-Zee (BZ) diagrams, and hence we can safely neglect the one-loop contributions. 
The contribution from the BZ-type diagrams can be classified into fermionic loops shown in Fig.~\ref{BZ:fermion} and bosonic loops shown in 
Figs.~\ref{BZ:gauge} and \ref{BZ:higgs}, respectively. 
The bosonic-loop contributions can further be decomposed into the ``gauge-loop'' and the ``scalar-loop'' ones, where the former involves just the gauge coupling while the latter involves also the scalar three-point couplings given in the potential. 
Details of the eEDM formulas are listed in Appendix~\ref{sec:a}, with some of the formulas adapted from the calculations given in Refs.~\cite{Abe:2013qla,Altmannshofer:2020shb}.

\subsection{nEDM}\label{sec:4:2}

The current bound on the nEDM is given by the nEDM Collaboration~\cite{Abel:2020pzs} as 
\begin{align}
\vert d_n\vert < 1.8\times10^{-26}~e\,\text{cm}
\end{align}
at 90\% CL.  We use the QCD sum rule to estimate its magnitude as~\cite{Abe:2013qla}
\begin{equation}
	d_n = 0.79d_d - 0.20d_u + \frac{e}{g_3(m_q)}\left(0.59d_d^C + 0.30d_u^C\right) ,
\end{equation}
where $g_3(m_q)$ is the QCD gauge coupling constant at the $m_q$ scale and $d_{d,u}^C$ are the chromo-EDMs (CEDMs) of the $d,u$ quarks. 
In our model, the constraint from the nEDM is much weaker than that from the eEDM, because 
there is no particular enhancement for quark Yukawa interactions as in the type-I 2HDM, see the discussion given at the end of Sec.~\ref{sec:2-1}.

We note that the other flavor constraints such as $B \to X_s\gamma$ can easily be avoided even for the case with the masses of $H_{1,2}^\pm$ to be ${\cal O}(100)$ GeV when 
$v_\xi$ and $v_\chi$ are taken to be ${\cal O}(10)$ GeV or smaller
which corresponds the case with $\tan\beta \gtrsim 10$ in the type-I 2HDM. 
See e.g., Ref.~\cite{Haller:2018nnx} for the flavor constraints in the 2HDMs.

%--------------------------------------------------------------------------------------------------------------------------------------------------------------------------
\subsection{Tevatron and LHC Measurement Constraints}\label{sec:4:3}

We also consider constraints from measurements at the Tevatron and LHC, including the Higgs signal strengths and direct searches for additional scalar bosons.  
A complete list of these constraints has been compiled in Refs.~\cite{Chiang:2018cgb,Chen:2022zsh} 
and summarized in Tables~\ref{HSS}-\ref{DS:6} in Appendix~\ref{sec:b}.

%=================================================================================================
\section{Global Fit and Benchmark Study}\label{sec:5}

\begin{table}[t]
\centering
\begin{tabular}{>{\centering\arraybackslash}p{4.5cm}|>{\centering\arraybackslash}p{2.5cm}|>{\centering\arraybackslash}p{4.5cm}|>{\centering\arraybackslash}p{3.5cm}}
Parameters & $v_\chi,v_\xi$ & $\lambda_2,\lambda_3,\lambda_4,\lambda_5,\Re\sigma_4,\Im\sigma_4$ & $\mu_{\phi\xi},\mu_{\chi\xi},\Re\mu_{\phi\chi}$ \\
\colrule
Prior Range (Uniform) & $[0,30]$~GeV & $[-10,10]$ & $[-5,5]$~TeV \\
\end{tabular}
\caption{\label{input:prior} Priors of the input parameters used in \texttt{HEPfit}.  }
\end{table}

We use the Bayesian-based Markov Chain Monte Carlo package \texttt{HEPfit}~\cite{DeBlas:2019ehy} to explore the parameter space of the minimal extension model defined in Sec.~\ref{sec:minimal}. 
The priors of the input parameters are summarized in Table~\ref{input:prior}. 
We impose the theoretical and experimental constraints discussed in Sec.~\ref{sec:3} and Sec.~\ref{sec:4}, respectively, and the world-average value of the electroweak $\rho$ parameter. 
For each parameter point, we fix the values of $\lambda_1$ and $v_\phi$ such that $m_h \simeq 125$~GeV and $v \simeq 246$~GeV are satisfied.

\begin{figure}[t]
\centering
\includegraphics[width=0.55\textwidth]{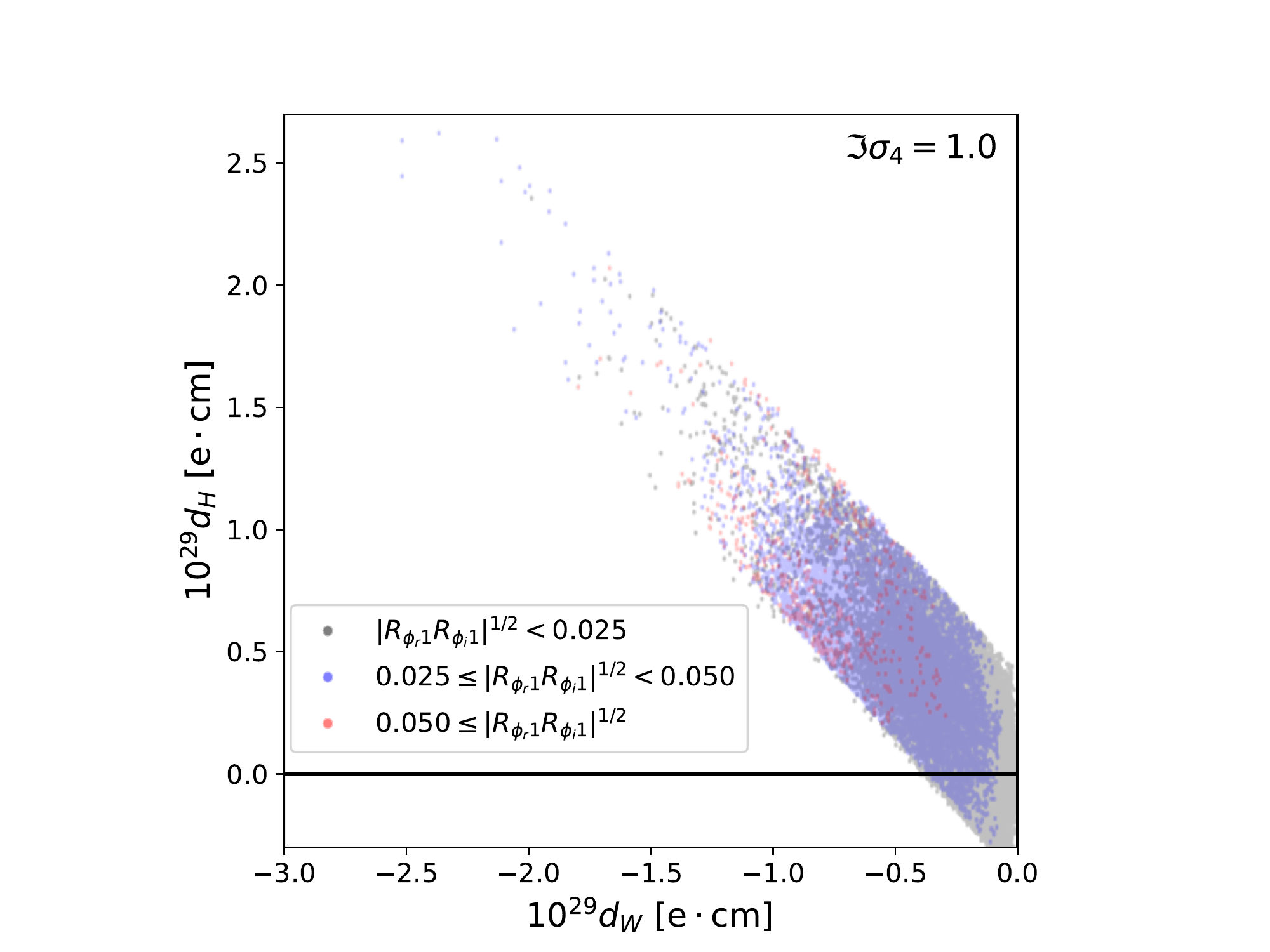}\hspace{-1.8cm}
\includegraphics[width=0.55\textwidth]{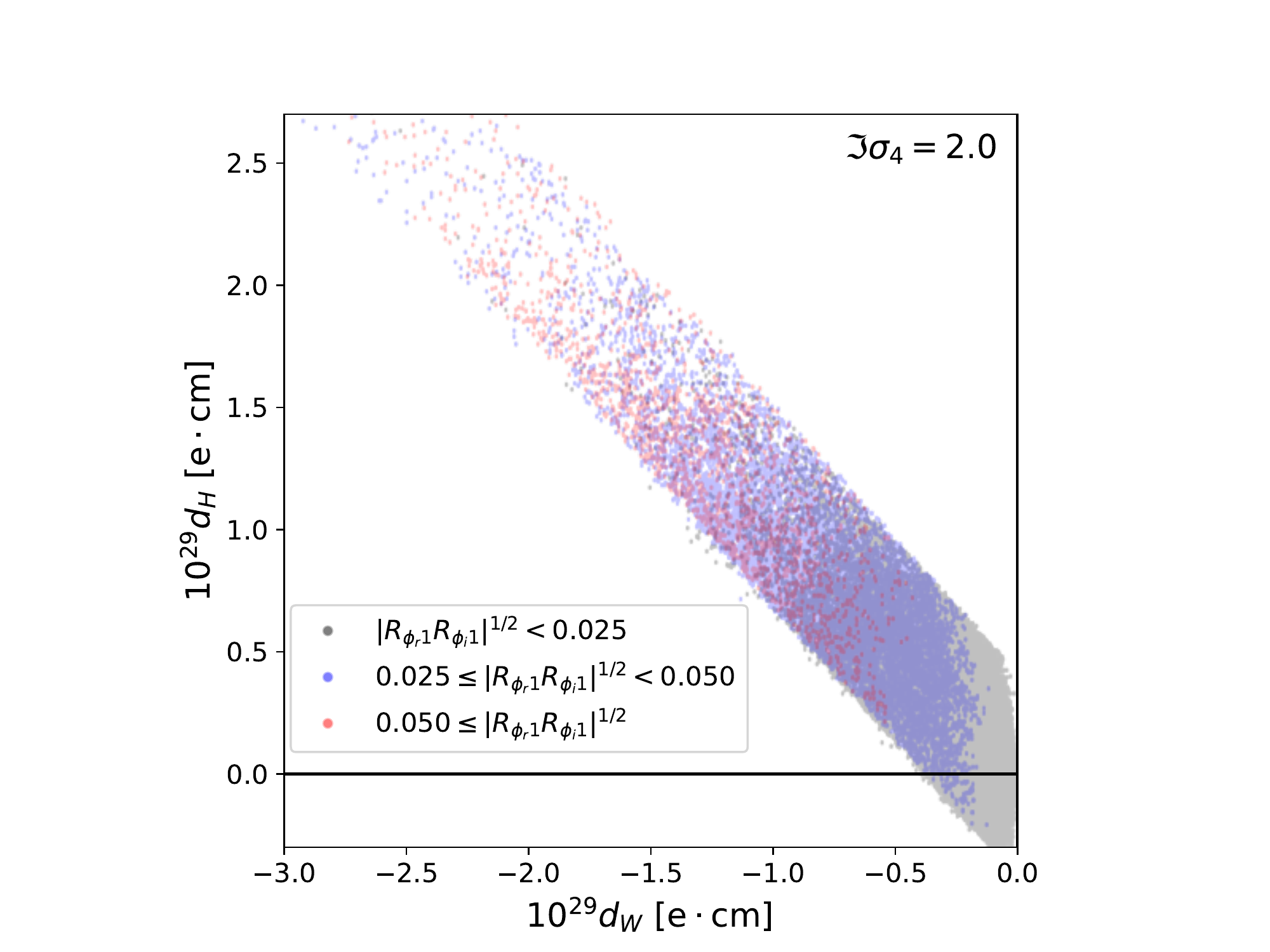}
\\
\includegraphics[width=0.55\textwidth]{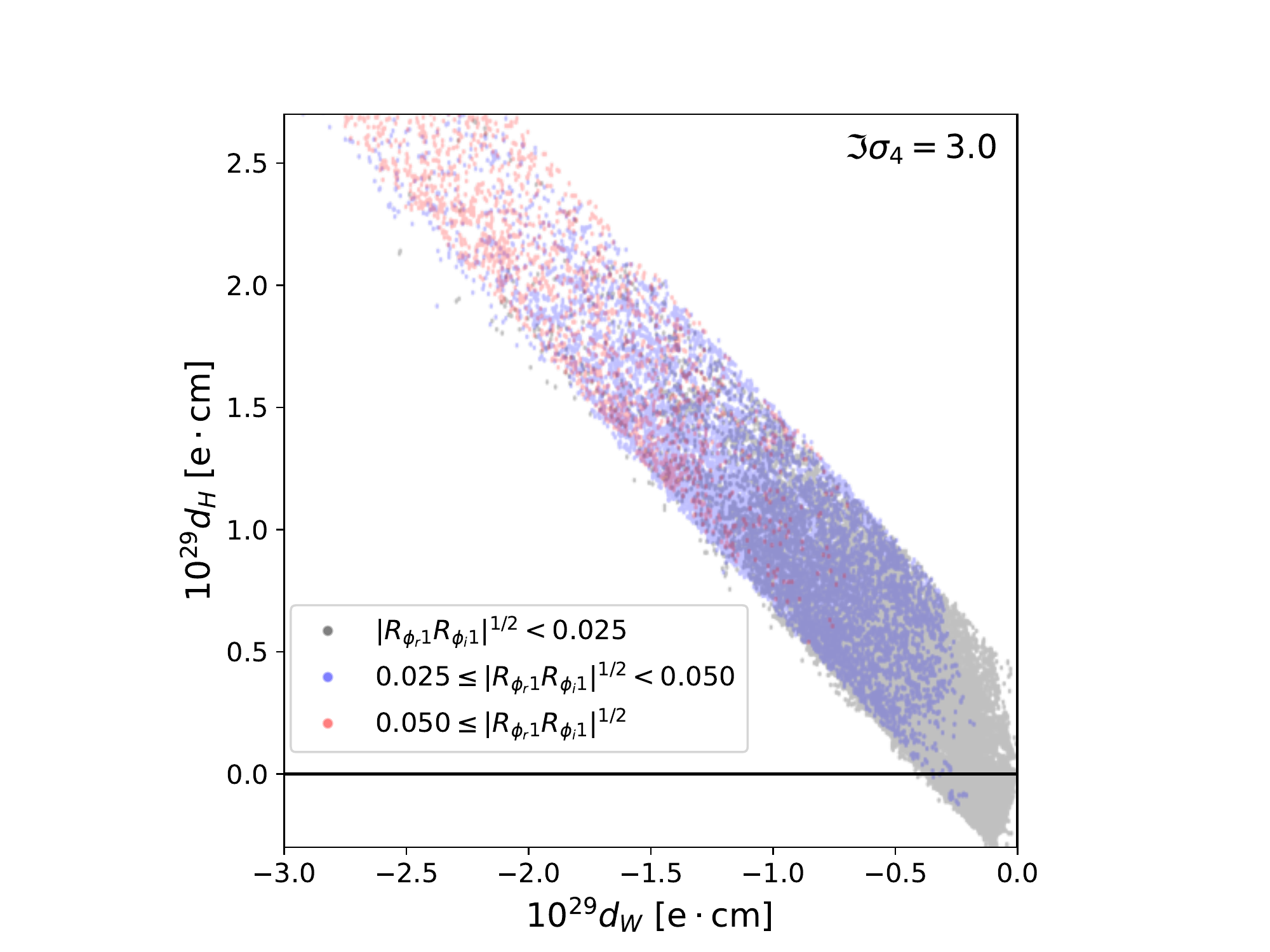}
\caption{\label{eEDM} Global fit distribution of the data with $m_{H_1}<1$~TeV and $\Im\sigma_4=1.0$ (top left), $\Im\sigma_4=2.0$ (top right), and $\Im\sigma_4=3.0$ (bottom) 
in the $d_W$-$d_H$ plane. The black, blue and red points respectively denote the data with $\left\vert R_{\phi_r1}R_{\phi_i1}\right\vert^{1/2}<0.025$, 
$0.025\leq\left\vert R_{\phi_r1}R_{\phi_i1}\right\vert^{1/2}<0.050$, and $0.050\leq\left\vert R_{\phi_r1}R_{\phi_i1}\right\vert^{1/2}$, where $R_{\phi_r1}$ and $R_{\phi_i1}$ are defined in Eq.~(\ref{eq:mixing}).
}
\end{figure}

We first present the global fit results for the eEDM.  
The current bounds from the nEDM turn out to be far weaker than the parameter ranges relevant to our later discussions, and hence we do not present them.  
From the fit results, we find that the contribution from fermion-loop diagrams is much smaller than that from the gauge-loop 
(denoted by $d_W$, defined by the sum of the diagrams shown in Fig.~\ref{BZ:gauge}) and the scalar-loop diagrams (denoted by $d_H$, defined by the sum of the diagrams shown in Fig.~\ref{BZ:higgs}), 
because the fermion-loop contribution is highly suppressed by the factor of $\sqrt{v_\chi^2+v_\xi^2}/v$ for each Yukawa coupling of the additional Higgs bosons. 
We thus show the correlation between $d_W$ and $d_H$ in Fig.~\ref{eEDM} for a fixed value of $\Im\sigma_4$, chosen to be 1 (upper-left), 2 (upper-right) and 3 (lower). We note that flipping the sign of $\Im\sigma_4$ would cause the distribution to reflect with respect to the origin.
Since we are particularly interested in the case where the additional Higgs bosons are not decoupled from the theory, we here impose 
the condition $m_{H_1}<1$~TeV. 
In this figure, we classify the predictions into three regions, with 
$0\leq \left\vert R_{\phi_r1}R_{\phi_i1}\right\vert^{1/2}< 0.025$ (black dots), 
$0.025\leq \left\vert R_{\phi_r1}R_{\phi_i1}\right\vert^{1/2} < 0.050$ (blue dots), and 
$0.050 \leq \left\vert R_{\phi_r1}R_{\phi_i1}\right\vert^{1/2}$ (red dots). 
Clearly, we see that the dots tend to appear at the upper-left region for larger $\left\vert R_{\phi_r1}R_{\phi_i1}\right\vert^{1/2}$, in which 
$|d_W|$ and $|d_H|$ become sizable, but the signs of these two are opposite. 
This means that a cancellation occurs between two contributions in order to satisfy the current limit on eEDM. \footnote{See also Refs.~\cite{Kanemura:2020ibp,Fuyuto:2019svr} 
for the other types of cancellations in the eEDM in 2HDMs. }
We also see that larger values of $|d_W|$ and $|d_H|$ tend to be obtained for larger $\Im\sigma_4$ because $\Im\sigma_4$ is the unique CPV source of the model, see Appendix~\ref{sec:a}.

\begin{figure}[t]
\includegraphics[width=0.55\textwidth]{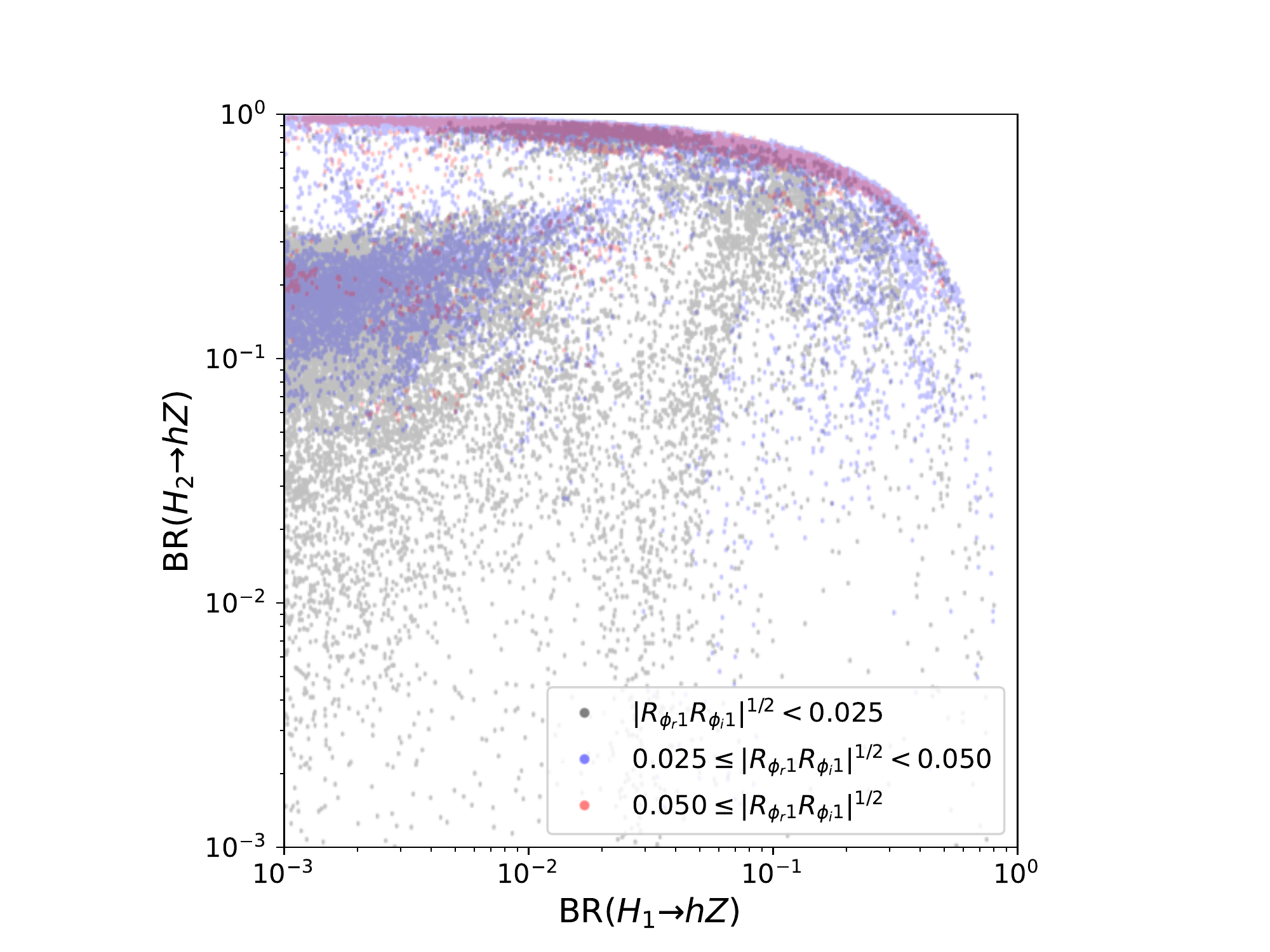}\hspace{-1.8cm}
\includegraphics[width=0.55\textwidth]{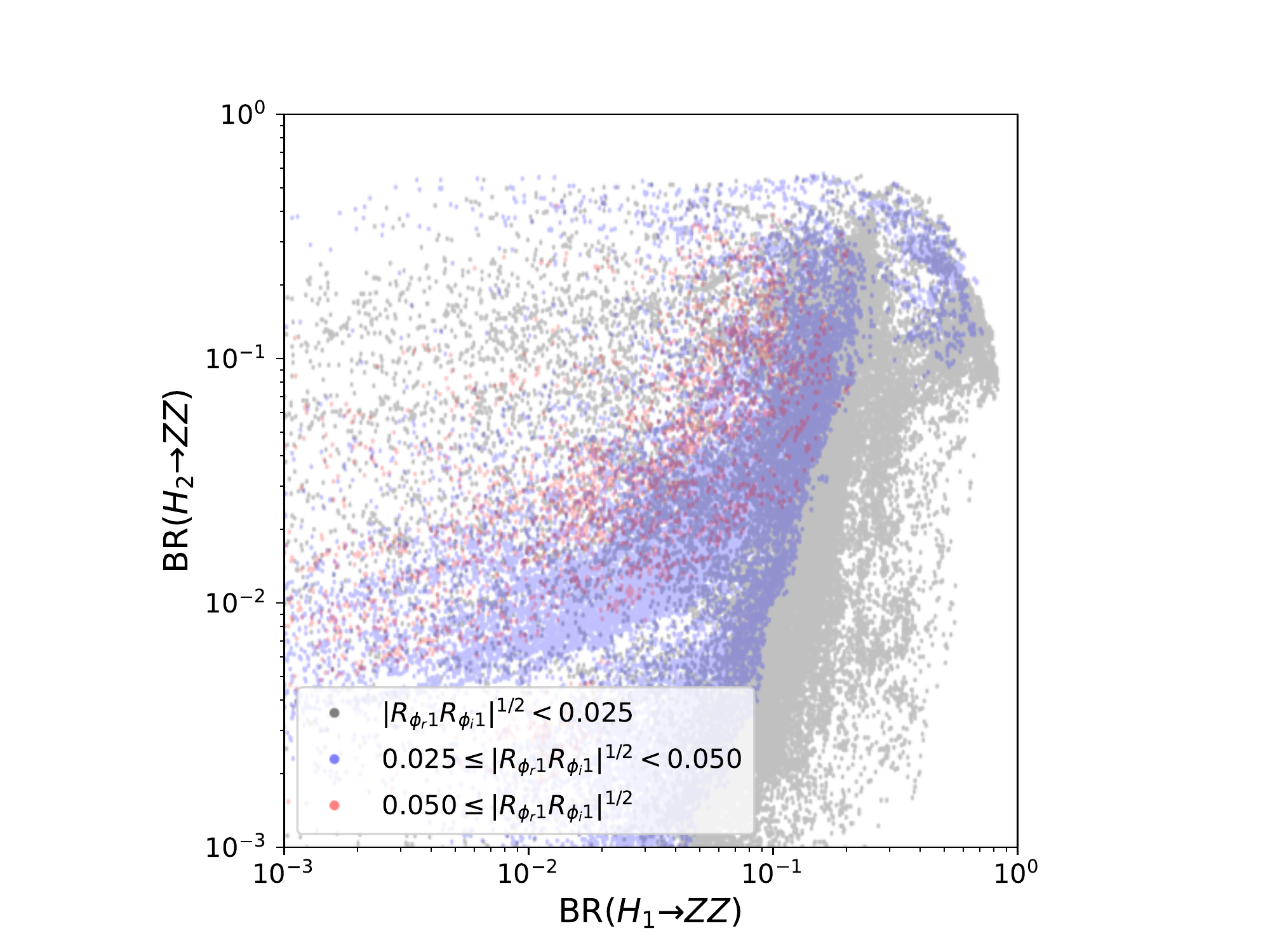}
\caption{\label{BR:hZ} Global fit distribution in the ${\rm BR}(H_1\to hZ)$-${\rm BR}(H_2\to hZ)$ plane (left) and the ${\rm BR}(H_1\to ZZ)$-${\rm BR}(H_2\to ZZ)$ plane (right) with $\Im\sigma_4=3.0$. 
The black, blue and red points respectively denote the data with $\left\vert R_{\phi_r1}R_{\phi_i1}\right\vert^{1/2}<0.025$, 
$0.025\leq\left\vert R_{\phi_r1}R_{\phi_i1}\right\vert^{1/2}<0.050$, and $0.050\leq\left\vert R_{\phi_r1}R_{\phi_i1}\right\vert^{1/2}$, where $R_{\phi_r1}$ and $R_{\phi_i1}$ are defined in Eq.~(\ref{eq:mixing}). 
}
\end{figure}

Next, we study the $H_{1,2}\to hZ$ decays. 
In the CP-conserving (CPC) limit, only one additional neutral Higgs boson, the CP-odd one, can decay into the $hZ$ state, so that a simultaneous observation of the Higgs bosons decaying into $hZ$ would be direct evidence of CP-mixed couplings.
The $H_3$ state is often much heavier than the other neutral states; so it is harder to produce in collider experiments, and thus we focus on the decays of $H_1$ and $H_2$.  In both plots of Fig.~\ref{BR:hZ}, we fix $\Im\sigma_4=3$.
In the left panel, we show the correlation between ${\rm BR}(H_1\to hZ)$ and ${\rm BR}(H_2\to hZ)$. 
Again, we separate the data into three subsets based on the value of $\left\vert R_{\phi_r1}R_{\phi_i1}\right\vert^{1/2}$. 
It can be seen that most of the dots tend to accumulate in the upper-right region for larger $\left\vert R_{\phi_r1}R_{\phi_i1}\right\vert^{1/2}$. 
Therefore,  both ${\rm BR}(H_1\to hZ)$ and ${\rm BR}(H_2\to hZ)$ become larger when the individual eEDM contributions are greater. 
Furthermore, ${\rm BR}(H_2\to hZ)$ is mostly greater than ${\rm BR}(H_1\to hZ)$, which implies that $H_2$ is often mostly CP-odd and $H_1$ mostly CP-even. 
This feature can also be seen from the ${\rm BR}(H_1\to ZZ)$-${\rm BR}(H_2\to ZZ)$ distribution shown in the right panel of Fig.~\ref{BR:hZ}, where ${\rm BR}(H_1\to ZZ)$ is mostly greater than ${\rm BR}(H_2\to ZZ)$. 
As we demonstrate more explicitly in the benchmark study, when $\vert\Im\sigma_4\vert$ increases, the enhanced CPV will make the two states further mix, which allows greater ${\rm BR}(H_1\to hZ)$ and ${\rm BR}(H_2\to ZZ)$. 
It is worth noting that the mass spectrum, e.g., $H_2$ is mostly CP-odd, is consistent with the findings of a prior global fit analysis presented in Ref.~\cite{Chiang:2018cgb}, in which the mass hierarchy of $m_{H_Q}>m_{H_T}>m_{H}$ or $m_{H}>m_{H_T}>m_{H_Q}$ is favored after accounting for all the theoretical and experimental constraints in the original $SU(2)_V$-symmetric GM model.  However, due to the explicit $SU(2)_V$ symmetry breakdown in the potential, it is unclear which Higgs boson belongs to which $SU(2)_V$ multiplet.  Nevertheless, the mass of the CP-odd state $H_T^0$ is expected to be between the two CP-even states, i.e., the mixture of $H_Q^0$ and $H$ states in the CPC limit.

\begin{figure}[t]
\centering
\includegraphics[width=0.7\textwidth]{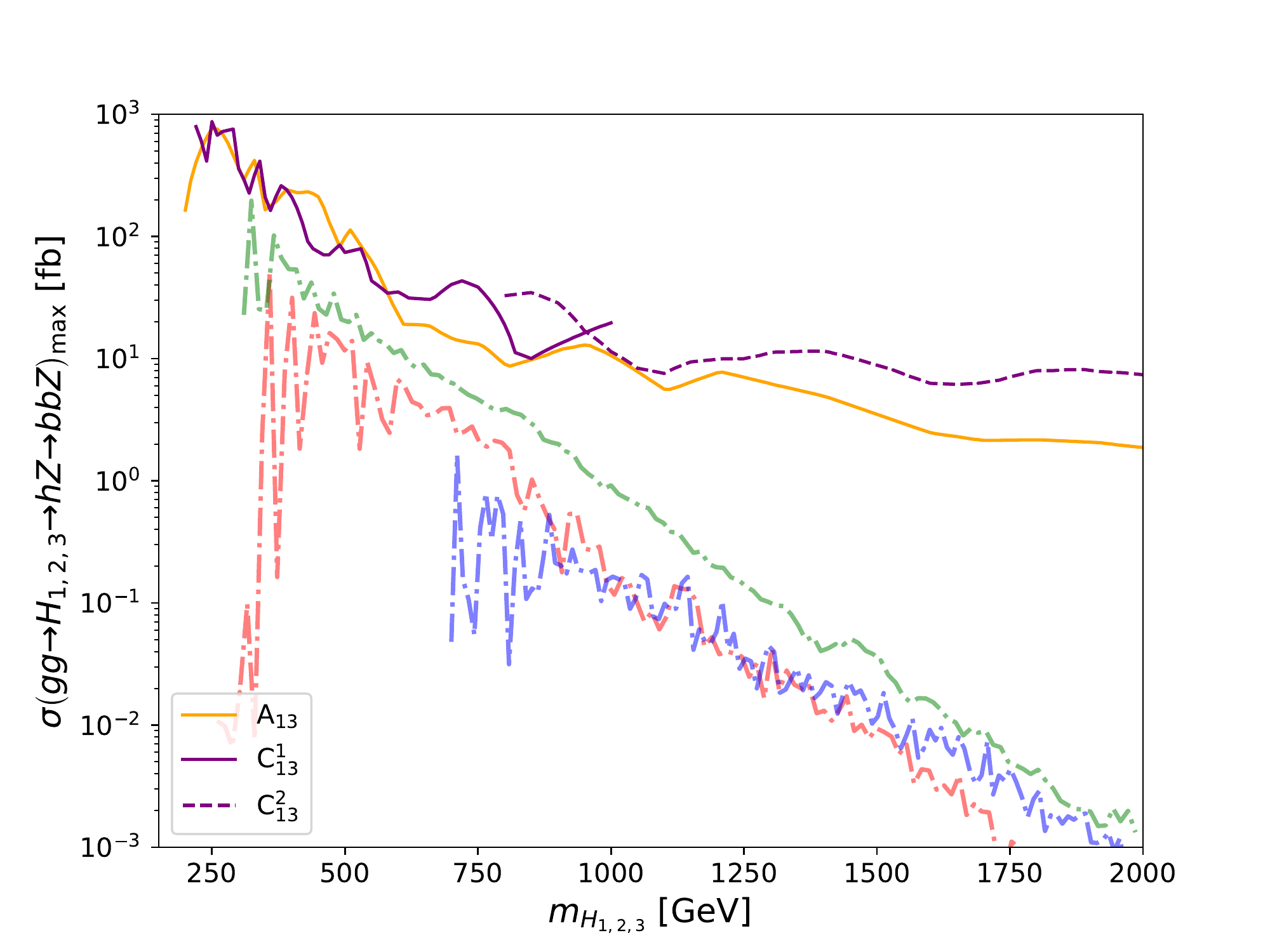}
\caption{\label{sigma:gg} Global fit upper bounds on $\sigma(gg\to H_1\to hZ\to bbZ)$ (red), $\sigma(gg\to H_2\to hZ\to bbZ)$ (green), and $\sigma(gg\to H_3\to hZ\to bbZ)$ (blue) at the 13-TeV LHC with respect to $m_{H_1,H_2,H_3}$. The 13-TeV ATLAS and CMS search bounds at 95\% CL are also given.
}
\end{figure}

Focusing on the $H_ihZ$ couplings, we study the current LHC sensitivity to the $gg\to H_i\to hZ\to bbZ$ processes.  We plot the global fit upper limits of $\sigma(gg\to H_{1,2,3}\to hZ\to bbZ)$ at the 13-TeV LHC with respect to $m_{H_1,H_2,H_3}$ in Fig.~\ref{sigma:gg}, in which we also show the current 13-TeV ATLAS and CMS search bounds at 95\% CL.  While the $H_3$-mediated process is mostly far below the current bounds, the $H_1$- and $H_2$-mediated processes can be very close to the bounds for masses below $750$~GeV.

\begin{table}[t]
\begin{tabular}{>{\centering\arraybackslash}p{2.1cm}||>{\centering\arraybackslash}p{1.2cm}|>{\centering\arraybackslash}p{1.2cm}|>{\centering\arraybackslash}p{1.2cm}|>{\centering\arraybackslash}p{1.2cm}|>{\centering\arraybackslash}p{1.2cm}|>{\centering\arraybackslash}p{1.2cm}|>{\centering\arraybackslash}p{1.2cm}|>{\centering\arraybackslash}p{1.4cm}|>{\centering\arraybackslash}p{1.4cm}|>{\centering\arraybackslash}p{1.4cm}}
Benchmarks & $v_\chi$ & $v_\xi$ & $\lambda_2$ & $\lambda_3$ & $\lambda_4$ & $\lambda_5$ & $\Re\sigma_4$ & $\mu_{\phi\xi}$ & $\mu_{\chi\xi}$ & $\Re\mu_{\phi\chi}$ \\
\toprule
1 & 7.11 & 7.51 & 0.319 & $-0.207$ & 0.045 & $-0.254$ & 0.234 & 1080 & $-11100$ & $-410$ \\
\colrule
2 & 8.81 & 8.90 & 0.639 & $-0.391$ & 0.169 & 0.527 & $-0.350$ & 680 & $-20700$ & $-290$ \\
\end{tabular}
\caption{\label{BPs} Input parameters in the two benchmarks.  The dimensionful parameters are all given in units of GeV. }\vspace{1cm}
\begin{tabular}{>{\centering\arraybackslash}p{2.1cm}||>{\centering\arraybackslash}p{1.2cm}|>{\centering\arraybackslash}p{1.2cm}|>{\centering\arraybackslash}p{1.2cm}|>{\centering\arraybackslash}p{1.2cm}|>{\centering\arraybackslash}p{1.2cm}|>{\centering\arraybackslash}p{1.2cm}}
Benchmarks & $m_{H_1}$ & $m_{H_2}$ & $m_{H_3}$ & $m_{H_1^\pm}$ & $m_{H_2^\pm}$ & $m_{H^{\pm\pm}}$ \\[1.02ex]
\toprule
1 & 475 & 477 & 1773 & 540 & 1773 & 598 \\
\colrule
2 & 562 & 578 & 1327 & 660 & 1324 & 748 \\
\end{tabular}
\caption{\label{scalar:mass} Masses of the additional Higgs bosons in GeV of the two benchmarks.}
\vspace{1cm}
\begin{tabular}{>{\centering\arraybackslash}p{2.1cm}||>{\centering\arraybackslash}p{1.2cm}|>{\centering\arraybackslash}p{1.2cm}|>{\centering\arraybackslash}p{1.2cm}|>{\centering\arraybackslash}p{1.2cm}}
Benchmarks & $\delta_{ff,gg}$ & $\delta_{\gamma\gamma}$ & $\delta_{WW}$ & $\delta_{ZZ}$ \\
\toprule
1 & $-0.005$ & $0.110$ & $0.015$ & $0.018$ \\
\colrule
2 & $-0.007$ & $0.150$ & $0.023$ & $0.028$ \\
\end{tabular}
\caption{\label{herr} Predictions of $\delta_{XX}^{}$ defined in Eq.~(\ref{eq:delta}) in the two benchmarks. }
\end{table}

In the following, we select two benchmarks with relatively large $\sigma(gg\to H_2\to hZ\to bbZ)$ at 14~TeV to perform a more in-depth study, 
with benchmark 1 having a more stringent upper bound on $\vert\Im\sigma_4\vert$ and benchmark 2 a weaker one, a result of their different $d_W$-$d_H$ cancellation patterns.  
In the later part of this paper, we will demonstrate that this difference between the two benchmarks will lead to distinguishable outcomes if more stringent bounds on the eEDM are imposed in the future.  Additionally, both of these benchmarks include additional Higgs bosons of sub-TeV masses and, as we will show later, they give rise to $\sigma\left(gg\to H_2\to hZ\to bbZ\right)$ of $\mathcal{O}(10^1)$~fb and $\sigma\left(gg\to H_1\to hZ\to bbZ\right)$ of $\mathcal{O}(10^0)$~fb, respectively, under the current constraints.  These findings should motivate ongoing searches in this regime.  We fix all the input parameters except for $\Im\sigma_4$, and also apply all the constraints mentioned earlier.  
As we will show later, $\vert\Im\sigma_4\vert$ is primarily constrained by eEDM for benchmark 1 and by the other theoretical and collider measurement constraints for benchmark 2. Such a difference is due to the level of cancellation between $d_W$ and $d_H$, which for benchmark 1 is characterized by $d_H/d_W\simeq -0.55$ and for benchmark 2 by $d_H/d_W\simeq -1.1$ as they scale with the varying $\Im\sigma_4$.
The two benchmarks are summarized in Table~\ref{BPs}. 
The mass spectra for the scalar bosons in the two benchmarks are presented in Table~\ref{scalar:mass}.  
We note that while $m_{H^{\pm\pm}}$ is independent of $\Im\sigma_4$, the other scalar masses can only be maximally changed by ${\cal O}(2 \%)$ with respect to the CPC limit.
In Table~\ref{herr}, we show the deviation in the branching ratios of the SM-like Higgs boson $h$ from the SM predictions, characterized by
\begin{align}
\delta_{XX}^{} \equiv \frac{ {\rm BR}(h\to X\bar{X})-{\rm BR}(h\to X\bar{X})_{\rm SM}}{{\rm BR}(h\to X\bar{X})_{\rm SM}}~~\text{with}~~X \in \{ f, g, \gamma, W, Z\}, \label{eq:delta}
\end{align}
where the values can only be maximally changed by below one percent level under the variation of $\Im\sigma_4$.  
Note that for the $ff$ and $gg$ channels, they are all modified by the same factor and thus of the same value.  Two remarks are in order.  First, the $WW$ and $ZZ$ deviations feature different behaviors in this model, which is in stark contrast to the original GM model where they should be identical.  The reason is due to the explicit violation of the custodial symmetry, leading to different $hWW$ and $hZZ$ coupling modifications at the tree level.  Second, while all the other deviations are below 3\%, the $\gamma\gamma$ channel can deviate from the SM prediction by up to $\sim10$\%.  While this certainly reflects the fact that the current measurement on $h\to\gamma\gamma$ does not quite agree with the SM prediction, it also shows that the effective $h\gamma\gamma$ coupling in our model is very sensitive to new physics contributions, including the charged Higgs bosons as well as the triplet-gauge couplings.  Thus, this could also serve as a promising probe of the model at the future LHC.

\begin{figure}[t]
\centering
\includegraphics[width=0.44\textwidth]{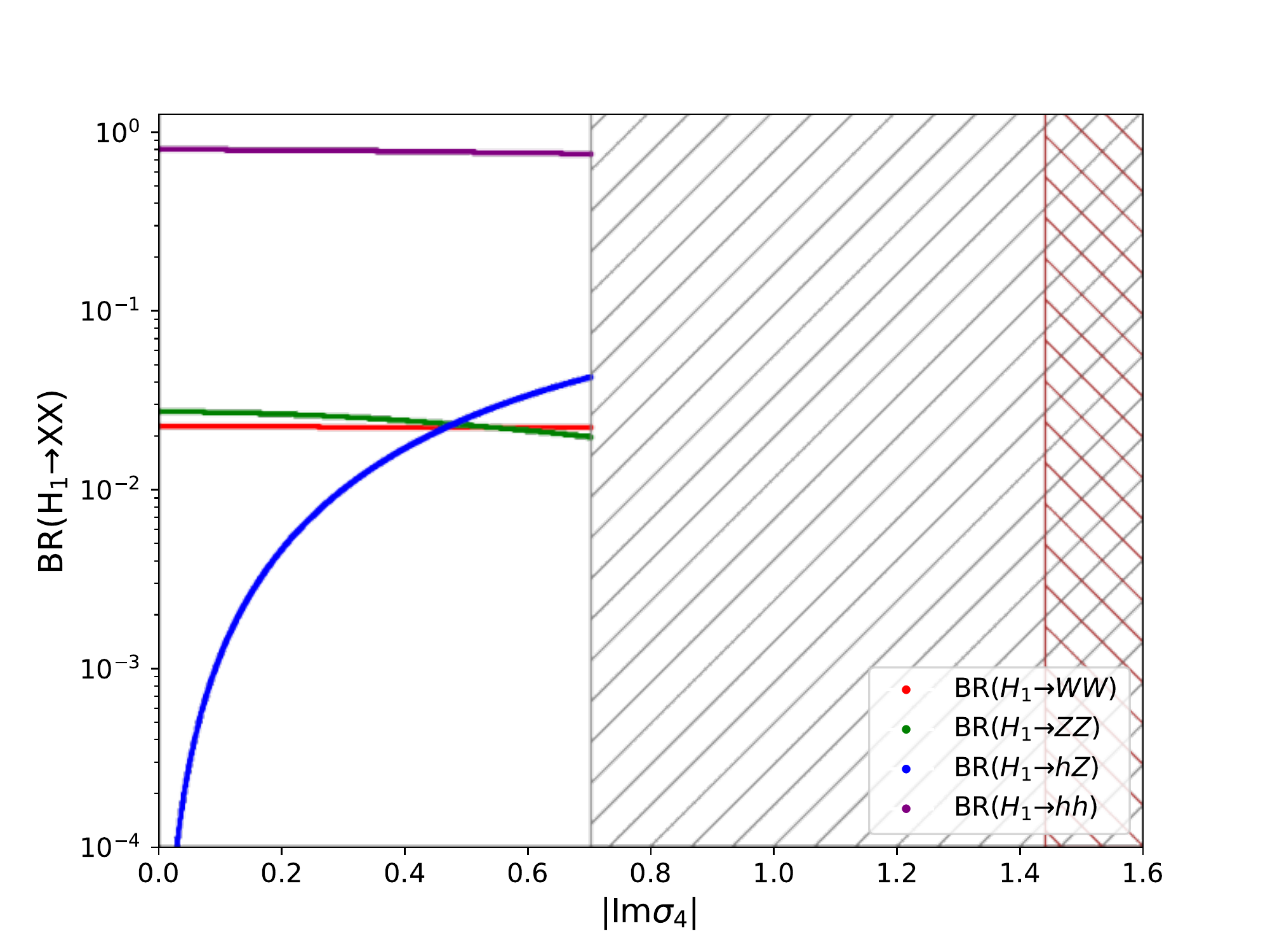}
\includegraphics[width=0.44\textwidth]{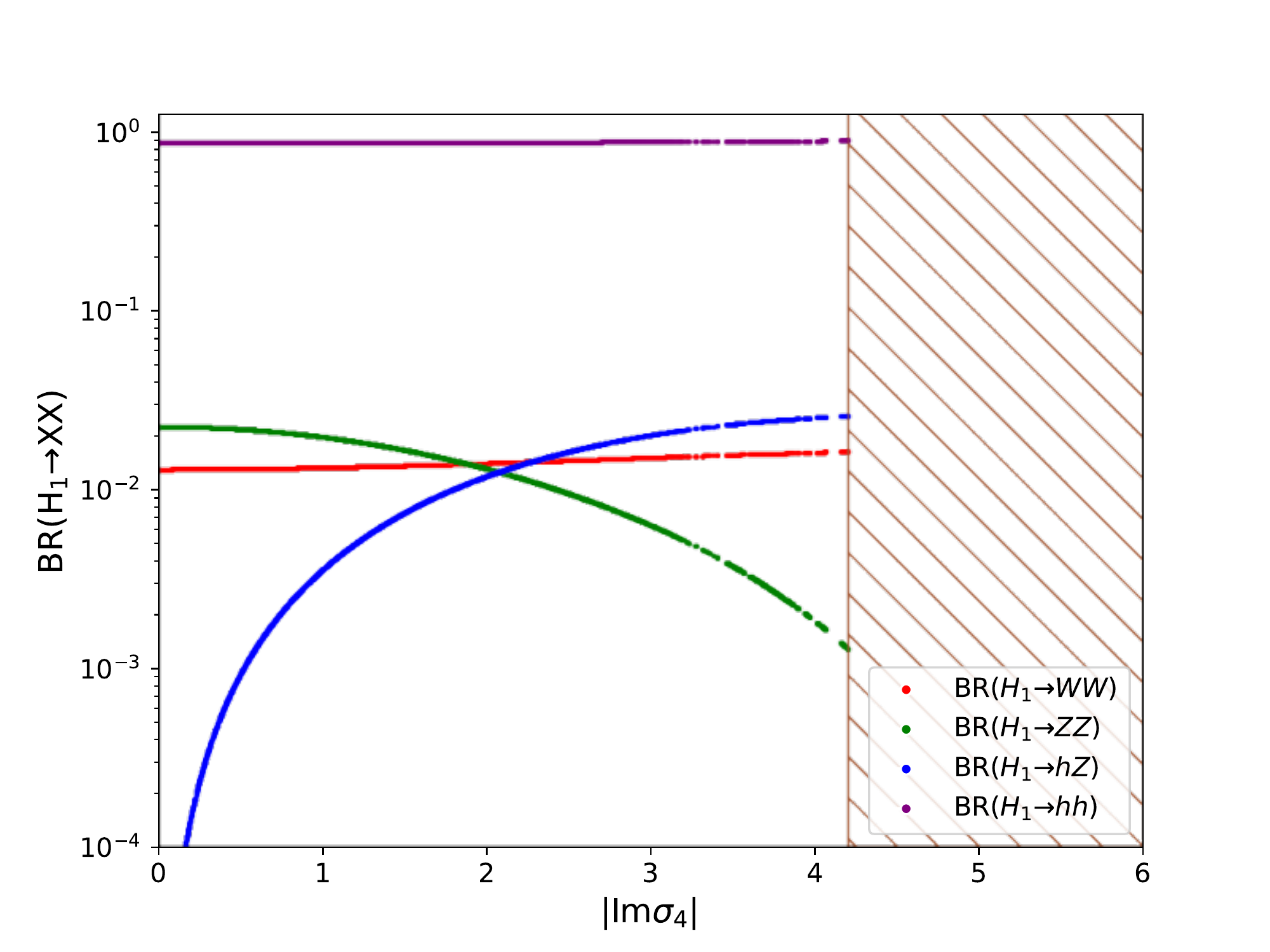}
\\
\includegraphics[width=0.44\textwidth]{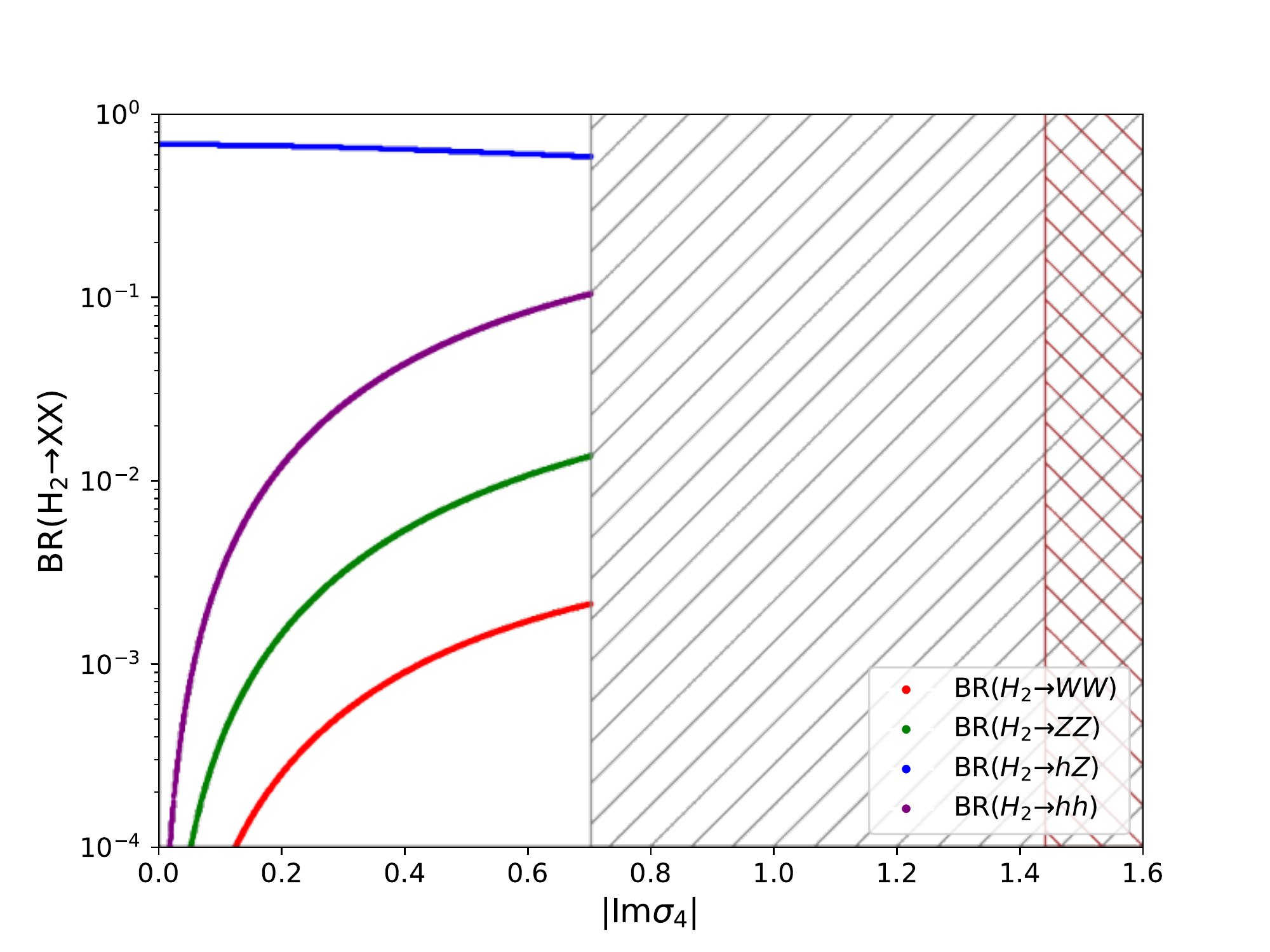}
\includegraphics[width=0.44\textwidth]{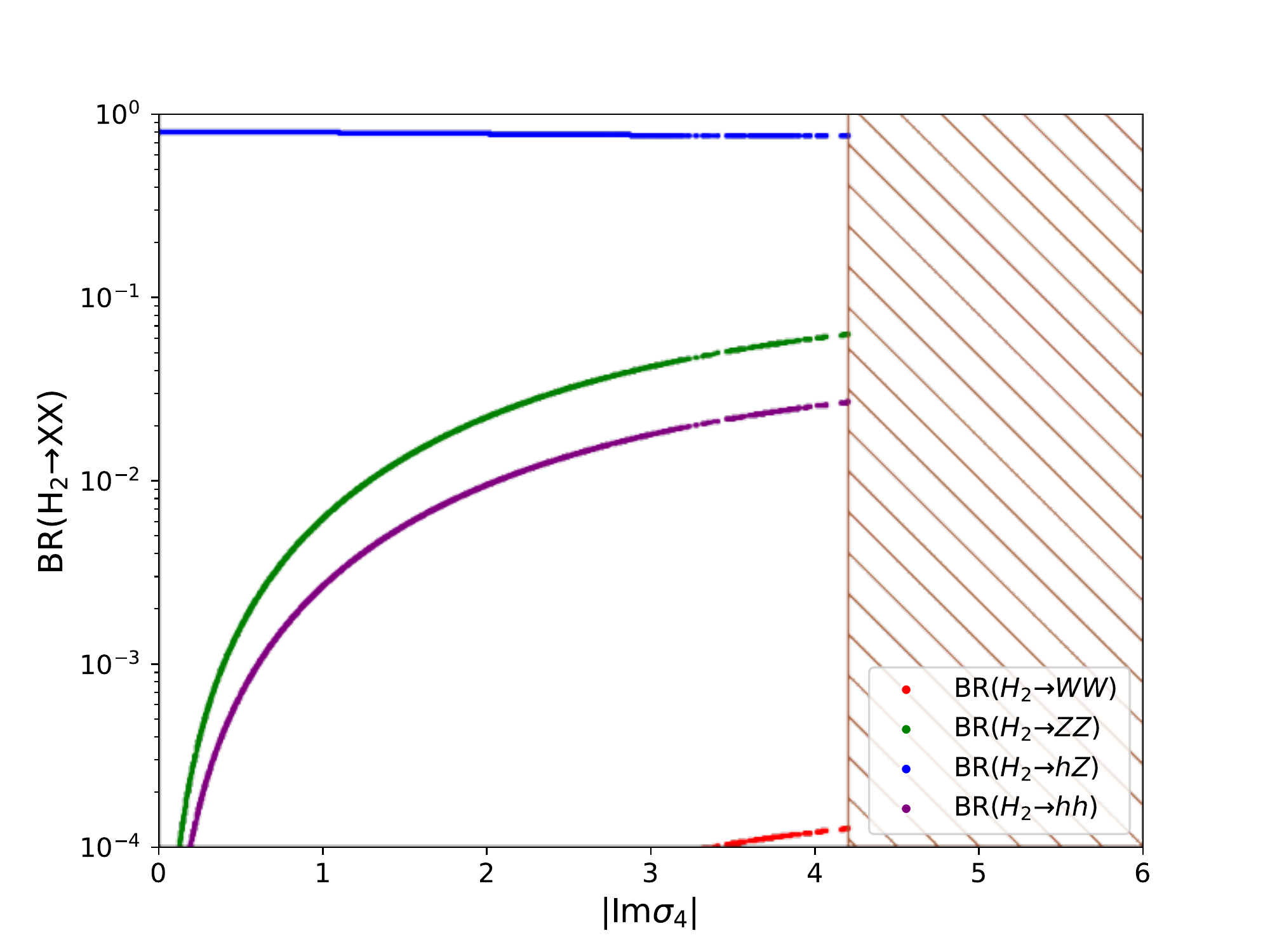}
\\
\includegraphics[width=0.44\textwidth]{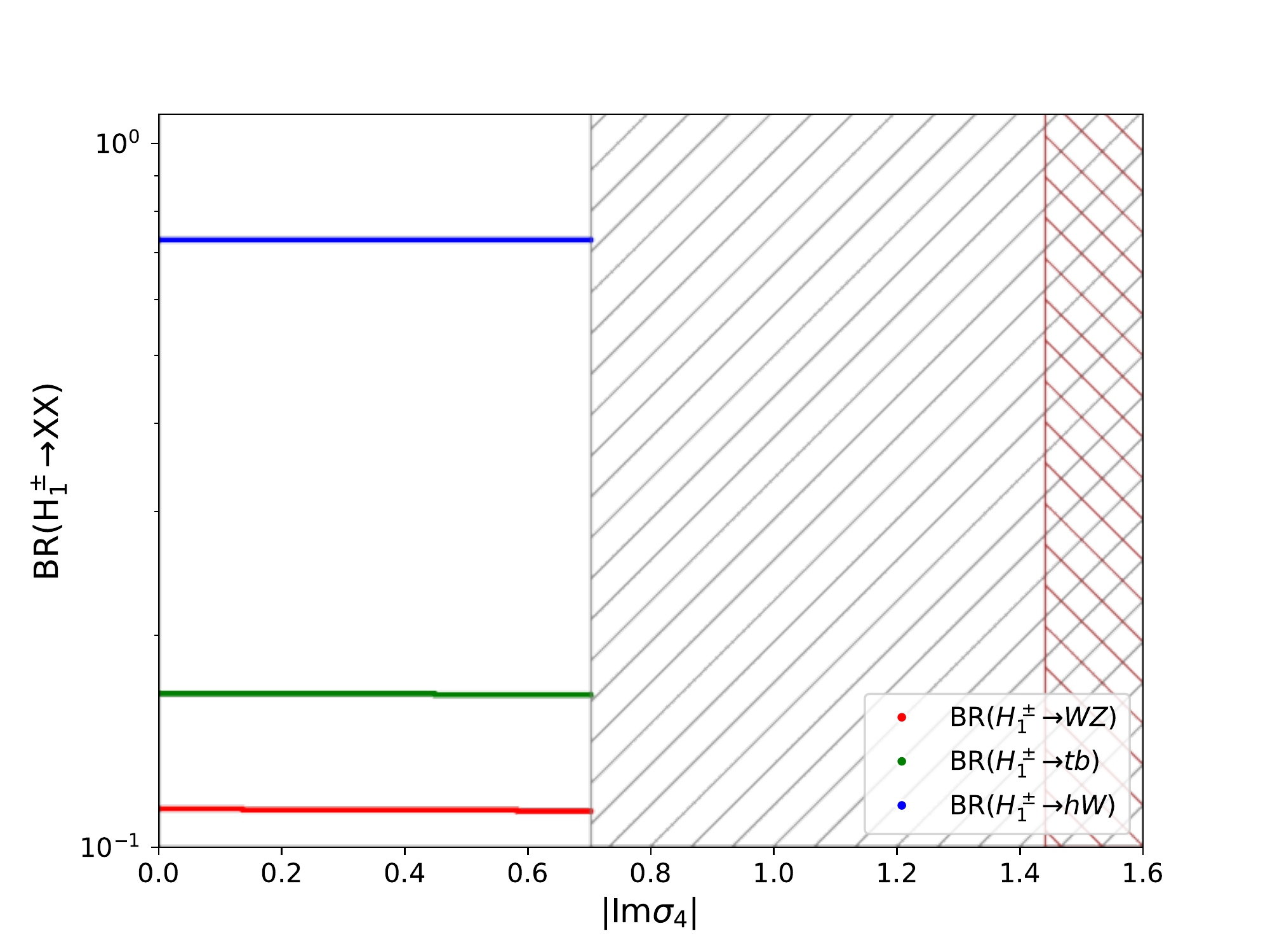}
\includegraphics[width=0.44\textwidth]{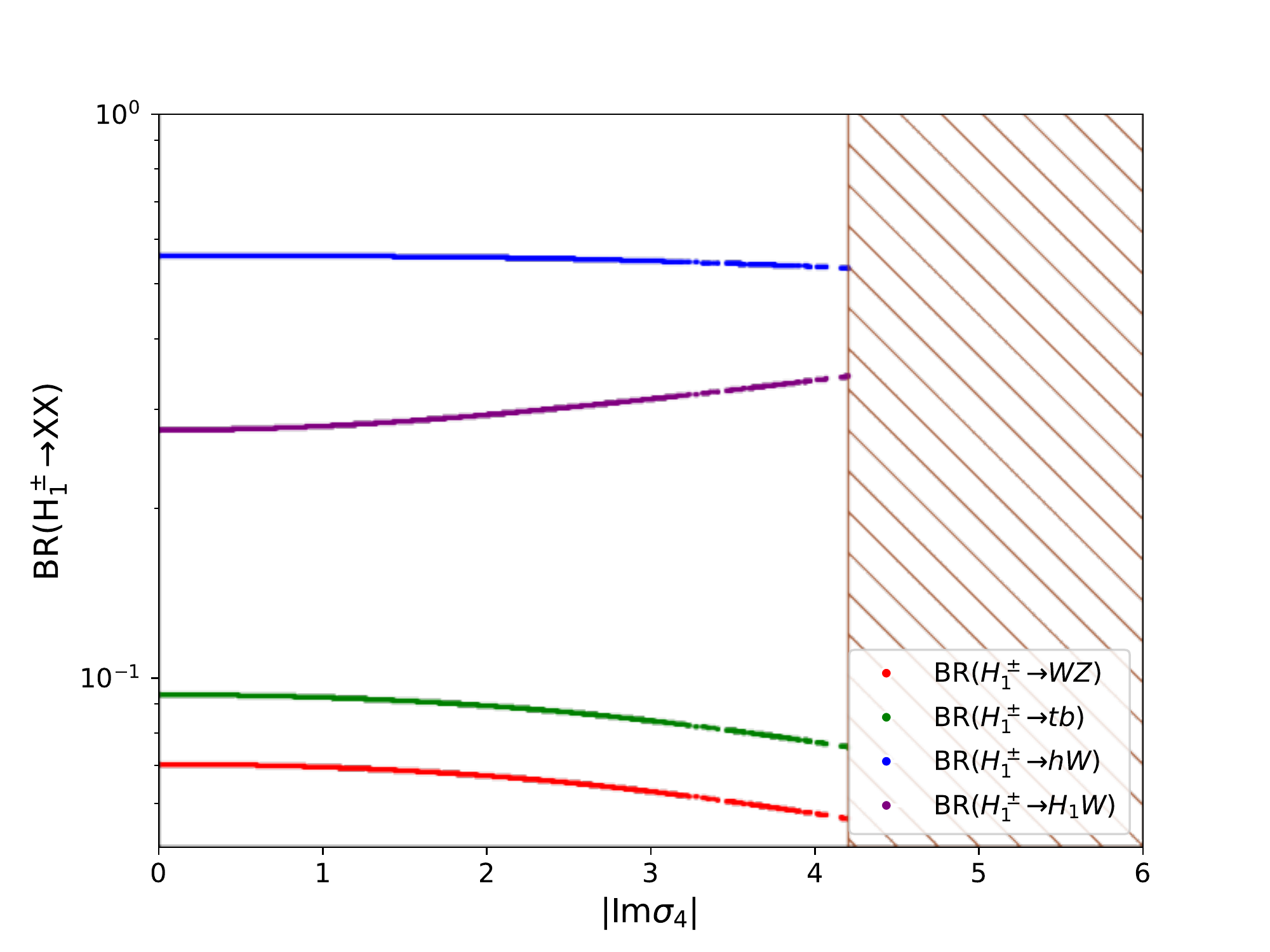}
\caption{\label{BP:BR} 
Branching ratios of the most dominant decay channels of $H_1$ (top row), $H_2$ (middle row) and $H_1^\pm$ (bottom row) for benchmark 1 (left column) and benchmark 2 (right column) as a function of $\vert\Im\sigma_4\vert$. The gray hatched regions are excluded by the eEDM constraint at 90\% CL, and the brown hatched regions by the other theoretical and collider measurement constraints at 95\% CL. The bound set by the eEDM constraint for benchmark 2 is beyond the plotting range.
}
\end{figure}

Fig.~\ref{BP:BR} depicts the branching ratios of the most dominant channels of $H_1$ (top plots), $H_2$ (middle plots), and $H_1^\pm$ (bottom plots) for the benchmark 1 (left plots) and the benchmark 2 (right plots), 
where the region shaded in gray is excluded by the eEDM constraint at 90\% CL, and the brown hatched region by the other theoretical and collider measurement constraints at 95\% CL.  The bound set by the eEDM constraint for benchmark 2 is way beyond the plotting range, and thus we do not show it. The fact that these two types of measurements have different constraining power for the two benchmarks clearly illustrates that the direct searches at colliders can indeed complement the EDM searches in probing the CPV.
In both of the benchmarks, the $hh$ channel is the most dominant for $H_1$, while the $hZ$ channel is the most dominant for $H_2$. 
We are particularly interested in the behavior of ${\rm BR}(H_{1,2}\to hZ)$. 
In either case, ${\rm BR}(H_1\to hZ)$ and ${\rm BR}(H_2\to hZ)$ respectively reach their minimum and maximum for the CPC limit, i.e., $\Im\sigma_4=0$.  As $\vert\Im\sigma_4\vert$ increases, there is more CP-mixing between $H_{1,2}$, causing ${\rm BR}(H_1\to hZ)$ to increase and ${\rm BR}(H_2\to hZ)$ to decrease.  We also remark that for both benchmarks, ${\rm BR}(H_1^\pm\to hW)$ always dominates, followed by ${\rm BR}(H_1^\pm\to tb)$ and ${\rm BR}(H_1^\pm\to WZ)$ for the benchmark 1 and by ${\rm BR}(H_1^\pm\to H_1W)$ and ${\rm BR}(H_1^\pm\to tb)$ for the benchmark 2.  
In particular, the fact that ${\rm BR}(H_1^\pm\to WZ)\sim\mathcal{O}(10^{-1})$ for the benchmark 1 can serve as a key signature to differentiate this model from the 2HDMs that do not afford such a decay mode at the tree level.

\begin{figure}[t]
\centering
\includegraphics[width=0.49\textwidth]{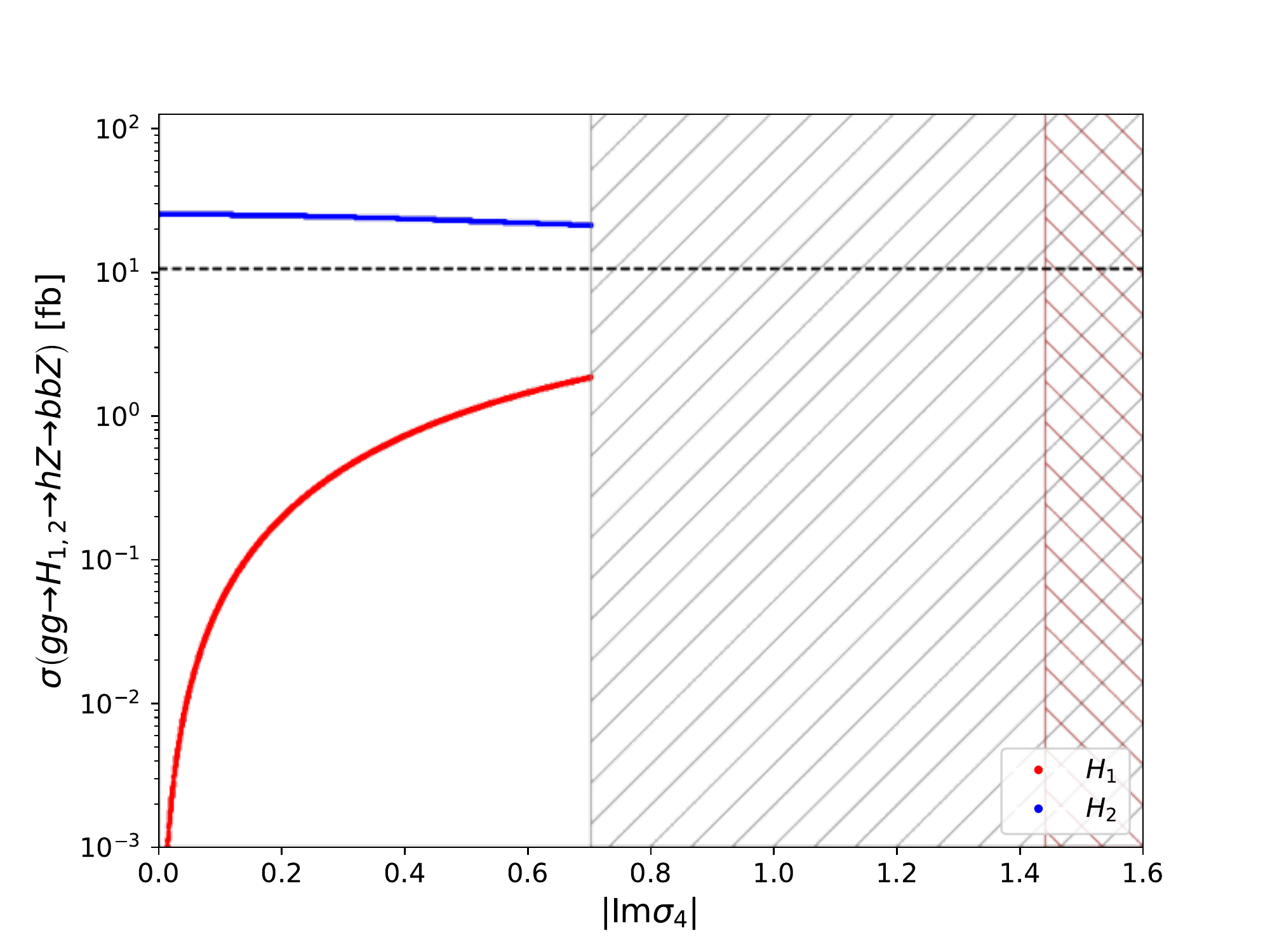}
\includegraphics[width=0.49\textwidth]{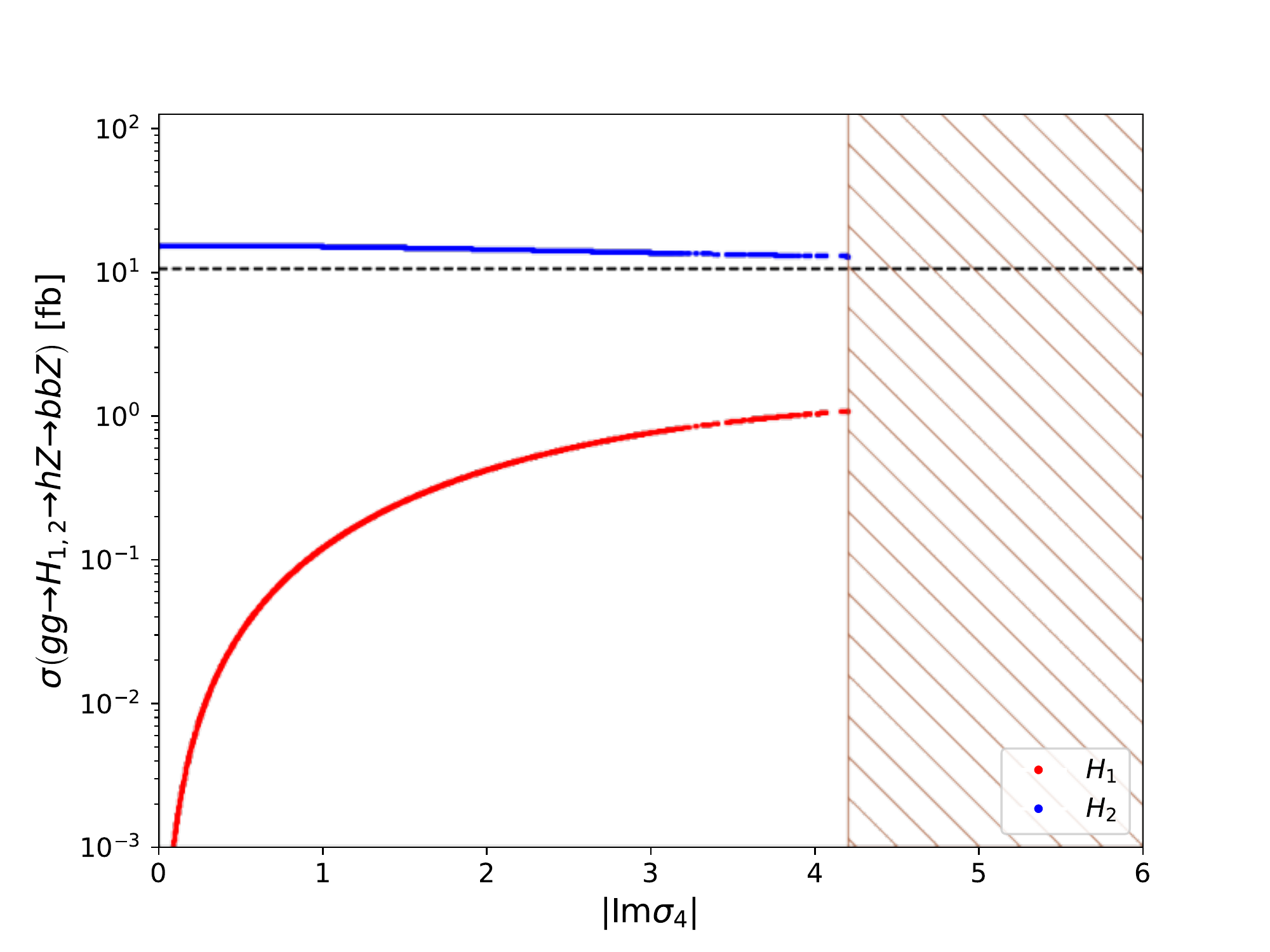}
\\
\caption{\label{BP:sigma} $\sigma(gg\to H_{1,2}\to hZ\to bbZ)$ at 14 TeV for benchmark 1 (left) and benchmark 2 (right) 
as a function of $\vert\Im\sigma_4\vert$. The gray hatched region is excluded by the eEDM constraint at 90\% CL, and the brown hatched region by the other theoretical and collider measurement constraints at 95\% CL. The dashed line represents the estimated 95\% CL limit at the HL-LHC.
}
\end{figure}

Fig.~\ref{BP:sigma} shows $\sigma(gg\to H_{1,2}\to hZ\to bbZ)$ for the two benchmarks, illustrating that the variation patterns are similar to those of ${\rm BR}(H_{1,2}\to hZ)$. Within the allowed range of $\vert\Im\sigma_4\vert$, $\sigma(gg\to H_{2}\to hZ\to bbZ)$ can reach above $\mathcal{O}\left(10^{1}\right)$~fb 
and $\sigma(gg\to H_{1}\to hZ\to bbZ)$ above $\mathcal{O}\left(10^{0}\right)$~fb for both of the benchmarks. We remark that rather than a horizontal band at the top of the plot, the 95\% CL bound extracted from Fig.~\ref{sigma:gg} is translated into the constraint on $\vert\Im\sigma_4\vert$.  Assuming a naive scaling from the current cross section upper bounds to the 14-TeV HL-LHC with an integrated luminosity of $3~{\rm ab}^{-1}$, the 95\% CL upper limit on the cross section in the mass regime of the two benchmarks is expected to reach $\sim10$~fb.  This limit, indicated by the dashed line in Fig.~\ref{BP:sigma}, will be able to probe the two benchmarks through the $H_2$ production channel.
Thus, it is promising to explore both $\sigma(gg\to H_{1}\to hZ\to bbZ)$ and $\sigma(gg\to H_{2}\to hZ\to bbZ)$ at the High-Luminosity LHC.

\begin{figure}[t]
\centering
\includegraphics[width=0.9\textwidth]{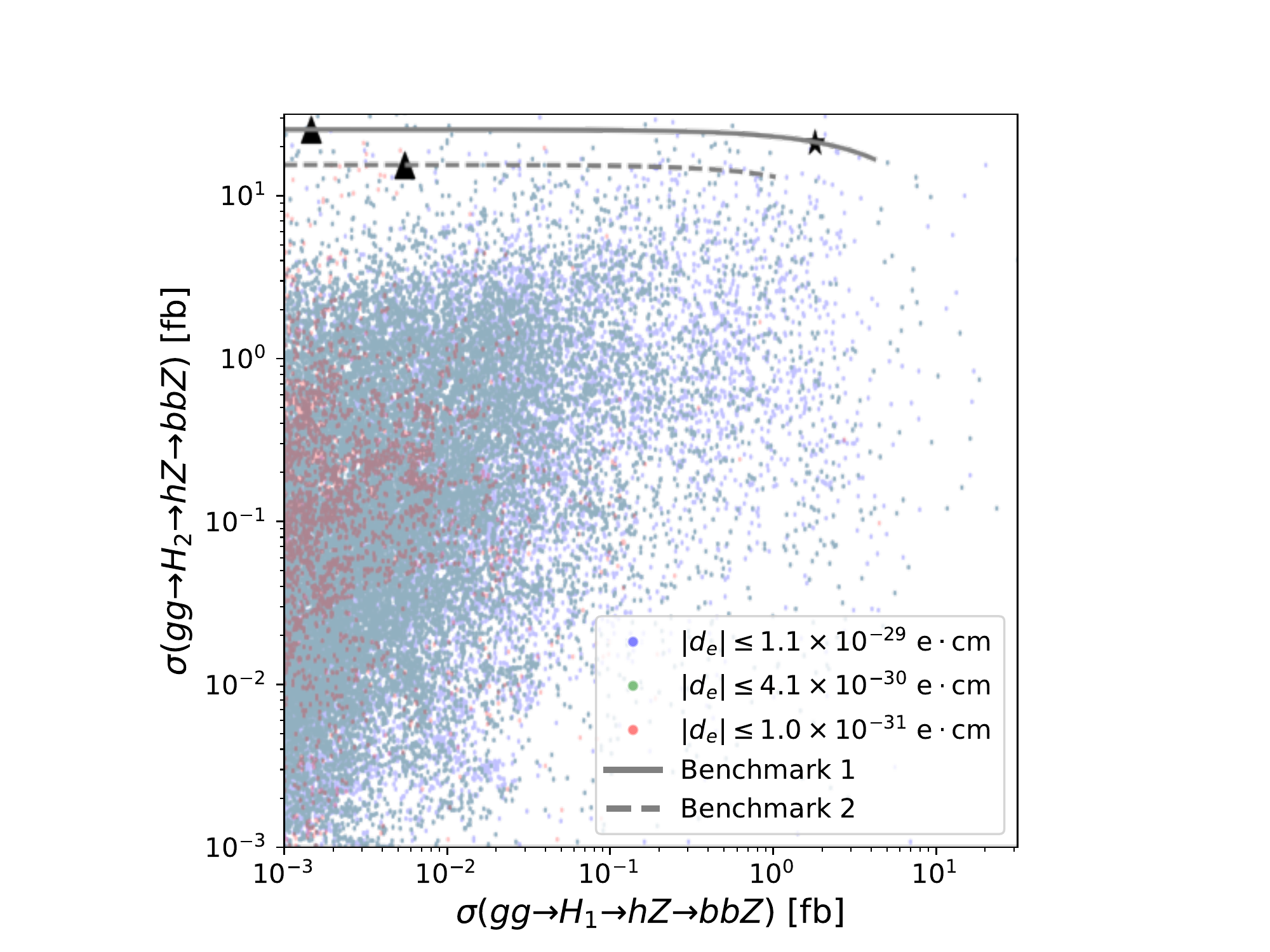}
\caption{\label{sigma:contour} Distribution in the $\sigma(gg\to H_1\to hZ\to bbZ)$-$\sigma(gg\to H_2\to hZ\to bbZ)$ plane at 14 TeV under the eEDM constraints given by the ACME Collaboration~\cite{ACME:2018yjb} (blue), by Ref.~\cite{Roussy:2022cmp}, and a future projection of $1.0\times10^{-31}~e~{\rm cm}$ at 90\% CL, respectively. The solid and dashed curves represent the trajectories of benchmark 1 and benchmark 2 with respect to $\vert\Im\sigma_4\vert$, respectively. The points on the contours to the right of the triangle have $\vert d_e\vert>4.1\times10^{-30}{\rm e\cdot cm}$, and those to the right of the star have $\vert d_e\vert>1.1\times10^{-29}{\rm e\cdot cm}$.
}
\end{figure}

Finally, we show the $\sigma(gg\to H_1\to hZ\to bbZ)$-$\sigma(gg\to H_2\to hZ\to bbZ)$ distribution at 14 TeV under the eEDM constraints given by the ACME Collaboration~\cite{ACME:2018yjb} (blue), by Ref.~\cite{Roussy:2022cmp}, and a future projection of $1.0\times10^{-31}~e~{\rm cm}$ at 90\% CL, respectively, in FIG.~\ref{sigma:contour}. From this plot, it is clear that $\sigma(gg\to H_2\to hZ\to bbZ)\gtrsim \sigma(gg\to H_1\to hZ\to bbZ)$ most of the time. It can also be seen that as the eEDM constraint becomes stricter, the allowed cross sections are more significantly bounded. If the constraint is pushed to the level of $10^{-31}~e~{\rm cm}$, most of the benchmarks will be constrained as $\sigma(gg\to H_1\to hZ\to bbZ)\lesssim\mathcal{O}(10^{-1})$~fb and $\sigma(gg\to H_2\to hZ\to bbZ)\lesssim\mathcal{O}(10^{0})$~fb. 
On the same plot, we also depict the trajectories of the two benchmarks as we vary $\vert\Im\sigma_4\vert$, where benchmark 1 (2) is represented by the solid (dashed) curve.  Along these trajectories, we mark the thresholds of $\vert d_e\vert=4.1\times10^{-30}{\rm e\cdot cm}$ (triangles) and $\vert d_e\vert=1.1\times10^{-29}{\rm e\cdot cm}$ (star) that rule out the points to their right.  Note that benchmark 2 consistently remains below the bound set by Ref.~\cite{Roussy:2022cmp}. It is evident that benchmark 1, owing to its cancellation nature, is more restricted by the projected $\sigma(gg\to H_1\to hZ\to bbZ)$ than benchmark 2.  Benchmark 2 serves as an example of various data points that scatter away from the main distribution under the different eEDM constraints.

%=================================================================================================
\section{Conclusions}\label{sec:6}

We have studied the extended GM model that explicitly violates the global $SU(2)_L\times SU(2)_R$ symmetry and contains one physical CPV phase in the Higgs potential.  
This corresponds to the minimal extension of the Higgs sector which gives a non-zero CPV phase, no quark FCNC and $\rho = 1$ at tree level in the scenario without imposing any new symmetries. 
In the most general form of the Higgs potential under the electroweak symmetry, 
we have derived the analytic expressions for the vacuum stability and the perturbative unitarity conditions as the theoretical constraints. 
In addition, we have presented the complete expressions for the contributions from Barr-Zee type diagrams to the eEDM and nEDM.

For the numerical analysis, we have considered the minimally extended GM model for simplicity, 
and have performed a global fit to the Tevatron and LHC measurements under the constraints from the uniqueness and stability of the vacuum, the perturbative unitarity, the eEDM and the nEDM. 
Our fit results have shown that the major contributions to the eEDM are the gauge-loop and charged-Higgs-loop diagrams. 
The size of each contribution can be larger than the current upper limit on the eEDM experiment, but the total contribution is within the bound due to the non-trivial cancellation.  
We then have studied the effects of CP-mixing for the neutral scalars $H_{1,2}$ on their decays into the $hZ$ and $ZZ$ final states.  
We have found that the lighter (heavier) Higgs boson $H_1$ ($H_2$) is often mostly CP-even (CP-odd).  
When $\vert\Im\sigma_4\vert$ increases, the enhanced CPV will make the two eigenstates further mix, thus allowing greater ${\rm BR}(H_1\to hZ)$ and ${\rm BR}(H_2\to ZZ)$.  
We have also studied $\sigma(gg\to H_{1,2,3}\to hZ\to bbZ)$ and found that while the $H_3$-mediated process is often far below the current LHC sensitivity, the $H_1$- and $H_2$-mediated processes can potentially be probed at the future LHC.

We have presented two benchmarks with larger $\sigma(gg\to H_{2}\to hZ\to bbZ)$ at 14 TeV and different $\vert\Im\sigma_4\vert$ upper bounds, and studied in depth the impacts of $\Im\sigma_4$ on their collider phenomenology.  One implication is that $m_{H^{\pm\pm}}$ is exactly invariant while the other scalar masses are approximately invariant as $\Im\sigma_4$ varies in these benchmarks.  Another implication is that in both benchmarks $\sigma(gg\to H_{2}\to hZ\to bbZ)$ can reach above $\mathcal{O}(10^{1})$~fb and $\sigma(gg\to H_{1}\to hZ\to bbZ)$ can above $\mathcal{O}(10^{0})$~fb at 14 TeV, while a rough projection shows that a 95\% CL upper limit of $\sim10$~fb on the production cross section can be achieved at the HL-LHC.  This implies that there is a great potential to explore both processes simultaneously, giving direct evidence of CPV in the model.  Moreover, the result that ${\rm BR}(H_1^\pm\to WZ)\sim\mathcal{O}(10^{-1})$ for benchmark 1 further serves as a signature to differentiate between this model and the 2HDMs.  We have also examined the deviations of the $h$ decay patterns from the SM predictions, and found that the $\gamma\gamma$ channel can deviate by up to $\sim10$\%, also a promising probe of the model.  Finally, we have also shown the influence of different eEDM constraints on the $\sigma(gg\to H_1\to hZ\to bbZ)$-$\sigma(gg\to H_2\to hZ\to bbZ)$ distribution at 14 TeV, and observed that if the constraint is pushed to the level of $10^{-31}~e~{\rm cm}$, most of the benchmarks will be constrained as $\sigma(gg\to H_1\to hZ\to bbZ)\lesssim\mathcal{O}(10^{-1})$~fb and $\sigma(gg\to H_2\to hZ\to bbZ)\lesssim\mathcal{O}(10^{0})$~fb.

%=================================================================================================
\section*{Acknowledgments}

The works of TKC and CWC were supported in part by the National Science and Technology Council of Taiwan under Grant Nos.~NSTC-108-2112-M-002-005-MY3 and 111-2112-M-002-018-MY3.  The work of KY was supported in part by the Grant-in-Aid for Early-Career Scientists, No.~19K14714.

%=================================================================================================
\clearpage
\appendix

%=================================================================================================
\section{Mass Formulas \label{sec:mass}}

We provide the explicit formulas for masses or mass matrices of the physical Higgs bosons based on the general Higgs potential given in Eq.~(\ref{pot_gen}) without imposing any assumptions.

First, the squared mass of the doubly-charged Higgs bosons $\chi^{\pm\pm}$ is given by 
\begin{align}
m_{\chi^{\pm\pm}}^2 &= -2 \rho_2 v_\chi^2 - \frac{\sigma_2}{2}v_\phi^2 -\sqrt{2} \mu_{\chi \xi}  v_\xi-\frac{v_\phi^2}{4 }\left(2\frac{\Re\mu_{\phi \chi}}{v_\chi} +\sqrt{2}\Re\sigma_4 \frac{v_\xi}{v_\chi} \right). 
\end{align}

Suppose $M_\pm$ and $M_0$ are respectively the Hermitian mass matrices for the singly-charged and neutral Higgs bosons in the basis of $(\tilde{H}_1^\pm,\tilde{H}_2^\pm)$ and $(\tilde{h},\tilde{H}_1,\tilde{H}_2,\tilde{H}_3)$ [see Eq.~(\ref{eq:masseigen1}) for the definition of these fields with a tilde].  Their matrix elements are given as follows: 
\begin{align}
\begin{split}
(M_\pm)_{11} 
=& 
-\frac{v^2}{4(v_\xi^2+v_\chi^2)}\left[\sigma_2   v_\chi^2 + \sqrt{2}\Re \sigma_4 v_\chi v_\xi -\sqrt{2}\mu_{\phi \xi}v_\xi + 2\Re\mu_{\phi \chi}v_\chi \right], \\
(M_\pm)_{22} 
=&
\frac{v_\xi^2+v_\chi^2}{2} \left(2 \rho_5  - \sqrt{2} \frac{\mu_{\chi\xi}}{v_\xi} \right) 
\\
&
-\frac{v_\phi^2 }{4(v_\xi^2+v_\chi^2)}\left[ v_\xi^2\left(\sigma_2 + 2\frac{\Re\mu_{\phi \chi}}{v_\chi}\right)   -\sqrt{2}v_\chi^2\frac{\mu_{\phi\xi}}{v_\xi}  
+ \sqrt{2}\Re \sigma_4 v_\xi v_\chi\left(2 + \frac{v_\xi^2}{v_\chi^2} + 2\frac{v_\chi^2}{v_\xi^2}  \right)  \right], \\
(M_\pm)_{12} 
=&
-\frac{v_\phi v}{4 (v_\xi^2+v_\chi^2)}v_\chi v_\xi \left[\sigma_2 + 2\frac{\Re\mu_{\phi \chi}}{v_\chi} + 
  \sqrt{2} \frac{\mu_{\phi\xi}}{v_\xi}  -\sqrt{2} \frac{v_\chi}{v_\xi}\Re \sigma_4 \right]  + i \frac{v_\phi v}{2\sqrt{2}}\Im \sigma_4, 
\end{split}
\end{align}
and 
\begin{align}
\begin{split}
(M_0)_{11} & = 2v_\phi^2\lambda , \\
(M_0)_{22} & = 4v_\chi^2(\rho_1 + \rho_2) - \frac{v_\phi^2}{4}\left(\sqrt{2}\frac{v_\xi}{v_\chi}\Re \sigma_4 + \frac{1}{2}\frac{\Re\mu_{\phi\chi}}{v_\chi} \right) , \\
(M_0)_{33} & = 8v_\xi^2\rho_3 - \frac{v_\chi^2}{2\sqrt{2}}\frac{\mu_{\chi\xi}}{v_\xi} - v_\phi^2 \frac{v_\chi}{\sqrt{2} v_\xi}\Re\sigma_4 + \frac{v_\phi^2}{2\sqrt{2}}\frac{\mu_{\phi\xi}}{v_\xi}, \\
(M_0)_{44} & = -\frac{v_\phi^2 + 8v_\chi^2}{4}\left(2\frac{\Re\mu_{\phi\chi}}{v_\chi} + \frac{\sqrt{2}v_\xi}{v_\chi} \Re \sigma_4 \right), \\
(M_0)_{12} & = \sqrt{2} v_\phi v_\chi \left(\sigma_1 + \sigma_2\right) + v_\phi v_\xi\Re \sigma_4 + \sqrt{2}v_\phi \Re\mu_{\phi\chi},  \\
(M_0)_{13} & = 2v_\phi v_\xi \sigma_3+\sqrt{2}v_\phi v_\chi \Re\sigma_4 -\frac{v_\phi \mu_{\phi\xi}}{\sqrt{2}}, \\
(M_0)_{14} & = 0, \\
(M_0)_{23} & = 2\sqrt{2}v_\chi v_\xi \rho_4 + \frac{v_\phi^2}{2}\Re\sigma_4 + v_\chi \mu_{\chi\xi}, \\
(M_0)_{24} & = 0, \\
(M_0)_{34} & = -\frac{v_\phi}{2}\sqrt{v_\phi^2 + 8v_\chi^2}\Im \sigma_4. 
\end{split}
\end{align}
It is clear that in the limit of $\Im \sigma_4 \to 0$, $(M_0)_{34}$ vanishes, and then the $(\tilde{H}_1,\tilde{H}_2,\tilde{H}_3)$ sector and $\tilde{H}_0$ decouple as a consequence of the restoration of CP invariance.

%=================================================================================================
\section{Vacuum Stability \label{sec:stability}}

In Ref.~\cite{Arhrib:2011uy}, the idea of parametrizing the field values using four parameters, $r$, $\gamma$, $\zeta$, and $\xi$, was first proposed.  We will follow the same notation for our discussion below.

When the field values are large, the scalar potential is dominated by the quartic terms, which are collectively given by
\begin{align}
V_{\rm quartic}
=&
\lambda (\phi^\dagger \phi)^2 
+\rho_1[\text{tr}(\chi^\dagger\chi)]^2+\rho_2\text{tr}(\chi^\dagger \chi\chi^\dagger \chi)
+\rho_3\text{tr}(\xi^4)
+\rho_4 \text{tr}(\chi^\dagger\chi)\text{tr}(\xi^2)
+\rho_5\text{tr}(\chi^\dagger \xi)\text{tr}(\xi \chi)\notag\\
&
+\sigma_1\text{tr}(\chi^\dagger \chi)\phi^\dagger \phi+\sigma_2 \phi^\dagger \chi\chi^\dagger \phi
+\sigma_3\text{tr}(\xi^2)\phi^\dagger \phi 
+ (\sigma_4 \phi^\dagger \chi\xi \tilde{\phi} + \text{H.c.}).  \label{eq:v4}
\end{align}

We first introduce the following parameterization for the component scalar fields:
\begin{align}
(\phi^+,\phi^0) = (0,r_0), \quad (\chi^{++},\chi^{+},\chi^0) = (r_1e^{i\theta_1},r_2e^{i\theta_2},r_3e^{i\theta_3}),~(\xi^+,\xi^0) = (r_4e^{i\theta_4},r_5), 
\end{align}
where $(r_i,\theta_j) \in \mathbb{R}$ with $i = 0,\dots, 5$ and $j=1,\dots,4$. 
We note that we have utilized the $SU(2)_L\times U(1)_Y$ invariance so that $\phi$ lies entirely in the real neutral component.  We also introduce the parameters:
\begin{align}
\zeta &= \frac{\text{tr}(\chi^\dagger\chi\chi^\dagger\chi)}{[\text{tr}(\chi^\dagger\chi)]^2}, ~~
\omega = \frac{\phi^\dagger \chi\chi^\dagger \phi}{(\phi^\dagger\phi)\text{tr}(\chi^\dagger\chi)},~~
\eta = \frac{\text{tr}(\chi^\dagger \xi)\text{tr}(\xi\chi)}{\text{tr}(\chi^\dagger\chi)\text{tr}(\xi^2)},~~
\delta = \frac{\phi^\dagger \chi\xi \tilde{\phi}}{(\phi^\dagger\phi)\sqrt{\text{tr}(\chi^\dagger\chi)\text{tr}(\xi^2)}}.
\end{align}
All the invariants in the potential can then be expressed in terms of the above-defined parameters as 
\begin{align}
\begin{split}
\phi^\dagger\phi &= r_0^2,\quad\text{tr}(\chi^\dagger\chi) = r_1^2 + r_2^2 + r_3^2 ,\quad \text{tr}(\xi^2) = 2r_4^2 + r_5^2 , \\
\zeta& = 1-\frac{4r_1^2r_3^2 -4r_1r_3r_2^2\cos\phi_0+r_2^4}{2(r_1^2+r_2^2+r_3^2)^2} ,\quad\omega = \frac{1}{2}-\frac{r_1^2-r_3^2}{2(r_1^2+r_2^2+r_3^2)}, \\
\eta &= \frac{r_1^2r_4^2+r_2^2r_5^2+r_3^2r_4^2+2\big[r_1r_2r_4r_5\cos\phi_1-r_1r_3r_4^2\cos\phi_2-r_2r_3r_4r_5\cos(\phi_1-\phi_2)\big]}{(r_1^2+r_2^2+r_3^2)(2r_4^2+r_5^2)} , \\
\delta &= \frac{r_2r_4e^{i(\phi_2-\phi_1)}+r_3r_5 }{\sqrt{2(r_1^2+r_2^2+r_3^2)(2r_4^2+r_5^2)}}e^{i\theta_3},
\end{split}
\end{align}
where $\phi_0 = \theta_1-2\theta_2+\theta_3$, $\phi_1 = \theta_1- \theta_2- \theta_4$ and $\phi_2 = \theta_1- \theta_3-2\theta_4$.
We note that only the $\delta$ parameter is complex, and its absolute value is expressed as 
\begin{equation}
	\vert\delta\vert = \Bigg[\frac{ r_2^2r_4^2+r_3^2r_5^2+2r_2r_3r_4r_5\cos(\phi_1 - \phi_2)}{2(r_1^2+r_2^2+r_3^2)(2r_4^2+r_5^2)}\Bigg]^{\frac{1}{2}}. 
\end{equation}
We then find the domain of each parameter: 
\begin{align}
\zeta\in[1/2,1],~~\omega\in[0,1],~~\eta\in[0,1],~~|\delta|\in\big[0,1/\sqrt{2}\big], 
\end{align}
where $\zeta$-$\omega$ and $\eta$-$|\delta|$ are correlated, as discussed below.

To examine the correlation $\zeta$ and $\omega$, we parameterize
\begin{equation}
	r_1 = \sqrt{\alpha_1+\alpha_3} ~,~ r_2 = \sqrt{\alpha_2} ~,~ r_3 = \sqrt{\alpha_1-\alpha_3} ~,
\end{equation}
with the domains $\alpha_1\in[0,\infty],\alpha_2\in[0,\infty],\alpha_3\in[-\alpha_1,\alpha_1]$. Then, we can further express
\begin{equation}
	\bar{\zeta} = \frac{4(\alpha_1^2-\alpha_3^2)-4\sqrt{\alpha_1^2-\alpha_3^2}\alpha_2\cos\phi+\alpha_2^2}{(2\alpha_1+\alpha_2)^2} ,~ \bar{\omega} = -\frac{\alpha_3}{2\alpha_1+\alpha_2}, 
\end{equation}
where $\bar{\zeta} \equiv 2(1 - \zeta)$ and $\bar{\omega} \equiv \omega - 1/2$. 
For a given set of $(\alpha_1,\alpha_2,\alpha_3)$, $\bar{\omega}$ is fixed 
and the maximum (minimum) of $\bar{\zeta}$, denoted by $\bar{\zeta}_+\,(\bar{\zeta}_-)$, is given at $\phi=\pi\,(0)$. 
Explicitly, 
\begin{equation}
	\bar{\zeta}_\pm(\omega,\alpha_1,\alpha_2) =  \Bigg(\frac{2\sqrt{\alpha_1^2-(2\alpha_1+\alpha_2)^2\omega^2}\pm\alpha_2}{2\alpha_1+\alpha_2}\Bigg)^2 .
\end{equation}
We thus find
\begin{align}
0 \leq \bar{\zeta} \leq 1-4\bar{\omega}^2 
~\Leftrightarrow~ 
-\frac{\sqrt{1-\bar{\zeta}}}{2} \leq \bar{\omega} \leq \frac{\sqrt{1-\bar{\zeta}}}{2}.
\end{align}
In terms of the original variables $\zeta$ and $\omega$, we obtain
\begin{align}
 \frac{1}{2}(1-\sqrt{2\zeta - 1}) \leq \omega \leq \frac{1}{2}(1+\sqrt{2\zeta - 1}). 
\end{align}

For the correlation of $\eta$ and $|\delta|$, we observe the relation
\begin{equation}
	\eta+2\vert\delta\vert^2 = \frac{(r_1^2+r_2^2+r_3^2)r_4^2+(r_2^2+r_3^2)r_5^2+2\big(r_1r_2r_4r_5\cos\phi_1-r_1r_3r_4^2\cos\phi_2\big)}{(r_1^2+r_2^2+r_3^2)(2r_4^2+r_5^2)}\in[0,1] ,
\end{equation}
which identifies a domain in the $\eta$-$\vert\delta\vert$ plane and implies that $\eta \in \big[ -2\vert\delta\vert^2,1-2\vert\delta\vert^2 \big]$. 
Combining this with the independent intervals of $\eta$ and $\vert\delta\vert$, we can derive the boundaries
\begin{equation}
0\leq \eta \leq 1-2\vert\delta\vert^2 .
\end{equation}

After identifying the domains of the field value parameters, we now turn to the quartic potential. 
Redefining 
\begin{equation}
\chi^\dagger\chi = r_0^2r^2\cos^2\gamma , ~\xi^\dagger\xi = r_0^2r^2\sin^2\gamma ,
\end{equation}
with $\gamma\in[0,\pi/2]$ and $r\in[0,\infty)$, we can rewrite the potential given in Eq.~(\ref{eq:v4}) as a quadratic function of $r^2$: 
\begin{align}
\begin{split}
\bar{V}_{\rm quartic} (r^2)
=&  (A_\rho t^4 - B_\rho t^2 + \rho_3) (r^2)^2 + C_\sigma r^2 + \lambda , \label{eq:v4_2}
\end{split}
\end{align}
where $\bar{V}_{\rm quartic} \equiv V_{\rm quartic}/r_0^4$, $t \equiv \cos\gamma \in[0,1]$, and 
\begin{align}
\begin{split}
A_\rho &\equiv \rho_1 + \zeta\rho_2 + \rho_3 - \rho_4 -\eta\rho_5, \\
B_\rho &\equiv 2\rho_3 - \rho_4 - \eta\rho_5, \\
C_\sigma &\equiv \left(\sigma_1 + \omega\sigma_2  -\sigma_3 \right)t^2	+ (\delta\sigma_4 + \text{H.c.}) t\sqrt{1-t^2} + \sigma_3.
\end{split}
\end{align}
The potential is minimized when the coefficient $C_\sigma$ is minimized, which is realized when $\omega\sigma_2$ and $(\delta \sigma_4 + \text{H.c.})$ are taken to have their minimum values for fixed values of $\zeta$ and $\eta$. We thus replace them with 
\begin{align}
\omega\sigma_2 \to \frac{\sigma_2-\vert\sigma_2\vert\sqrt{2\zeta-1}}{2},\quad 
\delta\sigma_4 \to -|\delta||\sigma_4| \to -\sqrt{\frac{1-\eta}{2}}|\sigma_4|, 
\end{align}
where we used the phase degrees of freedom of $\delta$ such that arg$(\delta\sigma_4)$ is fixed to $\pi$.  The coefficient $C_\sigma$ is then replaced as
\begin{align}
C_\sigma & \to \left(\sigma_1 + \frac{\sigma_2-\vert\sigma_2\vert\sqrt{2\zeta-1}}{2}  -\sigma_3 \right)t^2  -t\sqrt{\frac{(1-\eta)(1-t^2)}{2}}|\sigma_4| + \sigma_3. 
\end{align}
Therefore, the positivity of the potential should be examined in terms of the field parameters $t$, $\zeta$ and $\eta$ in the domains
\begin{equation}
	t\in[0,1],~\zeta\in[1/2,1],~\eta\in[0,1]. \label{eq:domain}
\end{equation}

From Eq.~(\ref{eq:v4_2}), it is clear that $\bar{V}_{\rm quartic}>0$ is ensured by requiring 
\begin{equation}
	\lambda > 0 ,\quad  A_\rho t^4 - B_\rho t^2 + \rho_3  > 0,\quad 
4(A_\rho t^4 - B_\rho t^2 + \rho_3)\lambda   > C_\sigma^2. \label{eq:vc10}
\end{equation}
Because the domain of $t$ is restricted to $[0,1]$, the second and third conditions are further analyzed as follows. 
We first focus on the second condition in Eq.~(\ref{eq:vc10}). 
At the endpoints $t=0,1$, we obtain
\begin{equation}
	\rho_3 > 0,\quad   \rho_1 + \zeta\rho_2 > 0, 
\end{equation}
where the second condition can be expressed as $\rho_1 + \text{min}(\rho_2, \rho_2/2) > 0$ because of $\zeta\in[1/2,1]$. 
If 
\begin{align}
A_\rho  > 0 \quad \& \quad B_\rho > 0 \quad \& \quad 0\leq \frac{B_\rho}{2A_\rho} \leq 1, 
\end{align}
the quadratic equation, $f(t^2) = A_\rho (t^2)^2 - B_\rho t^2 + \rho_3$, has the minimal value in $0\leq t^2 \leq 1$. 
We thus require
		\begin{equation}
			4A_\rho \rho_3 > B_\rho^2. 
		\end{equation}
Regarding the third condition in Eq.~(\ref{eq:vc10}), 
we obtain the conditions at the endpoints $t=0,1$
\begin{equation}
4\lambda \rho_3 > \sigma_3^2,\quad 4\lambda(\rho_1 + \zeta\rho_2) >  \left(\sigma_1 +  \frac{\sigma_2-\vert\sigma_2\vert\sqrt{2\zeta-1}}{2}\right)^2.
\end{equation}
For $0 < t < 1$, we require the third condition in Eq.~(\ref{eq:vc10}) within the domain given in Eq.~(\ref{eq:domain}). 
In practice, it is easier to just numerically minimize $G(t,\zeta,\eta)\equiv 4(A_\rho t^4 - B_\rho t^2 + \rho_3)\lambda   - C_\sigma^2$ and then check whether $G_{\rm min}(t,\zeta,\eta)>0$.

We note that we have checked the consistency of our derivation with the literature by reproducing the conditions given in Ref.~\cite{Hartling:2014zca} for the custodial symmetric case.

%=================================================================================================
\section{Formulas for BZ Diagram Contributions to the eEDM and nEDM}\label{sec:a}

We present the analytic formulas for the BZ diagram contributions to the fermion EDM $d_f$ defined in Eq.~(\ref{eq:edm}).  Calculations are done in the 't~Hooft-Feynman gauge.    We define the coefficients of the Lagrangian as follows: 
\begin{align}
{\cal L} = g_{V_1V_2S}^{} V_{1\mu} V_2^\mu S + g_{S_1S_2V}^{}(S_1 \overleftrightarrow{\partial}_\mu S_2) V^\mu + \lambda_{S_1S_2S_3}S_1S_2S_3 \cdots, 
\end{align}
with $S_i$ and $V_i$ being the generic symbols for a scalar and a gauge boson, respectively, and 
$(S_1 \overleftrightarrow{\partial}_\mu S_2)\equiv S_1 (\partial_\mu S_2) - (\partial_\mu S_1)S_2$. 
In addition, we introduce the notation $d_f^{\cal VH}(X,Y)$, where $X$ and $Y$ are the particles running in the loop with $X$ being the one to which the external photon attaches, and ${\cal V}$ $({\cal H})$ is a gauge (scalar) boson mediating between the external fermion line and the internal loop.

First, the contribution from Fig.~\ref{BZ:fermion} (a) is expressed as 
\begin{align}
	\frac{d_f^{VH_i}(F,F)}{\kappa_{\rm BZ}} &= -16I_fg_{Vff}^{}g_{VFF}^{} Q_FN_c^F\frac{m_F^2}{v_\phi}R_{\phi_r i} R_{\phi_i i}\int_0^1dz\left[ \frac{1 + I_F/I_f }{z}-2(1-z) \right] C_{FF}^{VH_{i}}(z) ,\label{eq:dev}
\end{align}
where $\kappa_{\rm BZ} = e/(16\pi^2)^2(m_f/v_\phi)$, $N_c^F = 1~(3)$ for $F$ being leptons (quarks), 
$g_{\gamma ff}=eQ_f$, $g_{Zff}=e(I_f/2-Q_fs_W^2)/(c_Ws_W)$, $c_W\equiv\cos\theta_W$, $s_W\equiv\sin\theta_W$, with $\theta_W$ being the weak angle, and
\begin{equation}
	C_{XY}^{\cal VH}(z) = C_0\left(0,0,0;m_{\cal V}^2,m_{\cal H}^2,\frac{(1-z)m_X^2+zm_Y^2}{z(1-z)}\right),
\end{equation}
with 
\begin{equation}
	C_0(0,0,0;m_1^2,m_2^2,m_3^2) = \frac{1}{m_1^2-m_2^2}\left[ \frac{m_1^2}{m_1^2-m_3^2}\log\left(\frac{m_3^2}{m_1^2}\right)-\frac{m_2^2}{m_2^2-m_3^2}\log\left(\frac{m_3^2}{m_2^2}\right) \right] .
\end{equation}
In our model, the contribution from Fig.~\ref{BZ:fermion} (b) vanishes if we neglect the Kobayashi-Maskawa (KM) phase.  As a conservative constraint on the CPV in our model, we have neglected the KM phase throughout this paper.

Next, we list the gauge-loop contributions shown in Fig.~\ref{BZ:gauge}.  Note that the $G^\pm$ and $G^0$ loop contributions are included in the $W^\pm$ and $Z$ boson loop diagrams, respectively.  They are given by 
\begin{align}
\frac{d_f^{VH_i}(W,W)}{\kappa_{\rm BZ}^{}} &= 
-4I_fg_{WWV}^{} g_{WWH_i}^{} \,g_{Vff}^{}  R_{\phi_i i} \int_0^1 dz\notag\\
& \hspace{-1.5cm}\times \left[5(1-z) - \frac{6}{z}  + \left(1-z -\frac{2}{z}\right)\left(1 - \frac{m_V^2}{m_W^2} \right) 
 + \frac{1-z}{2}\frac{m_{H_i}^2}{m_W^2}\left(2 - \frac{m_V^2}{m_W^2}\right) \right] C_{WW}^{VH_i}(z),\\
\frac{d_f^{WH_i^\pm}(W,Z)}{\kappa_{\rm BZ}^{}}&= -\frac{g}{2} g_{WWZ}^{} \Im \left( R_{\phi^\pm i} \, g_{WZH^+}^*\right)\int_0^1dz  \notag\\
& \hspace{-1.5cm}\times \left[5(1-z)-\frac{6}{z} + (z+3)\left(1 - \frac{m_Z^2}{m_W^2} \right) +  \frac{1-z}{2}\frac{m_{H_i^\pm}^2}{m_W^2}\left(2 - \frac{m_Z^2}{m_W^2}\right) \right]C_{WZ}^{WH_i^\pm}(z), \\
\frac{d_f^{ZH_j} (W,H_i^\pm)}{\kappa_{\rm BZ}} &= -4I_fg_{Zff}R_{\phi_i j} \Re(i g_{H_i^+ H_j W}^* g_{WZH_i^+})\int_0^1dz \left[\xi_{WH_i^\pm}^{H_j}(z)C_{WH_i^\pm}^{ZH_j}(z)\right], \\
\frac{d_f^{WH_j^\pm}(W,H_i)}{\kappa_{\rm BZ}} &= -\frac{g}{2}g_{WW{H_i}}^{}\Im\left(ig_{H_j^+ H_i W}^* R_{\phi^\pm j}\right) \int_0^1dz \left[\xi_{WH_i}^{H_j^\pm}(z)\, C_{WH_i}^{WH_j^\pm}(z)\right], \\
\frac{d_f^{WH_i^\pm}(W,H^{\pm\pm})}{\kappa_{\rm BZ}} &= -gg_{WWH^{--}}^{} \Im\left(ig_{H^{++}H_i^- W}^{} R_{\phi^\pm i}\right) \int_0^1dz \left[\xi_{WH^{\pm\pm}}^{H_i^\pm}(z)\,C_{WH^{\pm\pm}}^{WH_i^\pm}(z)\right], \\
\frac{d_f^{ZH_j}(H_i^\pm, W)}{\kappa_{\rm BZ}} &= 4I_f g_{Zff}^{} R_{\phi_i j} \Re(i g_{H_i^+ H_j W}^{} g_{WZH_i^+}^*)\int dz \left[\zeta_{H_i^\pm W}^{H_j}(z)\,C_{H_i^\pm W}^{ZH_j}(z)\right], \\ 
\frac{d_f^{WH_j^\pm}(H_i^\pm , Z)}{\kappa_{\rm BZ}} &= -\frac{g}{2}\Im(ig_{H_i^+ H_j^- Z}^{} g_{WZH_i^+}^{*} R_{\phi^\pm j}) 
\int dz \left[\zeta_{H_i^\pm Z}^{H_j^\pm}(z)\,C_{H_i^\pm Z}^{WH_j^\pm}(z)\right], \\
\frac{d_f^{WH_i^\pm }(H^{\pm\pm}, W)}{\kappa_{\rm BZ}} &= -2g g_{WWH^{--}}^{}\Im\left(i g_{H^{++} H^-_i W}^{}R_{\phi^\pm i}\right)\int dz\left[
\zeta_{H^{\pm\pm} W}^{H_i^\pm}(z)\,C_{H^{\pm\pm} W}^{WH_i^\pm}(z) \right],
\end{align}
where $g_{WW\gamma}=e$ and $g_{WWZ}=gc_W$, and
\begin{align} 
g_{WWH^{\pm\pm}}^{} & = g^2v_\chi , \\
g_{WZH_i^+} & = \frac{g^2}{c_W}(-R_{\chi^\pm i}v_\chi + R_{\xi^\pm i} v_\xi), \\
g_{WWH_i} & = \frac{g^2}{2}(R_{\phi_r i}v_\phi + 2\sqrt{2}R_{\chi_r i}v_\chi + 4R_{\xi_r i}v_\xi), \\
g_{H^{++} H^-_i W} &=  -igR_{\chi^\pm i}^*, \\
g_{H_i^+ H_j W} &=    -i\frac{g}{2}[R_{\phi^\pm i}\left(R_{\phi_r j} - iR_{\phi_i j}\right) + \sqrt{2}R_{\chi^\pm i}\left(R_{\chi_r j} - iR_{\chi_i j}\right) + 2R_{\xi^\pm i}R_{\xi_r j}], \\
g_{H^{++} H^{--} Z} &=  -2ie\frac{c_{2W}}{s_{2W}}, \\
g_{H_i^+ H_j^- Z} &=  -i\frac{g}{2c_W}[(c_W^2-s_W^2)R_{\phi^\pm i}R_{\phi^\pm j}^* - 2s_W^2R_{\chi^\pm i} R_{\chi^\pm j}^* + 2c_W^2R_{\xi^\pm i}R_{\xi^\pm j}^*], \\
g_{H^{++} H^{--} \gamma} &=  -2ie, \\
g_{H_i^\pm H_j^\mp \gamma} &=  -ie\delta_{ij}. 
\end{align}
In the above expression, we have introduced the functions
\begin{align} 
\xi_{VS_1}^{S_2}(z) &= \left(\frac{4-z}{z}+\frac{m_{S_1}^2-m_{S_2}^2}{m_V^2} \right)(1-z), \quad
\zeta_{S_1 V}^{S_2}(z) =3+z+\frac{m_{S_1}^2-m_{S_2}^2}{m_V^2}(1-z). 
\end{align}

Finally, the diagrams shown in Fig.~\ref{BZ:higgs} have the contributions:
\begin{align}
 \frac{d_f^{VH_k}(H_i^\pm , H_j^\pm )}{\kappa_{\rm BZ}} &= 8I_fg_{Vff} R_{\phi_i k}\Re\left(ig_{H_i^+H_j^-V}\lambda_{H_i^- H_j^+ H_k}\right) \int_0^1dz(1-z)C_{H_i^\pm H_j^\pm}^{VH_k}(z) ,\\
\frac{d_f^{VH_i}(H^{\pm\pm},H^{\pm\pm})}{\kappa_{\rm BZ}} &= 16I_fg_{Vff}^{}(ig_{H^{++}H^{--}V}^{})R_{\phi_i i}\lambda_{H^{++} H^{--} H_i}\int_0^1dz(1-z)C_{H^{\pm\pm} H^{\pm\pm}}^{VH_i}(z) ,\\
\frac{d_f^{WH_k^\pm}(H_i^\pm ,H_j)}{\kappa_{\rm BZ}} &= g\Im(ig_{H_i^+ H_j W}^* \lambda_{H_i^+ H_k^-  H_j}R_{\phi^\pm k}  ) \int_0^1 dz(1-z)C_{H_i^\pm H_j}^{WH_k^\pm}(z) ,\\
\frac{d_f^{WH_j^\pm}(H^{\pm\pm},H_i^\pm)}{\kappa_{\rm BZ}} &= 2g\Im\Big(ig_{H^{++} H_i^- W}^* \lambda_{H^{++} H_j^- H_i^-} R_{\phi^\pm j}\Big)\delta_{ij} \int_0^1 dz(1-z)C_{H^{\pm\pm}H_i^\pm}^{WH_j^\pm}(z) ,\\
\frac{d_f^{WH_j^\pm}(H_i^\pm, H^{\pm\pm})}{\kappa_{\rm BZ}} &= g\Im\Big(ig_{H^{++} H_i^- W}^*\lambda_{H^{++} H_j^- H_i^-}R_{\phi^\pm j} \Big)\delta_{ij} \int_0^1 dz(1-z)C_{H_i^\pm H^{\pm\pm}}^{WH_j^\pm}(z) .
\end{align}

The expression for the CEDMs, $d_q^C$, is given by~\cite{Abe:2013qla}
\begin{align}
d_q^C &\equiv \sum_{i=0}^3d_q^{gH_i}(f,f) \notag\\
&= \frac{m_q}{(16\pi^2)^2}4g_3^2(m_q)\frac{m_f^2}{v_\phi^2} R_{\phi_r i}R_{\phi_i i}\int_0^1dz\left\{ 2I_q\left[2(1-z) - \frac{1}{z}\right]-\frac{2I_f}{z} \right\}C_{ff}^{VH_i}(z).
\end{align}

Let us remark on the vanishment of all the above EDMs in the CPC limit, i.e., when $\Im \sigma_4 \to 0$.  In this limit, the mixing matrix
\begin{align}
R_{\varphi i} &\propto 
\begin{cases}
\delta_{0i}+\delta_{1i}+\delta_{2i},\quad \text{for}\quad \varphi = \phi_r,\chi_r,\xi_r,\\
\delta_{3i},\quad  \text{for}\quad \varphi = \phi_i, 
\end{cases}
\end{align}
and the matrix $R_{\varphi^\pm i}$ ($\varphi^\pm=\phi^\pm,\chi^\pm,\xi^\pm$) becomes purely real.  We can then prove that all the above EDMs vanish.

%=================================================================================================
\section{List of Experimental Data from the Tevatron and LHC}\label{sec:b}

In this appendix, we list in Tables~\ref{HSS} to \ref{DS:6} all the experimental measurements of Tevatron and LHC that we have taken into account in our global fit for the minimally extended GM model.

\begin{table}[htbp]
\centering
\begin{tabular}{>{\centering\arraybackslash}p{2.5cm}||>{\centering\arraybackslash}p{1.8cm}|>{\centering\arraybackslash}p{1.8cm}|>{\centering\arraybackslash}p{1.8cm}|>{\centering\arraybackslash}p{1.8cm}|>{\centering\arraybackslash}p{1.8cm}|>{\centering\arraybackslash}p{1.8cm}|>{\centering\arraybackslash}p{1.8cm}}
\toprule
Production & $b\overline{b}$ & $WW$ & $ZZ$ & $\tau\tau$ & $\gamma\gamma$ & $Z\gamma$ & $\mu\mu$ \\
\toprule
ggF$_8$ &--& \cite{ATLAS:2014aga,Chatrchyan:2013iaa}& \cite{Aad:2015vsa,Chatrchyan:2014nva}& \cite{Aad:2014eva,Khachatryan:2014jba}& \cite{Aad:2014eha,Khachatryan:2014ira}& \cite{Aad:2015gba,Chatrchyan:2013vaa} & \cite{Sirunyan:2018hbu}\\
ggF$_{13}$ &--& \cite{Aad:2020jym,Sirunyan:2018egh}& \cite{ATLAS-CONF-2020-027,Sirunyan:2017khh}& \cite{ATLAS-CONF-2020-027,Sirunyan:2017exp,CMS-PAS-HIG-18-001}& \cite{ATLAS-CONF-2020-027,Sirunyan:2018ouh}& \cite{Aaboud:2017uhw,Sirunyan:2018tbk,ATLAS:2020qcv,CMS-PAS-HIG-19-014} & \cite{ATLAS-CONF-2018-026,Sirunyan:2018hbu}\\
\cline{1-7}
VBF$_8$ &--& \cite{ATLAS:2014aga,Chatrchyan:2013iaa}& \cite{Aad:2015vsa,Chatrchyan:2014nva}& \cite{Aad:2014eva,Khachatryan:2014jba}& \cite{Aad:2014eha,Khachatryan:2014ira}& & \\
VBF$_{13}$& \cite{Aad:2020jym,CMS-PAS-HIG-16-003}& \cite{ATLAS-CONF-2020-045,Sirunyan:2018egh}& \cite{ATLAS-CONF-2020-027,Sirunyan:2017khh}& \cite{ATLAS-CONF-2020-027,Sirunyan:2017exp,CMS-PAS-HIG-18-001}& \cite{ATLAS-CONF-2020-027,Sirunyan:2018ouh}& \cite{ATLAS:2020qcv,CMS-PAS-HIG-19-014} & \\
\cline{1-7}
Vh$_8$ & \cite{Aad:2014xzb,Chatrchyan:2013zna}& \cite{Aad:2015ona,Chatrchyan:2013iaa}& \cite{Aad:2015vsa,Chatrchyan:2014nva}& \cite{Aad:2014eva,Khachatryan:2014jba}& \cite{Aad:2014eha,Khachatryan:2014ira}& & \\
Vh$_{13}$ & \cite{Aad:2020jym,Sirunyan:2017elk}& \cite{ATLAS-CONF-2018-004,Sirunyan:2018egh}& \cite{Sirunyan:2017khh,Sirunyan:2018cpi}& \cite{ATLAS-CONF-2020-027,Sirunyan:2017exp,CMS-PAS-HIG-18-001}& \cite{ATLAS-CONF-2020-027,Sirunyan:2018ouh}& \cite{ATLAS:2020qcv,CMS-PAS-HIG-19-014} &\\
\cline{1-7}
tth$_8$ & \cite{Aad:2015gra,Khachatryan:2014qaa}&--&--& \cite{Aad:2014eva,Khachatryan:2014jba}& \cite{Aad:2014eha,Khachatryan:2014ira}& & \\
tth$_{13}$ & \cite{Aad:2020jym,Sirunyan:2018ygk,Sirunyan:2018hoz}& \cite{Aaboud:2017jvq,Sirunyan:2018egh,Sirunyan:2018shy}& \cite{ATLAS-CONF-2020-027,Sirunyan:2018shy}& \cite{ATLAS-CONF-2020-027,Aaboud:2017jvq,Sirunyan:2017exp,Sirunyan:2018shy,CMS-PAS-HIG-18-001}& \cite{ATLAS-CONF-2020-027,Sirunyan:2018ouh}& \cite{ATLAS:2020qcv,CMS-PAS-HIG-19-014} & \\
\toprule
Vh$_2$ & \cite{Aaltonen:2013ipa,Abazov:2013gmz} \\
\colrule
tth$_2$ & \cite{Aaltonen:2013ipa} \\
\botrule
\end{tabular}
\caption{\label{HSS} Higgs signal strength constraints considered in this work. The Higgs decays are listed in separate columns. In each row, we give all LHC and Tevatron references of the used signal strengths, ordered by production mechanism and colliding energy.}
\end{table}

\begin{table}[htbp]
\centering
\begin{tabular}{>{\centering\arraybackslash}p{2.7cm}|>{\centering\arraybackslash}p{1.5cm}||>{\centering\arraybackslash}p{2.2cm}|>{\centering\arraybackslash}p{3.5cm}|>{\centering\arraybackslash}p{2cm}}
\toprule
Channel & $\sqrt{s}$ [${\rm TeV}$] & Experiment & Mass Range [${\rm TeV}$] & $\mathcal{L}~[{\rm fb}^{-1}]$ \\
\toprule
$tt\to\phi^0\to tt$ & 13 & ATLAS~\cite{ATLAS:2018alq} & [0.4,1] & 36.1 \\
\colrule
$bb\to\phi^0\to tt$ & 13 & ATLAS~\cite{ATLAS:2016btu} & [0.4,1] & 13.2 \\
\toprule
$bb\to\phi^0\to bb$ & 8 & CMS~\cite{CMS:2015grx} & [0.1,0.9] & 19.7 \\
\colrule
$gg\to\phi^0\to bb$ & 8 & CMS~\cite{CMS:2018kcg} & [0.33,1.2] & 19.7 \\
\colrule
$pp\to\phi^0\to bb$ & 13 & CMS~\cite{CMS:2016ncz} & [0.55,1.2] & 2.69 \\
\colrule
 \multirow{2}{*}{$bb\to\phi^0\to bb$} & \multirow{2}{*}{13} & ATLAS~\cite{ATLAS:2019tpq} & [0.45,1.4] & 27.8 \\
& & CMS~\cite{CMS:2018hir} & [0.3,1.3] & 35.7 \\
\toprule
\multirow{2}{*}{$gg\to\phi^0\to \tau\tau$} & \multirow{2}{*}{8}  & ATLAS~\cite{ATLAS:2014vhc} & [0.09,1] & 20 \\
& & CMS~\cite{CMS:2015mca} & [0.09,1] & 19.7 \\
\colrule
\multirow{2}{*}{$bb\to\phi^0\to \tau\tau$} & \multirow{2}{*}{8}  & ATLAS~\cite{ATLAS:2014vhc} & [0.09,1] & 20 \\
& & CMS~\cite{CMS:2015mca} & [0.09,1] & 19.7 \\
\colrule
\multirow{3}{*}{$gg\to\phi^0\to \tau\tau$} & \multirow{3}{*}{13}  & ATLAS~\cite{ATLAS:2017eiz} & [0.2,2.25] & 36.1 \\
& & ATLAS~\cite{ATLAS:2020zms} & [0.2,2.5] & 139 \\
& & CMS~\cite{CMS:2018rmh} & [0.09,3.2] & 35.9 \\
\colrule
\multirow{4}{*}{$bb\to\phi^0\to \tau\tau$} & \multirow{4}{*}{13}  & ATLAS~\cite{ATLAS:2017eiz} & [0.2,2.25] & 36.1 \\
& & ATLAS~\cite{ATLAS:2020zms} & [0.2,2.5] & 139 \\
& & CMS~\cite{CMS:2018rmh} & [0.09,3.2] & 35.9 \\
& & CMS~\cite{CMS:2019hvr} & [0.025,0.070] & 35.9 \\
\toprule
\multirow{2}{*}{$gg\to\phi^0\to\mu\mu$} & \multirow{2}{*}{13} & ATLAS~\cite{ATLAS:2019odt} & [0.2,1] & 36.1 \\
& & CMS~\cite{CMS:2019mij} & [0.13,0.6] & 35.9 \\
\colrule
\multirow{2}{*}{$bb\to\phi^0\to\mu\mu$} & \multirow{2}{*}{13} & ATLAS~\cite{ATLAS:2019odt} & [0.2,1] & 36.1 \\
& & CMS~\cite{CMS:2019mij} & [0.13,0.6] & 35.9 \\
\botrule
\end{tabular}
\caption{\label{DS:1} Neutral heavy scalar searches relevant to our model using the fermionic final states.}
\end{table}

\begin{table}[htbp]
\centering
\begin{tabular}{>{\centering\arraybackslash}p{6cm}|>{\centering\arraybackslash}p{1.5cm}||>{\centering\arraybackslash}p{2.2cm}|>{\centering\arraybackslash}p{3.5cm}|>{\centering\arraybackslash}p{2cm}}
\toprule
Channel & $\sqrt{s}$ [${\rm TeV}$] & Experiment & Mass Range [${\rm TeV}$] & $\mathcal{L}~[{\rm fb}^{-1}]$ \\
\toprule
$gg\to\phi^0\to\gamma\gamma$ & 8 & ATLAS~\cite{ATLAS:2014jdv} & [0.065,0.6] & 20.3 \\
\colrule
$pp\to\phi^0\to\gamma\gamma$ & 13 & ATLAS~\cite{ATLAS:2017ayi} & [0.2,2.7] & 36.7 \\
\colrule
$gg\to\phi^0\to\gamma\gamma$ & 13 & CMS~\cite{CMS:2016kgr} & [0.5,4] & 35.9 \\
\toprule
\multirow{2}{*}{$pp\to\phi^0\to Z\gamma\to(\ell\ell)\gamma$} & \multirow{2}{*}{8}  & ATLAS~\cite{ATLAS:2014lfk} & [0.2,1.6] & 20.3 \\
& & CMS~\cite{CMS:2016all} & [0.2,1.2] & 19.7 \\
\colrule
$gg\to\phi^0\to Z\gamma\big[\to(\ell\ell)\gamma\big]$ & 13 & ATLAS~\cite{ATLAS:2017zdf} & [0.25,2.4] & 36.1 \\
\colrule
$gg\to\phi^0\to Z\gamma\big[\to(qq)\gamma\big]$ & 13 & ATLAS~\cite{ATLAS:2018sxj} & [1,6.8] & 36.1 \\
\colrule
$gg\to\phi^0\to Z\gamma$ & 13 & CMS~\cite{CMS:2017dyb} & [0.35,4] & 35.9 \\
\toprule
$gg\to\phi^0\to ZZ$ & 8 & ATLAS~\cite{ATLAS:2015pre} & [0.14,1] & 20.3 \\
\colrule
$VV\to\phi^0\to ZZ$ & 8 & ATLAS~\cite{ATLAS:2015pre} & [0.14,1] & 20.3 \\
\colrule
\multirow{2}{*}{$gg\to\phi^0\to ZZ\big[\to(\ell\ell)(\ell\ell,\nu\nu)\big]$} & \multirow{2}{*}{13} & ATLAS~\cite{ATLAS:2017tlw} & [0.2,1.2] & 36.1 \\
& & ATLAS~\cite{ATLAS:2020tlo} & [0.2,2] & 139 \\
\colrule
\multirow{2}{*}{$VV\to\phi^0\to ZZ\big[\to(\ell\ell)(\ell\ell,\nu\nu)\big]$} & \multirow{2}{*}{13} & ATLAS~\cite{ATLAS:2017tlw} & [0.2,1.2] & 36.1 \\
& & ATLAS~\cite{ATLAS:2020tlo} & [0.2,2] & 139 \\
\colrule
$gg\to\phi^0\to ZZ\big[\to(\ell\ell,\nu\nu)(qq)\big]$ & 13 & ATLAS~\cite{ATLAS:2017otj} & [0.3,3] & 36.1 \\
\colrule
$VV\to\phi^0\to ZZ\big[\to(\ell\ell,\nu\nu)(qq)\big]$ & 13 & ATLAS~\cite{ATLAS:2017otj} & [0.3,3] & 36.1 \\
\colrule
$pp\to\phi^0\to ZZ\big[\to(\ell\ell)(qq,\nu\nu,\ell\ell)\big]$ & 13 & CMS~\cite{CMS:2018amk} & [0.13,3] & 35.9 \\
\colrule
$pp\to\phi^0\to ZZ\big[\to(qq)(\nu\nu)\big]$ & 13 & CMS~\cite{CMS:2018ygj} & [1,4] & 35.9 \\
\botrule
\end{tabular}
\caption{\label{DS:2} Neutral heavy scalar searches relevant to our model using the $\gamma\gamma$, $Z\gamma$, and $ZZ$ final states, with $\ell=e,\mu$.}
\end{table}

\begin{table}[htbp]
\centering
\begin{tabular}{>{\centering\arraybackslash}p{6cm}|>{\centering\arraybackslash}p{1.5cm}||>{\centering\arraybackslash}p{2.2cm}|>{\centering\arraybackslash}p{3.5cm}|>{\centering\arraybackslash}p{2cm}}
\toprule
Channel & $\sqrt{s}$ [${\rm TeV}$] & Experiment & Mass Range [${\rm TeV}$] & $\mathcal{L}~[{\rm fb}^{-1}]$ \\
\toprule
$gg\to\phi^0\to WW$ & 13 & ATLAS~\cite{ATLAS:2015iie} & [0.3,1.5] & 20.3 \\
\colrule
$VV\to\phi^0\to WW$ & 13 & ATLAS~\cite{ATLAS:2015iie} & [0.3,1.5] & 20.3 \\
\colrule
$gg\to\phi^0\to WW\big[\to(e\nu)(\mu\nu)\big]$ & 13 & ATLAS~\cite{ATLAS:2017uhp} & [0.25,4] & 36.1 \\
\colrule
$VV\to\phi^0\to WW\big[\to(e\nu)(\mu\nu)\big]$ & 13 & ATLAS~\cite{ATLAS:2017uhp} & [0.25,3] & 36.1 \\
\colrule
$(gg+VV)\to\phi^0\to WW\to(\ell\nu)(\ell\nu)$ & 13 & CMS~\cite{CMS:2016jpd} & [0.2,1] & 2.3 \\
\colrule
$gg\to\phi^0\to WW\big[\to(\ell\nu)(qq)\big]$ & 13 & ATLAS~\cite{ATLAS:2017jag} & [0.3,3] & 36.1 \\
\colrule
$VV\to\phi^0\to WW\big[\to(\ell\nu)(qq)\big]$ & 13 & ATLAS~\cite{ATLAS:2017jag} & [0.3,3] & 36.1 \\
\colrule
$pp\to\phi^0\to WW\big[\to(\ell\nu)(qq,\ell\nu)\big]$ & 13 & CMS~\cite{CMS:2019bnu} & [0.2.3] & 35.9 \\
\colrule
$VV\to\phi^0\to WW\big[\to(\ell\nu)(qq,\ell\nu)\big]$ & 13 & CMS~\cite{CMS:2019bnu} & [0.2.3] & 35.9 \\
\toprule
$pp\to\phi^0\to VV$ & 8 & CMS~\cite{CMS:2015hra} & [0.145,1] & 24.8 \\
\botrule
\end{tabular}
\caption{\label{DS:3} Neutral heavy scalar searches relevant to our model using the $WW$ and $VV$ final states, with $V=W,Z$ and $\ell=e,\mu$.}
\end{table}

\begin{table}[htbp]
\centering
\begin{tabular}{>{\centering\arraybackslash}p{6cm}|>{\centering\arraybackslash}p{1.5cm}||>{\centering\arraybackslash}p{2.2cm}|>{\centering\arraybackslash}p{3.5cm}|>{\centering\arraybackslash}p{2cm}}
\toprule
Channel & $\sqrt{s}$ [${\rm TeV}$] & Experiment & Mass Range [${\rm TeV}$] & $\mathcal{L}~[{\rm fb}^{-1}]$ \\
\toprule
$gg\to\phi^0\to hh$ & 8 & ATLAS~\cite{ATLAS:2015sxd} & [0.26,1] & 20.3 \\
\colrule
$pp\to\phi^0\to hh\to(bb)(bb)$ & 8 & CMS~\cite{CMS:2015jal} & [0.27,1.1] & 17.9 \\
\colrule
$pp\to\phi^0\to hh\to(bb)(\gamma\gamma)$ & 8 & CMS~\cite{CMS:2016cma} & [0.26,1.1] & 19.7 \\
\colrule
$gg\to\phi^0\to hh\to(bb)(\tau\tau)$ & 8 & CMS~\cite{CMS:2015uzk} & [0.26,0.35] & 19.7 \\
\colrule
$pp\to\phi^0\to hh\big[\to(bb)(\tau\tau)\big]$ & 13 & CMS~\cite{CMS:2017yfv} & [0.35,1] & 18.3 \\
\colrule
\multirow{2}{*}{$pp\to\phi^0\to hh\to(bb)(bb)$} & \multirow{2}{*}{13}  & ATLAS~\cite{ATLAS:2018rnh} & [0.26,3] & 36.1 \\
& & CMS~\cite{CMS:2018qmt} & [0.26,1.2] & 35.9 \\
\colrule
$pp\to\phi^0\to hh\big[\to(bb)(\gamma\gamma)\big]$ & \multirow{2}{*}{13}  & ATLAS~\cite{ATLAS:2018dpp} & [0.26,1] & 36.1 \\
$pp\to\phi^0\to hh\to(bb)(\gamma\gamma)$ & & CMS~\cite{CMS:2018tla} & [0.25,0.9] & 35.9 \\
\colrule
\multirow{2}{*}{$pp\to\phi^0\to hh\to(bb)(\tau\tau)$} & \multirow{3}{*}{13}  & ATLAS~\cite{ATLAS:2018uni} & [0.26,1] & 36.1 \\
& & CMS~\cite{CMS:2017hea} & [0.25,0.9] & 35.9 \\
$pp\to\phi^0\to hh\big[\to(bb)(\tau\tau)\big]$ & & CMS~\cite{CMS:2018kaz} & [0.9,4] & 35.9 \\
\colrule
$pp\to\phi^0\to hh\to(bb)(VV\to\ell\nu\ell\nu)$ & 13 & CMS~\cite{CMS:2017rpp} & [0.26,0.9] & 35.9 \\
\colrule
$gg\to\phi^0\to hh\to(\gamma\gamma)(WW)$ & 13 & ATLAS~\cite{ATLAS:2018hqk} & [0.26,0.5] & 36.1 \\
\colrule
$pp\to\phi^0\to hh$ & 13 & CMS~\cite{CMS:2018ipl} & [0.25,3] & 35.9 \\
\botrule
\end{tabular}
\caption{\label{DS:4} Neutral heavy scalar searches relevant to our model using the $hh$ final state, with $V=W,Z$ and $\ell=e,\mu$.}
\end{table}

\begin{table}[htbp]
\centering
\begin{tabular}{>{\centering\arraybackslash}p{6cm}|>{\centering\arraybackslash}p{1.5cm}||>{\centering\arraybackslash}p{2.2cm}|>{\centering\arraybackslash}p{3.5cm}|>{\centering\arraybackslash}p{2cm}}
\toprule
Channel & $\sqrt{s}$ [${\rm TeV}$] & Experiment & Mass Range [${\rm TeV}$] & $\mathcal{L}~[{\rm fb}^{-1}]$ \\
\toprule
$gg\to\phi^0\to hZ\to(bb)Z$ & 8 & ATLAS~\cite{ATLAS:2015kpj} & [0.22,1] & 20.3 \\
\colrule
$gg\to\phi^0\to hZ\to(bb)(\ell\ell)$ & 8 & CMS~\cite{CMS:2015flt} & [0.225,0.6] & 19.7 \\
\colrule
$gg\to\phi^0\to hZ\to(\tau\tau)Z$ & 8 & ATLAS~\cite{ATLAS:2015kpj} & [0.22,1] & 20.3 \\
\colrule
$gg\to\phi^0\to hZ\to(\tau\tau)(\ell\ell)$ & 8 & CMS~\cite{CMS:2015uzk} & [0.22,0.35] & 19.7 \\
\colrule
\multirow{3}{*}{$gg\to\phi^0\to hZ\to(bb)Z$} & \multirow{3}{*}{13}  & ATLAS~\cite{ATLAS:2017xel} & [0.2,2] & 36.1 \\
& & CMS~\cite{CMS:2018xvc} & [0.22,0.8] & 35.9 \\
& & CMS~\cite{CMS:2018ljc} & [0.8,1] & 35.9 \\
\colrule
\multirow{3}{*}{$bb\to\phi^0\to hZ\to(bb)Z$} & \multirow{3}{*}{13}  & ATLAS~\cite{ATLAS:2017xel} & [0.2,2] & 36.1 \\
& & CMS~\cite{CMS:2018xvc} & [0.22,0.8] & 35.9 \\
& & CMS~\cite{CMS:2018ljc} & [0.8,1] & 35.9 \\
\colrule
$gg\to\phi^0\to hZ\to(\tau\tau)(\ell\ell)$ & 13 & CMS~\cite{CMS:2019kca} & [0.22,0.4] & 35.9 \\
\toprule
$pp\to\phi^0\to {\phi^0}'Z\to(bb)(\ell\ell)$ & 8 & CMS~\cite{CMS:2016xnc} & [0.13,1] & 19.8 \\
\colrule
$gg\to\phi^0\to {\phi^0}'Z\to(bb)Z$ & 13 & ATLAS~\cite{ATLAS:2018oht} & [0.13,0.8] & 36.1 \\
\colrule
$bb\to\phi^0\to {\phi^0}'Z\to(bb)Z$ & 13 & ATLAS~\cite{ATLAS:2018oht} & [0.13,0.8] & 36.1 \\
\botrule
\end{tabular}
\caption{\label{DS:5} Neutral heavy scalar searches relevant to our model using the $hZ$ and ${\phi^0}'Z$ final states, with $\ell=e,\mu$.}
\end{table}

\begin{table}[htbp]
\centering
\begin{tabular}{>{\centering\arraybackslash}p{7cm}|>{\centering\arraybackslash}p{1.5cm}||>{\centering\arraybackslash}p{2.2cm}|>{\centering\arraybackslash}p{3.5cm}|>{\centering\arraybackslash}p{2cm}}
\toprule
Channel & $\sqrt{s}$ [${\rm TeV}$] & Experiment & Mass Range [${\rm TeV}$] & $\mathcal{L}~[{\rm fb}^{-1}]$ \\
\toprule
$pp\to\phi^\pm\to \tau^\pm\nu$ & 8 & ATLAS~\cite{ATLAS:2014otc} & [0.18,1] & 19.5 \\
\colrule
$pp\to\phi^\pm\to \tau^\pm\nu$ & 8 & CMS~\cite{CMS:2015lsf} & [0.18,0.6] & 19.7 \\
\colrule
\multirow{3}{*}{$pp\to\phi^\pm\to \tau^\pm\nu$} & \multirow{3}{*}{13}  & ATLAS~\cite{ATLAS:2018gfm} & [0.15,2] & 36.1 \\
& & CMS~\cite{CMS:2016szv} & [0.18,3] & 12.9 \\
& & CMS~\cite{CMS:2019bfg} & [0.08,3] & 35.9 \\
\toprule
$pp\to\phi^\pm\to tb$ & 8 & ATLAS~\cite{ATLAS:2015nkq} & [0.2,0.6] & 20.3 \\
\colrule
$pp\to\phi^+\to t\overline{b}$ & 8 & CMS~\cite{CMS:2015lsf} & [0.18,0.6] & 19.7 \\
\colrule
\multirow{2}{*}{$pp\to\phi^\pm\to tb$} & \multirow{2}{*}{13} & ATLAS~\cite{ATLAS:2018ntn} & [0.2,2] & 36.1 \\
& & CMS~\cite{CMS:2020imj} & [0.2,3] & 35.9 \\
\toprule
$pp\to\phi^{\pm\pm}\phi^{\mp}\to(W^\pm W^\pm)(W^\mp Z)$ & 13 & ATLAS~\cite{ATLAS:2021jol} & [0.2,0.6] & 139 \\
\toprule
$pp\to\phi^{\pm\pm}\phi^{\mp\mp}\to(W^\pm W^\pm)(W^\mp W^\mp)$ & 13 & ATLAS~\cite{ATLAS:2021jol} & [0.2,0.6] & 139 \\
\toprule
$WZ\to\phi^\pm\to WZ\big[\to(qq)(\ell\ell)\big]$ & 8 & ATLAS~\cite{ATLAS:2015edr} & [0.2,1] & 20.3 \\
\colrule
\multirow{4}{*}{$WZ\to\phi^\pm\to WZ\big[\to(\ell\nu)(\ell\ell)\big]$} & \multirow{4}{*}{13}  & ATLAS~\cite{ATLAS:2018iui} & [0.2,0.9] & 36.1 \\
& & CMS~\cite{CMS:2017fgp} & [0.2,0.3] & 15.2 \\
& & CMS~\cite{CMS:2018ysc} & [0.3,2] & 35.9 \\
& & CMS~\cite{CMS:2021wlt} & [0.2,3] & 137 \\
\toprule
$pp\to\phi^{\pm\pm}\phi^{\mp\mp}\to (W^\pm W^\pm)(W^\mp W^\mp)$ & 13 & ATLAS~\cite{ATLAS:2018ceg} & [0.2,0.7] & 36.1 \\
\toprule
$VV\to\phi^{\pm\pm}\to W^\pm W^\pm\big[\to(\ell^\pm\nu)(\ell^\pm\nu)\big]$ & 8 & CMS~\cite{CMS:2014mra} & [0.2,0.8] & 19.4 \\
\colrule
\multirow{2}{*}{$VV\to\phi^{\pm\pm}\to W^\pm W^\pm\big[\to(\ell^\pm\nu)(\ell^\pm\nu)\big]$}  & \multirow{2}{*}{13} & CMS~\cite{CMS:2017fhs} & [0.2,1] & 35.9 \\
& & CMS~\cite{CMS:2021wlt} & [0.2,3] & 137 \\
\botrule
\end{tabular}
\caption{\label{DS:6} Singly and doubly charged heavy scalar searches relevant to our model, with $V=W,Z$ and $\ell=e,\mu$.}
\end{table}

%=================================================================================================
\clearpage

%=================================================================================================
\end{document}